\pdfoutput=1

\documentclass{article}
\usepackage{preprint,times}

\usepackage{amsmath,amsfonts,bm}

\def\eqref#1{equation~\ref{#1}}

\def\1{\bm{1}}

\DeclareMathAlphabet{\mathsfit}{\encodingdefault}{\sfdefault}{m}{sl}
\SetMathAlphabet{\mathsfit}{bold}{\encodingdefault}{\sfdefault}{bx}{n}

\usepackage{hyperref}
\usepackage{url}
\usepackage{graphicx}
\usepackage{soul}
\usepackage{wrapfig}
\usepackage{tabularx} 
\usepackage{booktabs}
\usepackage{multirow}
\newcolumntype{Y}{>{\raggedright\arraybackslash}X}
\usepackage{listings}
\usepackage{xcolor}
\usepackage{placeins}
\usepackage[most]{tcolorbox}
\usepackage[table]{xcolor}
\usepackage{makecell}

\definecolor{lstframe}{HTML}{2C5F8C}
\definecolor{lstkw}{HTML}{0B5394}
\definecolor{lstcomment}{HTML}{6A7787}
\definecolor{lststring}{HTML}{8B5E00}

\lstdefinestyle{pmjpython}{%
  language=Python,
  basicstyle=\ttfamily\footnotesize,
  keywordstyle=\color{lstkw}\bfseries,
  commentstyle=\color{lstcomment}\itshape,
  stringstyle=\color{lststring},
  showstringspaces=false, breaklines=true, breakatwhitespace=false,
  columns=fullflexible, keepspaces=true, upquote=false,
  aboveskip=2pt, belowskip=2pt,
}

\lstdefinestyle{pmjyaml}{%
  basicstyle=\ttfamily\footnotesize,
  keywordstyle=\color{lstkw}\bfseries,
  commentstyle=\color{lstcomment}\itshape,
  stringstyle=\color{lststring},
  showstringspaces=false, breaklines=true,
  columns=fullflexible, keepspaces=true, upquote=false,
  morecomment=[l]{\#}, morestring=[b]",
  aboveskip=2pt, belowskip=2pt,
}

\lstdefinestyle{pmjbash}{%
  language=bash,
  basicstyle=\ttfamily\footnotesize,
  keywordstyle=\color{lstkw}\bfseries,
  commentstyle=\color{lstcomment}\itshape,
  showstringspaces=false, breaklines=true,
  columns=fullflexible, keepspaces=true, upquote=false,
  aboveskip=2pt, belowskip=2pt,
}

\newtcolorbox{lstbox}[2][]{%
  enhanced, breakable,
  colback=white, colframe=lstframe,
  colbacktitle=lstframe, coltitle=white,
  fonttitle=\bfseries\small,
  arc=1.5mm, boxrule=0.6pt,
  title={#2},
  left=6pt, right=6pt, top=3pt, bottom=3pt,
  #1
}

\title{PowerMarketJax: A JAX Benchmark Suite for Multi-Agent Reinforcement Learning in Power Markets}

\author{%
  \parbox{\dimexpr\textwidth-2\tabcolsep\relax}{\centering
    Zhanhua Pan$^{1\dagger}$, Xin Qin$^{1\dagger}$, Xiao Liu$^{2}$, Zhilong Cao$^{1}$, Jianhong Wang$^{3\ddagger}$, Dawei Qiu$^{1\ddagger}$\thanks{Corresponding author: \texttt{dawei.qiu@ntu.edu.sg}. $^{\dagger}$These authors contributed equally. $^{\ddagger}$Co-supervisors.}\\[10pt]
    \normalfont\small
    $^{1}$Nanyang Technological University, Singapore.
    $^{2}$Cornell University, USA.
    $^{3}$University of Bristol, UK.}
}

\preprintfinalcopy
\begin{document}
\addtocontents{toc}{\protect\setcounter{tocdepth}{-1}}

\maketitle
\lhead{Preprint.}

\begin{abstract}
Power markets are a natural testbed for multi-agent reinforcement learning (MARL), where multiple self-interested participants repeatedly submit bids. A market-clearing mechanism then determines dispatch and prices subject to power grid constraints and market settlement rules. However, existing MARL environments typically focus on a single market setting, implement simplified clearing mechanisms, or rely on CPU-based optimization solvers that slow large-scale training and limit the systematic study of bidding strategies and market behavior. We introduce PowerMarketJax, a benchmark suite for MARL across five power markets: day-ahead wholesale, real-time balancing, ancillary services, peer-to-peer double auctions, and local flexibility. Each environment implements its own clearing, pricing, and settlement rules while providing a common framework for learning and evaluation. 
We find that learned bidding behavior depends strongly on the market design: independent learners can miss better strategies when gains require many agents to change together, when more profitable strategies lie beyond a region of lower profit, or when profits disappear as more agents adopt the same strategy.
PowerMarketJax implements both market simulation and policy training in JAX, allowing the entire pipeline to run on the GPU with 1\,024 $\times$ 1\,200 parallelisms across both environments and market participants, achieving up to 33$\times$ speedup over CPU-based baselines. Our open-source benchmark is available at: \url{https://github.com/powermarketjax/PowerMarketJax}.
\end{abstract}

\section{Introduction}
\label{sec.intro}
Unlike most commodities, electricity cannot be stored efficiently at scale, so power markets are needed to continuously match supply and demand through mechanisms that determine which resources produce and at what price. Historically, these markets involved a relatively small set of large generators and mainly traded energy, making bidding strategies easier to characterize~\citep{david2000-survey}. This structure is now changing. Renewable generation, flexible demand, and energy storage have expanded both who participates and what is traded~\citep{matamala2025-fcr}: alongside energy, markets now trade reserves, balancing, and ancillary services across multiple timescales and grid levels, each with its own clearing, pricing, and settlement rules~\citep{zhu2025-spotmarket}. This growing diversity of participants, products, and market mechanisms makes bidding behavior increasingly difficult to understand, creating an urgent need to study how participants bid and interact across different market settings. The repeated interactions among many market participants make power markets a natural setting for multi-agent reinforcement learning (MARL)~\citep{liang2020-ddpgabm}.

In a MARL formulation, each market participant can be modeled as an agent that observes market information, submits bids, receives clearing outcomes, and updates its bidding strategy over time. Since a participant's profit depends on both its own bid actions and the actions of others through market clearing, bidding in power markets is inherently a multi-agent decision problem. MARL enables participants to learn such bidding strategies through repeated market interactions without requiring explicit knowledge of competitors' exact strategies~\citep{du2021-nash}.

A number of environments have been developed for learning-based bidding in power markets, but important gaps remain for systematic MARL study. Most focus on a single market study, limiting their use as general benchmarks for MARL-based bidding~\citep{yeh2023-sustaingym,harder2025assume,elhelou2023-opengridgym,salazar2026marlem}. A second gap is computational burden. An environment must solve a market-clearing optimization subject to grid physical constraints after every round of bids. Existing implementations typically rely on CPU solvers, making the millions of clearing problems required during MARL training a major computational bottleneck, especially with many agents and parallel environments~\citep{wolgast2024-clearingapprox}. To reduce this computational cost, some environments simplify pricing or settlement rules, but these simplifications can remove important market features that shape bidding incentives~\citep{wolgast2024-clearingapprox}. As a result, it remains difficult to systematically study how bidding strategies emerge and differ across power markets.

Here, we introduce PowerMarketJax, a power market MARL benchmark implemented entirely in JAX~\citep{jax2018github}. It parallelizes both optimization-based market environments and multi-agent policy computation on the GPU. By compiling clearing and learning into a single computation graph, PowerMarketJax avoids repeated CPU solver calls and host-device synchronization at each market step, enabling large-scale exploration of bidding strategies across different markets. This design achieves up to 33$\times$ higher throughput than CPU-based implementations and runs 1\,024 parallel 1\,200-participant market environments at over 27\,000 environment steps per second (Appendix~\ref{app:speed}).

Our contributions are summarized as follows. First, we present a unified MARL benchmark covering five power markets: day-ahead wholesale, real-time balancing, ancillary services, peer-to-peer energy trading, and local flexibility. The benchmark spans different clearing mechanisms, timescales, and grid levels while preserving each market's clearing, pricing, and settlement rules under a common interface. Second, we implement both market-clearing optimization and MARL policy training entirely in JAX, enabling GPU-based parallel computation across environments and agents for large-scale training and evaluation. Third, using this benchmark, we empirically show that market design strongly shapes what independent learners can discover from their own profit signals: profitable outcomes may require coordinated changes by many agents, lie beyond locally unfavorable regions, or disappear as more agents adopt the same strategy. These findings reveal learning challenges that arise from the interaction between strategic incentives and physically constrained power markets.

\section{Related Work}
\label{sec.literature}
\textbf{Power Market RL Benchmarks.}
A growing number of environments support RL for bidding in power markets. At the transmission level, SustainGym~\citep{yeh2023-sustaingym} studies storage bidding in a network-constrained real-time market. ASSUME~\citep{harder2025assume} supports RL-based bidding in European power markets. At the distribution level, OpenGridGym~\citep{elhelou2023-opengridgym} studies competing bidding agents under distribution network constraints. MARLEM~\citep{salazar2026marlem} considers MARL-based storage participation in local energy markets. Collectively, these works demonstrate the potential of (MA)RL for learning bidding strategies under different market mechanisms and physical constraints.

However, existing environments typically focus on a single market, limiting systematic study of bidding strategies across diverse power markets. PowerMarketJax instead covers five representative markets spanning transmission and distribution levels. It further provides a unified framework for agent--market interaction while preserving each market's clearing, pricing, settlement, and physical constraints. Finally, existing environments commonly separate CPU-based market simulation from GPU-based policy learning, creating repeated CPU--GPU data transfers and making market clearing a major bottleneck for MARL training. PowerMarketJax implements both market clearing and agent policy computation in JAX, enabling parallel execution on the GPU.

\textbf{JAX-based RL Benchmarks.}
A growing number of RL benchmarks have been implemented in JAX, motivated by the need for fast and highly parallel simulation~\citep{rutherford2023-jaxmarl}. Isaac Gym~\citep{makoviychuk2021-isaacgym} showed that moving environment simulation onto the GPU can reduce CPU--GPU communication overhead. JAX~\citep{jax2018github} further supports accelerator-native RL through compilation and vectorization. These advances have enabled high-throughput JAX environments across video games~\citep{koyamada2023-pgx,radji2025-octax}, physics~\citep{freeman2021-brax}, robot learning~\citep{zakka2025-mujocoplayground}, and autonomous driving~\citep{gulino2023-waymax}.

Recent work has also brought JAX-based acceleration to economic and trading environments. EconoJax~\citep{ponse2024-econojax} simulates agent-based markets with order matching, JAX-LOB~\citep{frey2023-jaxlob} accelerates limit-order-book simulation. These environments primarily rely on predefined if--else rule-based matching logic rather than constrained optimization. In contrast, power markets provide a distinctive setting where strategic economic decisions are tightly coupled to a real-world physical grid, requiring market outcomes to satisfy both market rules and engineering constraints. PowerMarketJax targets this setting by implementing optimization-based market clearing and strategic bidding entirely in JAX across five representative power markets.

\section{Background} 
\label{sec.bg}
\begin{wrapfigure}{r}{0.60\textwidth}
    \centering
    \includegraphics[width=\linewidth]{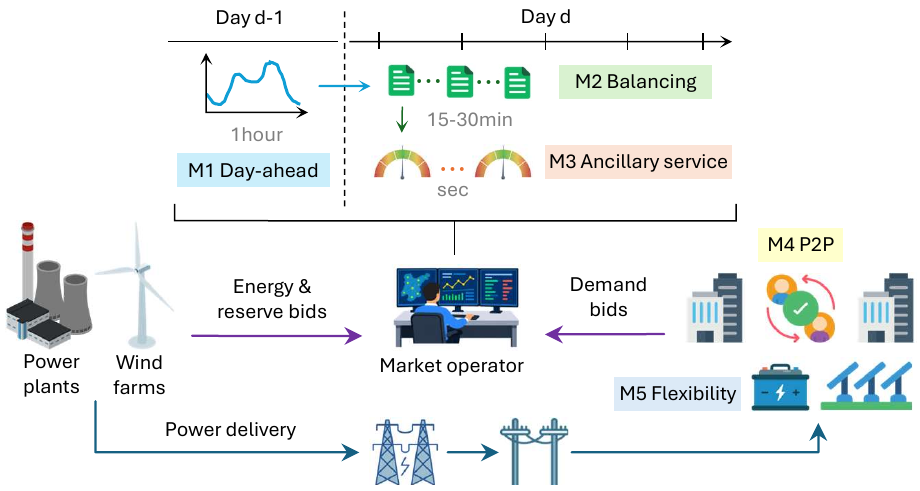}
    \caption{Five power markets in PowerMarketJax across different timescales and grid levels.}
    \label{fig:powermarket}
\end{wrapfigure}

Power systems maintain supply-demand balance through a range of markets for energy, reserves, and flexibility operating at different timescales and grid levels. Following the market structures and flexibility mechanisms discussed in~\citet{kirschen2026fundamentals}, this benchmark considers five representative settings: day-ahead wholesale, real-time balancing, ancillary services, peer-to-peer (P2P) energy trading, and local flexibility, as illustrated in Figure \ref{fig:powermarket}. They cover dispatch scheduling against day-ahead forecasts, adjustment to real-time imbalances, procurement of system reserves, local energy exchanges among prosumers, and distribution network support from flexible resources.

These markets share a common decision structure but differ in their mechanisms. In each market, self-interested participants such as generators, storages, and consumers submit bids based on their asset states and public market signals, without knowing competitors' private information. Their bids jointly determine market clearing, prices, and profits. However, the five markets differ in their traded products, information structures, physical constraints, and clearing and settlement rules. These differences shape participants' incentives and bidding behavior, ultimately affecting market outcomes.

\textbf{M1: Day-ahead wholesale} (Day $d-1$, transmission level).
Generation must be scheduled before delivery day $d$ because thermal units require time to start up and are subject to time-coupling constraints such as minimum up/down times. The day-ahead market clears supply--demand bids using forecasts for day $d$, determining unit commitment, generation schedules, and locational marginal prices (LMPs) that reflect transmission constraints. Cleared schedules then establish the baseline for subsequent real-time balancing.

\textbf{M2: Real-time balancing} (Day $d$, transmission level).
Actual demand and renewable generation may deviate from day-ahead forecasts because weather and consumption patterns cannot be predicted exactly one day in advance. The real-time market redispatches available resources over short intervals, typically every 15--30 minutes on day $d$, to maintain supply-demand balance under updated system conditions. Deviations from day-ahead schedules are settled at real-time prices, which can be more volatile and therefore expose participants to greater price risk.

\textbf{M3: Ancillary services} (Day $d$, transmission level).
Scheduled energy alone cannot ensure reliable operation because system disturbances such as a sudden generator outage can unbalance the system within seconds, far faster than the 15--30-minute energy dispatch interval. The ancillary services market procures capabilities such as frequency regulation and operating reserves, with requirements on availability and response speed. In this market, service providers are paid for reserved capacity, so their revenue can depend on maintaining available capacity rather than solely on energy production.

\textbf{M4: Peer-to-peer energy trading} (Day $d$, distribution level).
Prosumers with rooftop solar and storage can trade energy directly with nearby consumers instead of relying solely on upstream retailers. The P2P market matches buy bids and sell offers within a local energy community. The market-clearing price is determined by the intersection of the aggregated supply and demand curves, while matched quantities are determined from the submitted buy and sell quantities. Unmatched demand is supplied by the upstream grid at the higher retail price, while unmatched supply is exported to the grid at the lower feed-in tariff, creating an economic incentive for local peer-to-peer trading.

\textbf{M5: Local flexibility} (Day $d$, distribution level) 
Even when sufficient electricity is available from the transmission system, distribution networks may become overloaded or local voltages may violate operating limits. To address these issues, the distribution system operator (DSO) procures flexibility from local resources, paying them to increase or decrease generation or consumption at specified locations, times, and durations. Under the pay-as-bid rule considered here, accepted flexibility providers are paid their submitted bid prices rather than a common clearing price.

\section{PowerMarketJax: Power Market Environments in JAX}
\label{sec.framework}
\begin{figure}[h!]
    \centering
    \includegraphics[width=1.00\textwidth]{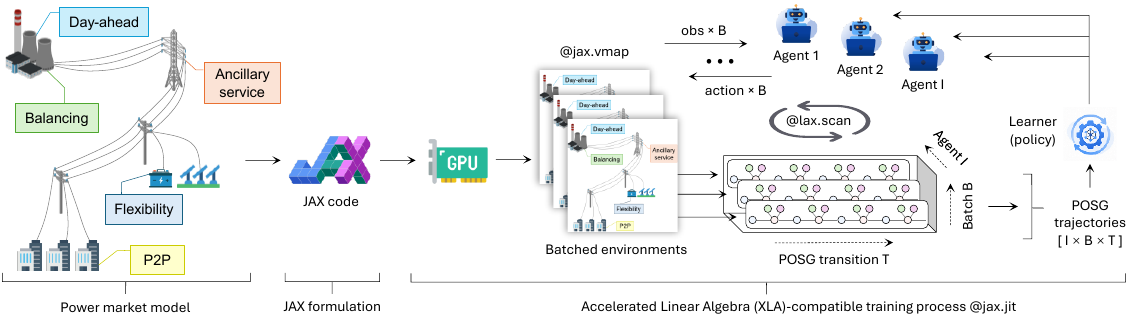}
    \caption{Execution paradigm of PowerMarketJax.}
    \label{fig:paradigm} 
\end{figure}

PowerMarketJax is a suite of five power market environments implemented in JAX, with each market formulated as a MARL environment, as illustrated in Figure~\ref{fig:paradigm}. The environments span day-ahead wholesale (M1), real-time balancing (M2), ancillary services (M3), peer-to-peer energy trading (M4), and local flexibility (M5). The markets preserve their distinct clearing, pricing, settlement, and physical constraints while following a common agent--market interaction interface.

Despite their different market rules, all five environments follow the same step sequence. Agents first observe the information available to them and submit their market actions. The market then clears according to its specific mechanism, the resulting dispatches and prices are settled, and the environment returns the next states and rewards. We define the reward as the participant profit under the corresponding settlement rule. Together, these operations define one transition of the corresponding POSG and are implemented through a common \texttt{step} interface. Further details of the five environments and their accelerator-based execution are provided below.

\subsection{Five market environments as POSGs}
We formulate each market environment as a general-sum partially observable stochastic game (POSG)~\citep{hansen2004-posg}. For each environment, we describe the market step, agent observations and actions, the clearing mechanism, reward, and state transition. Detailed market models and complete POSG formulations are provided in Appendix~\ref{app:da}--\ref{app:flex}.

\textbf{M1: Day-ahead wholesale.} 
One step corresponds to one day-ahead market clearing, in which all 24 hours of the following day are cleared jointly. Each generator agent observes its initial commitment status, initial power output, and residual minimum up- and down-time requirements, together with the public day-ahead demand forecast. Its action consists of the prices submitted across all bid segments and delivery periods. Given the joint bids, the market-clearing mechanism determines unit commitment, dispatches, and LMPs, and each generator receives its daily profit as the reward.

\textbf{M2: Real-time balancing.} 
One step corresponds to one 30-minute real-time market clearing during the delivery day. Each generator agent observes its previous realized dispatch, the day-ahead commitment and schedule for the corresponding hour, the associated day-ahead LMP, and the public realized net-demand conditions. Its action consists of the prices submitted across all bid segments for the current real-time interval. Given the joint bids from all generators, the market-clearing mechanism redispatches the committed generators subject to network and ramping constraints and determines the real-time LMPs. Each generator receives its realized interval profit under the two-settlement rule as the reward, where the day-ahead schedule remains settled at the day-ahead LMP and only deviations are settled at the real-time LMP. The realized dispatch of the current interval is carried forward as the previous dispatch for the next interval.

\textbf{M3: Ancillary services.} 
One step corresponds to one 30-minute joint energy--reserve market clearing during the delivery day. Each generator agent observes its previous realized dispatch, the day-ahead commitment and schedule for the corresponding hour, the associated day-ahead LMP, the public realized net-demand conditions, and the published reserve requirements. Its action consists of the energy bid prices across all bid segments together with one reserve bid price for each reserve product. Given the joint bids of both energy and reserve, the market-clearing mechanism co-optimizes energy dispatch and reserve procurement subject to network, ramping, response time, and reserve capacity constraints, and determines both real-time LMPs and reserve prices. Each generator then receives its realized interval profit as the reward, including day-ahead and real-time energy settlements and reserve capacity payments. The realized dispatch of the current interval is carried forward as the previous dispatch for the next interval.

\textbf{M4: Peer-to-peer energy trading.} 
One step corresponds to one 15-minute P2P trading interval. Each prosumer agent observes its own battery state of charge, net energy position, and the public retail and export prices, where a positive net position corresponds to a buy bid and a negative net position to a sell offer. It does not observe the private states or contemporaneous bids and asks of other prosumers. Its trading quantity is fixed before the auction based on its local generation, demand, and battery operation, so its action is the price submitted as an ask when selling or a bid when buying. Given the joint submissions, the double-auction mechanism sorts sellers and buyers, determines the matched quantities and uniform local clearing price, and settles any unmatched surplus or deficit with the upstream grid. Each prosumer then receives its realized interval profit as the reward, accounting for both P2P and residual grid transactions. The battery state of charge evolves to the next interval, while generation and demand update the prosumer's next net energy position.

\textbf{M5: Local flexibility.} 
One step corresponds to one 1-hour local flexibility procurement interval. Each aggregator agent observes its own battery state of charge, photovoltaic generation, local demand, and available flexibility, together with the public voltage and line-congestion requirements published by the DSO. It does not observe the private operating states, delivery costs, or contemporaneous bids of competing aggregators. Its action is a price--quantity bid specifying the requested payment and the maximum flexibility quantity it is willing to provide. Given the joint offers, the DSO clears the market by minimizing procurement cost subject to network voltage, line flow, and flexibility delivery constraints, and may partially accept a bid. Each aggregator then receives its realized interval profit as the reward under the pay-as-bid rule, with accepted flexibility paid at the submitted bid price. The accepted flexibility changes the battery state of charge and therefore determines the flexibility available in the next interval, while photovoltaic generation, local demand, and network conditions evolve to form the next market state.

\subsection{Clearing and learning in a single computation graph}
Despite their different clearing and settlement mechanisms, all five environments share the same accelerator-based execution framework, with market simulation and policy learning running within a single compiled computation graph.

Every operation within a training iteration, including market clearing, runs inside a single compiled computation graph on the device. \texttt{reset} and \texttt{step} are pure functions: each step sequentially applies action submission, market clearing, settlement, and observation construction, while the environment state is represented as a fixed-structure \texttt{pytree} that is passed as input and returned as output. Each step is compiled with \texttt{jit}, vectorized across parallel environments with \texttt{vmap}, and repeated over a fixed rollout horizon using \texttt{lax.scan}. The policy is evaluated within the same graph: during rollout, it maps observations to actions that are passed directly to the environment step. Policy updates are implemented with a second \texttt{lax.scan} over epochs and minibatches. A complete training iteration, consisting of rollout and update, is compiled as a single function. Further details of the compiled execution graph are provided in Appendix~\ref{app:api:graph}.

Market clearing remains inside the compiled graph because each clearing mechanism is expressed with fixed shape operations and iteration counts. The optimization-based markets use a common primal--dual interior-point solver for linear programs, with a fixed number of Newton iterations. The day-ahead market additionally includes binary commitment decisions. Rather than invoking a mixed-integer solver inside the compiled graph, we use a three-stage procedure. We first relax the commitment problem to a linear program, round the resulting commitment decisions, and then re-solve the continuous dispatch problem with the rounded commitment fixed (Appendix~\ref{app:da:clearing}). It is noted that the P2P market does not require an optimization solver. Its sorting, matching, and settlement operations are implemented directly using JAX Primitives~\citep{jax2018github}.

We validate these market-clearing implementations in three ways (Appendix~\ref{app:speed:validation}). First, clearing and settlement results are compared against a NumPy reference implementation. Second, dual prices produced by the JAX linear-program solver are compared with those obtained from HiGHS~\citep{huangfu2018-highs}, an independent high-performance optimization solver. Third, the three-stage procedure is compared with the exact mixed-integer solution from HiGHS, with the commitment decisions kept binary. The resulting optimality gap is reported in Appendix~\ref{app:speed:gap}.

We parallelize computation across both environments and agents. Each environment has its own market state and random key, and \texttt{vmap} batches multiple environments for parallel execution. Within each environment, agents remain coupled through the market-clearing mechanism, since their submitted actions jointly determine the market outcome. For policy evaluation, agents are also vectorized, so the actions of all agents across all parallel environments are computed simultaneously on the GPU in a single batched policy evaluation.

\section{Results}
\label{sec.results}
\textbf{Systems and data.}
The three transmission-level markets (M1--M3) use the British \texttt{case29gb} system\footnote{\url{https://webhomes.maths.ed.ac.uk/OptEnergy/NetworkData/reducedGB/}} with 29 buses and 66 generators as the main test case (Appendix~\ref{app:results:da:data}). They use one year of Great Britain demand and day-ahead forecast data, with 329 days for training and 36 held-out days for evaluation. M2 and M3 inherit the commitment and schedules obtained from the day-ahead market M1. Additional experiments on the 73-unit RTS-GMLC system\footnote{\url{https://github.com/GridMod/RTS-GMLC}} and the 151-unit Australian NEM system\footnote{\url{https://github.com/akxen/egrimod-nem-dataset}} are reported in Appendices~\ref{app:results:da} to~\ref{app:results:as}. At the distribution level, M4 uses 1,200 Belgian households with PV from the Fluvius smart meter dataset\footnote{\url{https://opendata.fluvius.be/}}. M5 uses the 129-bus SwissDN feeder~\citep{zapparoli2025-swissdn} with battery fleets of 24 and 34 aggregators under its 2040 and 2050 scenarios. Detailed market formulations, system parameters, and data descriptions are provided in Appendices~\ref{app:market} and~\ref{app:results}.

\textbf{MARL algorithms.}
We evaluate two independent MARL algorithms, IPPO~\citep{dewitt2020-independent} and SAC~\citep{haarnoja2018-sac} (the independent version), each with parameter sharing (PS) and agent-specific parameters (NoPS), giving four learners: IPPO-PS, IPPO-NoPS, SAC-PS, and SAC-NoPS. Each agent observes only its own available information and maximizes its own market profit, without a centralized critic or inter-agent communication. We use common default hyperparameters across markets and parallel environment rollouts on the GPU. Full training configurations and market-specific hyperparameters are provided in Appendix~\ref{app:train}.

\textbf{Baselines and evaluation.}
We compare the learned policies against truthful bidding and the corresponding untrained networks. Truthful bidding provides a competitive reference, while the untrained network is used to measure the effect of policy training. We further compare the learned policies with market-specific predefined non-learning strategies, including tests over different generator bidding levels in M1--M3, a fixed battery arbitrage schedule in M4, and unilateral flexibility-price deviations in M5. M1--M4 use 3 random seeds per experiment, while M5 uses 3--10 depending on the configuration. Evaluation uses held-out data and averages results across seeds. Full evaluation protocols and additional results are reported in Appendix~\ref{app:results}.

All experiments run on a workstation with RTX 4500 Ada GPUs, described in Appendix~\ref{app:speed}. The market models and data, the evaluation protocol and further results are in Appendices~\ref{app:market} to~\ref{app:results}.

\subsection{Day-ahead wholesale market}
\label{sec.results.da}
\begin{wrapfigure}{r}{0.64\textwidth}
    \centering
    \includegraphics[width=\linewidth]{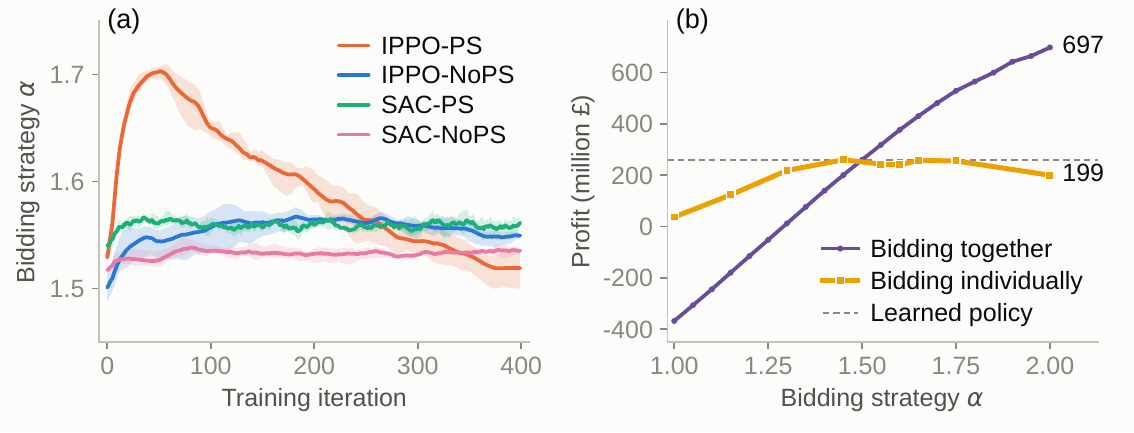}
    \caption{M1 day-ahead wholesale market (a) learning curves, (b) profit evaluation under joint and unilateral bidding strategies.}
    \label{fig.da}
\end{wrapfigure}

We start with the day-ahead market (M1) on \texttt{case29gb}. Each generator $i$ bids at $\alpha_i$ times its marginal generation cost, where $\alpha_i \geq 1$. Here, $\alpha_i=1$ corresponds to truthful bidding, while $\alpha_i>1$ represents bidding above marginal cost. Figure~\ref{fig.da}(a) shows that all four learners end near a bidding strategy around $\alpha_i=1.5$. Over the 36 evaluation days, IPPO-PS and IPPO-NoPS learn to raise market prices with $\alpha_i=1.50$ and $\alpha_i=1.56$, respectively, and thereby increase the total generator profits by \pounds 643 and \pounds 917 million. The other two test systems show the same effect on a smaller scale, with $\alpha_i$ reaching 1.07--1.29 (Appendix~\ref{app:results:da:eval}).

However, the learned policies still leave substantial joint profit unrealized. As shown in Figure~\ref{fig.da}(b), when all 66 generators use the same $\alpha$, total profit reaches the learned policy level near $\alpha=1.5$ and rises to \pounds 697 million at $\alpha=2$. In contrast, when one generator's $\alpha$ is increased at a time while the others keep their learned IPPO-PS bids fixed, the resulting profit never exceeds the learned-policy level and is highest around $\alpha=1.5$--$1.7$. The learners therefore settle near a level where raising the bid further does not benefit an individual generator, even though all generators could gain substantially by raising their bids together.

This gap arises for two reasons. First, the joint gain is not visible in each agent's individual learning signal. An independent learner observes only how its own bidding strategy affects its own profit. For example, cheap nuclear units almost never set the market price (Appendix~\ref{app:results:da:supply}), so raising their bids alone has little effect on price. Second, independent exploration rarely reaches the high-profit joint bidding region. The sampled $\alpha$ values remain concentrated around 1.5, making it unlikely that all units simultaneously explore the higher-bid region where the joint gain occurs.

\subsection{Real-time balancing and ancillary services markets}
\label{sec.results.rt}
\begin{wrapfigure}{r}{0.64\textwidth}
    \centering
    \includegraphics[width=\linewidth]{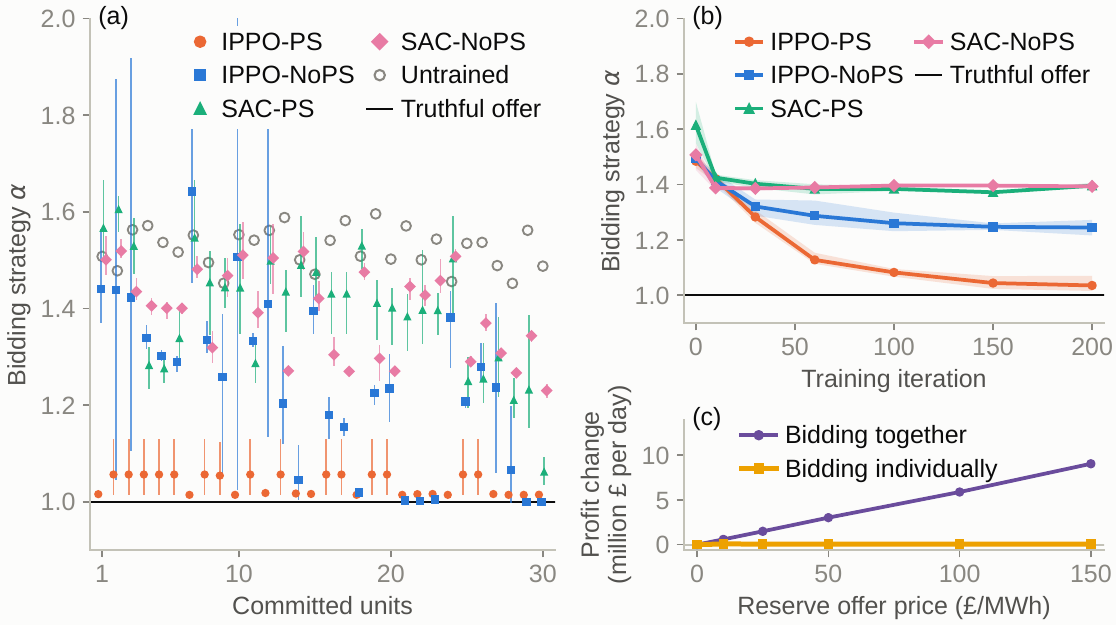}
    \caption{M2 real-time balancing market (a) bidding strategy, (b) learning curves, and M3 ancillary services market (c) profit under joint and unilateral bidding strategies.}
    \label{fig.rt}
\end{wrapfigure}

We next consider the real-time balancing (M2) and ancillary services (M3) markets. In M2, raising a bid alone is rarely profitable because a generator can lose output to its competitors. Accordingly, $\alpha_i$ of the 30 day-ahead committed generators decreases during training under all four learners, reaching 1.04 under IPPO-PS and 1.24--1.40 under the other three methods (Figure~\ref{fig.rt}(a,b)). This contrasts with the day-ahead market, where bidding above cost can increase an individual generator's profit.

The difference between individual and joint incentives is even clearer in M3 (Figure~\ref{fig.rt}(c)). Raising the reserve bid of a single generator produces little gain: across the 29 committed generators, the largest profit increase is only \pounds0.08 million per day. However, when all 66 generators raise their reserve bids together to \pounds150/MWh, total profit increases by \pounds9.05 million per day. However, the learned policies remain far below this joint gain, earning only \pounds0.56--\pounds3.69 million per day more than truthful bidding, and none outperforms its corresponding non-learning strategies.

These results show how markets shape the learning signal available to each independent learner. In M2, raising a bid can reduce a generator's own profit, pushing the learned bid toward cost. In M3, the individual profit signal is nearly zero, making the much larger joint gain difficult to discover.

\subsection{Peer-to-peer energy market}
\label{sec.results.p2p}
We next consider the P2P energy market (M4), where 1,200 households with PV and a battery trade in a double auction every 15 minutes. The learned policies improve household economics relative to truthful bidding. IPPO-NoPS increases household profit on all evaluation days, while IPPO-PS achieves the highest average profit, reducing the daily net cost per household from €1.20 under truthful bidding to €0.69--0.77 (Appendix~\ref{app:results:p2p:eval}).

\begin{figure}[h]
    \centering
    \includegraphics[width=\linewidth]{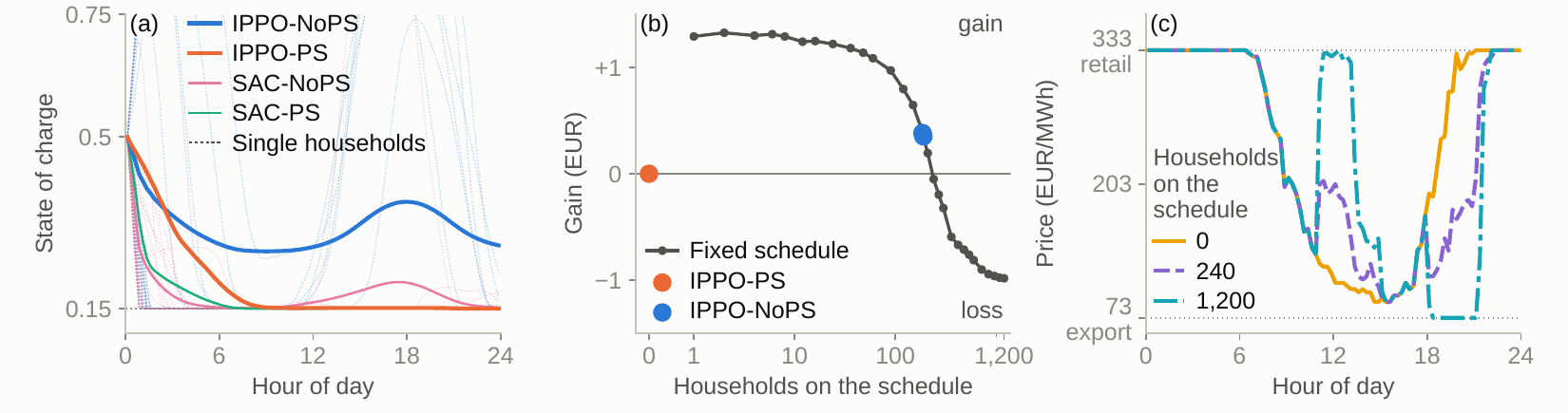}
    \caption{P2P energy market (a) battery state of charge, (b) household gain, (c) local clearing price.}
    \label{fig.p2p}
\end{figure}

The learners, however, adopt different battery strategies (Figure~\ref{fig.p2p}(a)). IPPO-NoPS and SAC-NoPS charge during the midday PV surplus and discharge during the evening demand peak. IPPO-PS and SAC-PS keep their batteries close to the minimum state of charge after selling their initial energy. To understand these differences, we let an increasing number of households follow a fixed schedule that charges from 11:00--15:00 and discharges from 18:00--22:00. A single household gains about €1.13 per day, but this gain decreases as more households adopt the same strategy and becomes negative at around 210--240 households (black line in Figure~\ref{fig.p2p}(b)), detailed in Appendix~\ref{app:results:p2p:crowding}.

The decline is caused by the resulting change in market prices (Figure~\ref{fig.p2p}(c)). Simultaneous charging raises the midday clearing price, while simultaneous discharging lowers the evening price, reducing the price differentials on which the arbitrage relies. Once enough households participate, the remaining differential no longer covers efficiency losses and degradation costs. Thus, a strategy that is profitable for one household may become unprofitable when widely adopted. Figure~\ref{fig.p2p}(b) further suggests that PS can make market outcomes requiring heterogeneous behavior harder to learn: IPPO-NoPS leads only a subset of households to arbitrage, while IPPO-PS leads none to do so.

\subsection{Local flexibility market}
\label{sec.results.flex}
\begin{figure}[h]
    \centering
    \includegraphics[width=\linewidth]{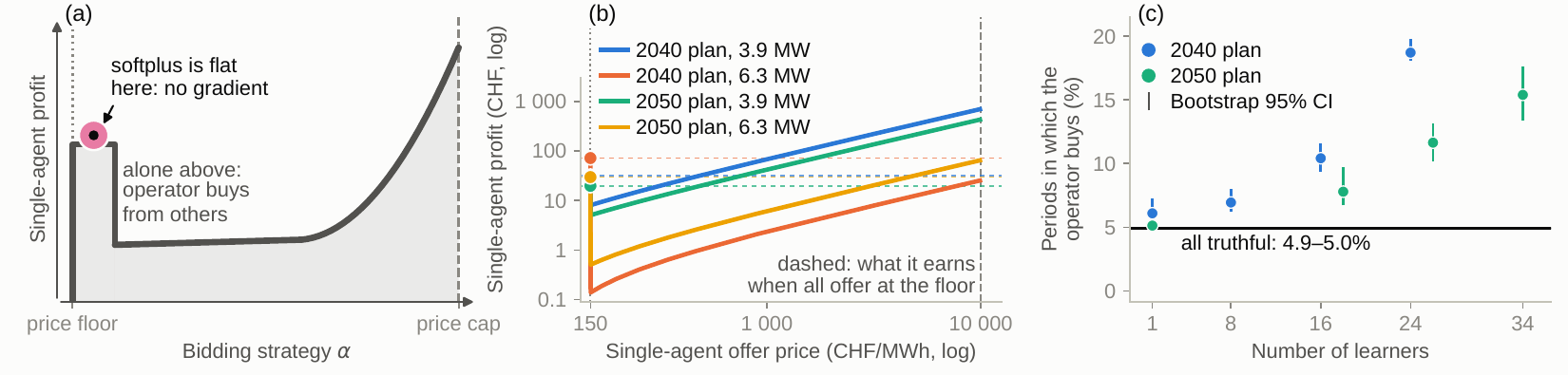}
    \caption{M5 local flexibility market (a) schematic return versus own offer price, (b) measured return in four system configurations, (c) procurement frequency versus the number of SAC-PS learners (seed mean with
    95\% bootstrap CI).}
    \label{fig.flex}
\end{figure}

Finally, we consider the local flexibility market (M5), where the DSO procures flexibility from battery aggregators to relieve network congestion. The market is pay-as-bid with a price floor. All four learners remain close to the floor. Yet a single aggregator can earn more by raising its offer sufficiently far above the floor; in the two 3.9~MW configurations, its return exceeds the floor-level return
again at prices around 3.2 times the floor (Figure~\ref{fig.flex}(a,b)).

The reason is the shape of the individual return. A small bidding price increase from an aggregator causes the operator to procure from other cheaper aggregators instead, sharply reducing the deviating aggregator's award and profit. Only at much higher prices does its return recover. Gradient-based learners therefore encounter a local barrier: small moves away from the floor reduce profit, while the more profitable region lies much farther away. The price parameterization further weakens exploration near the floor because the softplus mapping becomes nearly flat.

The learners instead increase profit through their planned charging decisions. Planned charging raises the baseline flow on the congested line and therefore increases the flexibility requirement that the operator subsequently procures. With SAC-PS learners, the share of periods with procurement rises from 5.0\% to 18.7\% with 24 aggregators and from 4.9\% to 15.4\% with 34 aggregators (Figure~\ref{fig.flex}(c)). A baseline monitor that excludes this learner-induced charging largely removes the effect, causing both procurement and learner returns to fall sharply. Thus, M5 exposes two distinct challenges for MARL in markets: profitable deviations may lie beyond a local learning barrier, and agents can change the demand for the service they are paid to provide.

\section{Conclusion}
\label{sec.conclusion}
PowerMarketJax shows that MARL benchmarks for power markets can preserve market-design fidelity while remaining computationally scalable. By integrating market clearing, settlement, policy evaluation, and learning into a GPU-accelerated JAX pipeline, the benchmark supports fast multi-agent rollouts while retaining the mechanisms and physical constraints that shape strategic bidding. The five markets provide a common POSG framework for evaluating learned bidding behavior and market outcomes across diverse market designs.

\textbf{Discussion.}
The experiments show that learned bidding behavior depends strongly on market design and the profit signal available to each agent. Across the five markets, learners may miss profitable joint outcomes, move toward truthful bidding when individual deviations are unprofitable, lose arbitrage gains when many agents adopt the same strategy, or fail to reach more profitable strategies beyond lower-profit regions. More broadly, these results show that market mechanisms shape the learning landscape itself: they determine which incentives are applied to individual agents and how the actions of many learners feed back into prices, dispatch, and procurement. This reflects challenges for MARL beyond maximizing return, including coordinated exploration, endogenous market responses, and heterogeneous agent behavior. Such learning-based simulations can also help stress-test market designs and reveal strategic incentives before new mechanisms are deployed.

\textbf{Limitations.}
PowerMarketJax is a research benchmark rather than a deployment-ready market simulator or bidding system, and the five environments simplify aspects of real market operation, network modeling, participant constraints, and market rules. Our experiments cover only the considered systems and independent PPO and SAC variants, so the learned policies should not be interpreted as predictions of real participant behavior or as equilibrium strategies. Future work could extend the benchmark to richer market designs, participant models, uncertainty, and broader classes of MARL algorithms.

\section*{Acknowledgement}
Dawei Qiu is supported by the Nanyang Technological University (NTU) Start-Up Grant project \#026749-00001 “Market Design for Low-Carbon Power Systems: Towards a Reliable, Affordable, and Resilient Transition”. Jianhong Wang is supported by the Engineering and Physical Sciences Research Council (EPSRC) [Grant Ref: EP/Y028732/1].

\newpage
\bibliography{refs,lit}
\bibliographystyle{preprint}

\newpage
\appendix
\addtocontents{toc}{\protect\setcounter{tocdepth}{2}}
\renewcommand{\contentsname}{Appendix Contents}
\tableofcontents

\clearpage
\section{Market Models and Partially Observable Stochastic Game}
\label{app:market}

\subsection{M1: Day-ahead Wholesale Market}
\label{app:da}
The day-ahead market is a centralized auction held on day $d-1$ that clears all $T$ hourly delivery periods of day $d$ simultaneously, with $\Delta^{\mathrm{da}} = 1~\text{hour}$. Market participants submit their strategic bids, which the market operator clears through a security-constrained unit commitment (SCUC) problem. The resulting schedules are priced and settled using locational marginal prices (LMPs). Section \ref{app:da:clearing} presents the market clearing model, section \ref{app:da:price} introduces the pricing and settlement rules, and section \ref{app:da:mg} formulates the partially observable stochastic game (POSG) for strategic bidding in day-ahead wholesale market.

\subsubsection{Market model}
\label{app:da:clearing}
Each generator agent $i$ submits, for each delivery period $t$, a stepwise supply curve consisting of $K$ segments. Each segment $k$ has quantity
\begin{equation}
    G_{i,k} = \frac{\bar{p}_i - \underline{p}_i}{K}
\end{equation}
and bid price $\pi_{i,k,t}$, where $\underline{p}_i$ and $\bar{p}_i$ are the minimum and maximum outputs of the unit. Bid prices are required to be non-decreasing across segments, i.e., $\pi_{i,k,t} \le \pi_{i,k+1,t}$. Technical parameters, including ramp rates, minimum up- and down-times, start-up costs, and no-load costs, are registered with the market operator and are not part of the per-period bid. Demand is inelastic and treated as exogenous.

The day-ahead market jointly determines unit commitment and dispatch over all $T$ hourly delivery periods by solving
\allowdisplaybreaks
\begin{subequations}
\label{eq:da-model}
\begin{equation}
    \min_{\{g,\,u,\,v,\,w,\,s\}} \; \sum_{t=1}^{T} \bigg[ \sum_{i,k} \pi_{i,k,t}\, g_{i,k,t} + \sum_i \Delta^{\mathrm{da}}\, \mathrm{NL}_i\, u_{i,t} + \mathrm{VOLL} \sum_n s_{n,t} \bigg] + \sum_{i,t} S_i\, v_{i,t},
    \label{eq:da-obj}
\end{equation}
subject to
\begin{gather}
    p_{i,t} = \underline{p}_i \, u_{i,t} + \sum_k g_{i,k,t}, \label{eq:da-output} \\
    0 \le g_{i,k,t} \le G_{i,k}\, u_{i,t}, \label{eq:da-segment} \\
    P_{n,t} = \sum_{i \,:\, \mathrm{bus}(i) = n} p_{i,t} + s_{n,t} - d_{n,t}, \label{eq:da-injection} \\
    \sum_n P_{n,t} = 0 \quad (: \lambda_t), \label{eq:da-balance} \\
    -F \le \mathrm{PTDF} \cdot P_t \le F \quad (: \mu^{+}_{t},\, \mu^{-}_{t}), \label{eq:da-flow} \\
    0 \le s_{n,t} \le d_{n,t} \qquad (: \rho_{n,t}), \label{eq:da-shed} \\
    -R^{\mathrm{dn}}_i \Delta^{\mathrm{da}} - \bar{p}_i\, w_{i,t} \;\le\; p_{i,t} - p_{i,t-1} \;\le\; R^{\mathrm{up}}_i \Delta^{\mathrm{da}} + \underline{p}_i\, v_{i,t}, \label{eq:da-ramp} \\
    u_{i,t} - u_{i,t-1} = v_{i,t} - w_{i,t}, \label{eq:da-transition} \\
    v_{i,t} + w_{i,t} \le 1, \label{eq:da-exclusive} \\
    \sum_{\tau = t - \mathrm{UT}_i + 1}^{t} v_{i,\tau} \le u_{i,t}, \label{eq:da-min-up} \\
    \sum_{\tau = t - \mathrm{DT}_i + 1}^{t} w_{i,\tau} \le 1 - u_{i,t}, \label{eq:da-min-down} \\
    u_{i,t},\,v_{i,t},\,w_{i,t} \in \{0,1\}. \label{eq:da-binary}
\end{gather}
\end{subequations}

The objective (\ref{eq:da-obj}) minimizes the total offered operating cost over the delivery day. The first term represents the cost of accepted energy offers, the second accounts for the no-load cost of committed generators, and the third penalizes involuntary load shedding. The start-up cost $S_i$ is incurred whenever generator $i$ transitions from off to on. Here, $\Delta^{\mathrm{da}}=1~\mathrm{h}$ is the duration of each day-ahead delivery period, $\mathrm{NL}_i$ is the no-load cost, and $\mathrm{VOLL}$ is the value of lost load, which is set sufficiently high so that load shedding is used only as a last resort.

Constraints (\ref{eq:da-output}) and (\ref{eq:da-segment}) define the output of each generator. The binary variable $u_{i,t}$ indicates whether generator $i$ is committed. When $u_{i,t}=1$, its output consists of the minimum generation level $\underline{p}_i$ plus the accepted quantities $g_{i,k,t}$ from its offer segments. Each accepted segment is bounded by $G_{i,k}$, so the total output remains within $[\underline{p}_i, \, \bar{p}_i]$. When $u_{i,t}=0$, all segment quantities and the generator output are zero.

Constraints (\ref{eq:da-injection})--(\ref{eq:da-flow}) impose the network constraints. Equation (\ref{eq:da-injection}) defines the net injection $P_{n,t}$ at bus $n$ as local generation plus load shedding minus exogenous demand $d_{n,t}$. Constraint (\ref{eq:da-balance}) enforces system-wide supply--demand balance. Under the DC power-flow approximation, constraint (\ref{eq:da-flow}) maps the nodal injections to transmission-line flows through the power transfer distribution factor matrix $\mathrm{PTDF}\in\mathbb{R}^{N^{\mathrm{line}}\times N^{\mathrm{bus}}}$ and limits each line flow by its rating $F$. Network losses are not modeled.

Constraint (\ref{eq:da-shed}) bounds load shedding at each bus by the local demand. The variable $s_{n,t}$ provides emergency feasibility when available generation and transmission capacity are insufficient to serve all demand. Because load shedding carries the large penalty $\mathrm{VOLL}$ in the objective, it is used only when lower-cost feasible alternatives are unavailable.

Constraint (\ref{eq:da-ramp}) limits the change in generator output between consecutive delivery periods according to the upward and downward ramp rates $R_i^{\mathrm{up}}$ and $R_i^{\mathrm{dn}}$. The start-up term $\underline p_i v_{i,t}$ allows a unit that starts in period $t$ to move from zero output to at least its minimum generation level, while the shut-down term $\bar p_i w_{i,t}$ allows a unit that shuts down to reduce its output to zero.

Constraints (\ref{eq:da-transition}) and (\ref{eq:da-exclusive}) describe commitment transitions. The binary variables $v_{i,t}$ and $w_{i,t}$ denote start-up and shut-down events, respectively. Equation (\ref{eq:da-transition}) links these events to changes in commitment status, while equation (\ref{eq:da-exclusive}) prevents a generator from starting up and shutting down in the same period.

Constraints (\ref{eq:da-min-up}) and (\ref{eq:da-min-down}) enforce the minimum up- and down-times $\mathrm{UT}_i$ and $\mathrm{DT}_i$. After a start-up, the unit must remain committed for at least $\mathrm{UT}_i$ periods; after a shut-down, it must remain offline for at least $\mathrm{DT}_i$ periods. The initial commitment status $u_{i,0}$, initial output $p_{i,0}$, and any residual minimum up- or down-time requirements from the previous day are treated as exogenous initial conditions.

\textbf{Solution method.} Optimization (\ref{eq:da-model}) is a security-constrained unit commitment (SCUC) problem formulated as a mixed-integer linear program (MILP). Exact MILP solvers rely on dynamic branch-and-bound procedures that are difficult to batch and compile efficiently on accelerators. We therefore use a relax-round-resolve procedure. First, the binary commitment variables are relaxed to $u_{i,t} \in [0,1]$, and the resulting linear program is solved. 
Second, the relaxed commitment is rounded by setting $u^{\mathrm{int}}_{i,t} = 1$ if $u_{i,t} > \varepsilon$ and $u^{\mathrm{int}}_{i,t} = 0$ otherwise, from which the start-up and shut-down indicators $v^{\mathrm{int}}_{i,t}$ and $w^{\mathrm{int}}_{i,t}$ are determined.
Third, these commitment decisions are fixed and the continuous dispatch problem is re-solved as a security-constrained economic dispatch (SCED). Dispatch schedules are obtained from this final solve, while LMPs are computed from the dual variables of its power balance and transmission constraint. This separation follows the common practice of determining commitment first and obtaining dispatch and prices with commitment fixed. 
The optimality gap relative to the exact solution is evaluated offline and reported in Appendix~\ref{app:speed:gap}.

\subsubsection{Pricing and settlement}
\label{app:da:price}
Given the fixed commitment obtained from the solution procedure above, LMPs are computed from the dual variables of the final SCED problem. The LMP at bus $n$ at period $t$ is the marginal cost of serving one additional MWh at that bus and is given by 
\begin{subequations}
\label{eq:da-settle}
\begin{equation}
    \mathrm{LMP}_{n,t} = \lambda_t - \sum_l \big( \mu^{+}_{l,t} - \mu^{-}_{l,t} \big)\, \mathrm{PTDF}_{l,n} - \rho_{n,t},
    \label{eq:da-lmp}
\end{equation}
where the first term is the system marginal energy price, which is common across the transmission network; the second term is the congestion component, which is zero when no transmission constraint is binding and causes nodal prices to differ under congestion; and the third term can be non-zero only when the upper bound on load curtailment is binding. There is no loss component because transmission losses are not modeled.

Cleared quantities $p_{i,t}$ are settled at the LMP of the bus where unit $i$ is located. The revenue, cost, and profit of unit $i$ over the market day are
\begin{align}
    \mathrm{revenue}_i &= \sum_t \Delta^{\mathrm{da}} \cdot \mathrm{LMP}_{\mathrm{bus}(i),t}\; p_{i,t}, \\
    \mathrm{cost}_i &= \sum_t \big[ \Delta^{\mathrm{da}} \cdot \mathrm{TC}_i(p_{i,t}) + \Delta^{\mathrm{da}} \cdot \mathrm{NL}_i\, u_{i,t} \big] + \sum_t S_i\, v_{i,t}, \\
    \mathrm{profit}_i &= \mathrm{revenue}_i - \mathrm{cost}_i,
\end{align}
\end{subequations}
where $\mathrm{TC}_i(p)=\int_0^p \mathrm{MC}_i(x)\,dx$ is the true variable generation cost rate of unit $i$, and
$\mathrm{MC}_i(p)=a_i p^2+b_i p+c_i$ is its true marginal cost. Setting $a_i=0$ recovers a linear marginal cost curve. The true variable generation cost $\mathrm{TC}_i(\cdot)$ is used only to evaluate participant profit and does not enter market clearing; the clearing problem instead uses submitted energy offer prices together with the registered no-load and start-up costs. Settlement does not include uplift payment, so a committed unit may earn negative profit when its energy market revenue does not recover its start-up and no-load costs.

\subsubsection{Partially Observable Stochastic Game}
\label{app:da:mg}
We formulate repeated participation in the day-ahead wholesale market as a general-sum POSG
\begin{equation}
    \mathcal{G}^{\mathrm{DA}} = \left\langle \ \mathcal{I}, \, \mathcal{S}, \{\mathcal{O}_i\}_{i\in\mathcal{I}}, \{\mathcal{A}_i\}_{i\in\mathcal{I}}, \, P, \, \{r_i\}_{i\in\mathcal{I}}, \, \gamma \ \right\rangle,
    \label{eq:da-posg}
\end{equation}
where $\mathcal{I}$ is the set of strategic generator agents, $\mathcal{S}$ is the market state space, $\mathcal{O}_i$ and $\mathcal{A}_i$ are the observation and action spaces of agent $i$, $P$ is the state transition kernel, $r_i$ is the individual profit-based reward, and $\gamma \in [0,1]$ is the discount factor. One episodic market step corresponds to one day-ahead market clearing, in which all $T$ delivery periods of the following day are cleared jointly.

\paragraph{State.}
At market day $d$, the underlying state contains the system information required to clear the following delivery day,
\begin{equation}
    s_d = \left(\, x^{\mathrm{sys}}_d, \{x_{i,d}\}_{i\in\mathcal{I}} \,\right) \in \mathcal{S},
    \label{eq:da-state}
\end{equation}
where $x^{\mathrm{sys}}_d$ contains the day-ahead forecast of system demand $d_{n,t,d}$ and other exogenous system conditions such as renewable generation. The generator state is
\begin{equation}
    x_{i,d} = \left( u^{0}_{i,d}, p^{0}_{i,d}, \tau^{\mathrm{up}}_{i,d}, \tau^{\mathrm{dn}}_{i,d} \right),
    \label{eq:da-obs}
\end{equation}
where $u^{0}_{i,d}$ and $p^{0}_{i,d}$ are the commitment status and power output of generator $i$ entering the delivery day, and $\tau^{\mathrm{up}}_{i,d}$ and $\tau^{\mathrm{dn}}_{i,d}$ denote any residual minimum up- and downtime requirements carried over from the previous day. Other system and generator parameters, such as the PTDF matrix $\mathrm{PTDF}$, transmission limits $F$, generation capacity limits $(\underline{p}_i,G_{i,k})$, and ramp rates $(R_i^{\mathrm{up}},R_i^{\mathrm{dn}})$, are fixed environment parameters known to the market operator.

\paragraph{Observation.}
Each generator $i$ observes its own private information and the public information available before bidding,
\begin{equation}
    o_{i,d} \in \mathcal{O}_i(s_d) = \left( x_{i,d}, x^{\mathrm{pub}}_d \right),
    \label{eq:da-observation}
\end{equation}
where $x^{\mathrm{pub}}_d$ includes the public day-ahead system information of demand forecast $d_{n,t,d}$. An agent does not observe the private generation costs or contemporaneous offers of competing generators before market clearing. Hence, $o_{i,d}$ does not fully reveal $s_d$, making the bidding problem partially observable.

\paragraph{Action.}
The action of generator $i$ is a single markup $\alpha_{i,d}$ on its truthful offer $\pi^{0}_{i,1} \le \cdots \le \pi^{0}_{i,K}$, in which each segment is priced at the unit's true marginal cost $\mathrm{MC}_i$, giving its complete day-ahead price bid over all segments and delivery periods,
\begin{equation}
    a_{i,d} = \alpha_{i,d} \in \mathcal{A}_{i} = [1, \bar{\alpha}], \qquad \pi_{i,k,t,d} = \alpha_{i,d}\, \pi^{0}_{i,k}, \quad k=1,\ldots,K;\; t=1,\ldots,T,
    \label{eq:da-action}
\end{equation}
where $\bar{\alpha}$ is the largest admissible markup. The segment quantities $G_{i,k}$ are fixed as defined in Section~\ref{app:da:clearing}, while the agent chooses the markup. Admissible bids satisfy the market price bounds and the monotonicity rule
\begin{equation}
    \underline{\pi} \le \pi_{i,1,t,d} \le \pi_{i,2,t,d} \le \cdots \le \pi_{i,K,t,d} \le \overline{\pi}, \qquad \forall t .
    \label{eq:da-action-constraint}
\end{equation}
The joint action $a_d=(a_{1,d},\ldots,a_{|\mathcal{I}|,d})$ therefore represents the set of supply bids submitted by all strategic generators for the following delivery day.

\paragraph{Market clearing and outcome.}

Given state $s_d$ and joint action $a_d$, the market operator applies the day-ahead clearing model and pricing rules described in Sections~\ref{app:da:clearing}--\ref{app:da:price}. We denote this mapping by
\begin{equation}
    z_d = \mathcal{C}^{\mathrm{DA}}(s_d,a_d),
\label{eq:da-clearing-map}
\end{equation}
where the market outcome $z_d$ contains the commitment decisions $u_{i,1...T,d}$, accepted generation quantities $g_{i,k,1...T,d}$, dispatch schedules $p_{i,1...T,d}$, load curtailment $s_{n,1...T,d}$, and nodal LMPs $\mathrm{LMP}_{n,1...T,d}$ for all $T$ delivery periods. Through this clearing mechanism, the payoff of one agent depends not only on its own offer but also on the bids submitted by the other agents.

\paragraph{Reward.}
Each agent $i$ receives its realized market profit as reward. Using the settlement defined in Section~\ref{app:da:price},
\begin{align}
    r_{i,d} &= \mathrm{profit}_{i,d} = \mathrm{revenue}_i - \mathrm{cost}_i \nonumber \\ &= \sum_{t=1}^{T} \Delta^{\mathrm{da}} \, \mathrm{LMP}_{\mathrm{bus}(i),t,d} \ p_{i,t,d} \nonumber \\ &- \sum_{t=1}^{T} \left[\, \Delta^{\mathrm{da}}\,\mathrm{TC}_i(p_{i,t,d}) + \Delta^{\mathrm{da}}\,\mathrm{NL}_i u_{i,t,d} \,\right] - \sum_{t=1}^{T} S_i \, v_{i,t,d}.
    \label{eq:da-reward}
\end{align}
Thus, submitted bidding prices affect the reward only through the resulting market-clearing outcomes; the true variable generation cost $\mathrm{TC}_i(\cdot)$ is used only to evaluate the realized cost and then profit. Since each generator maximizes its own profit rather than a common system-wide reward, the resulting POSG is general-sum.

\paragraph{State transition.}
After the delivery periods cleared at market step $d$, the terminal operating condition of each generator is carried into the next market step. Specifically, the initial commitment status and power output for the next day are
\begin{equation}
    u^{0}_{i,d+1} = u_{i,T,d},
    \qquad
    p^{0}_{i,d+1} = p_{i,T,d},
\end{equation}
while the residual minimum up- or downtime requirements $\tau^{\mathrm{up}}_{i,d+1}$ and $\tau^{\mathrm{dn}}_{i,d+1}$ are updated according to the terminal commitment trajectory. Hence, the next generator state
\begin{equation}
    x_{i,d+1} = \left( u^{0}_{i,d+1}, p^{0}_{i,d+1}, \tau^{\mathrm{up}}_{i,d+1}, \tau^{\mathrm{dn}}_{i,d+1} \right)
\end{equation}
depends on the commitment and dispatch outcomes produced by the joint bids at market step $d$.

The exogenous system state $x^{\mathrm{sys}}_{d+1}$, including the next day-ahead forecasts of demand, evolves according to the underlying data process. The overall state transition is therefore
\begin{equation}
    s_{d+1} \sim P(\,\cdot \mid s_d,a_d),
    \label{eq:da-state-transition}
\end{equation}
where the dependence on $a_d$ arises through the cleared commitment and dispatch decisions carried into the next market step.

\paragraph{Agent objective.}
Each generator uses a policy $\sigma_i(a_{i,d}\mid o_{i,d})$ to choose its day-ahead bid and seeks to maximize its expected discounted profit over an episode of $H$ market days,
\begin{equation}
    J_i(\sigma_i,\sigma_{-i}) = \mathbb{E} \left[ \sum_{d=1}^{H} \gamma^{d-1} r_{i,d} \right].
    \label{eq:da-objective}
\end{equation}
where $\sigma_{-i}$ denotes the joint policies of all other generators. Because market clearing depends on the joint bids of all generators, the expected return of agent $i$ depends on both its own policy $\sigma_i$ and the policies of the other agents $\sigma_{-i}$. Each agent therefore learns its bidding policy through repeated market interactions without observing competitors' private information or policies.

\subsection{M2: Real-Time Balancing Market}
\label{app:rt}
The real-time balancing market consists of a sequence of auctions during delivery day $D$, with one auction for each real-time interval $\Delta^{\mathrm{rt}} = 30~\text{mins}$. It takes the commitment and schedules established by the day-ahead wholesale market as fixed and redispatches available resources against realized system demand. Only deviations from the day-ahead schedules are settled at real-time LMPs, while the day-ahead schedules remain settled at day-ahead LMPs. This section presents the real-time balancing market clearing model in subsection \ref{app:rt:clearing}, the two-settlement rule in subsection \ref{app:rt:settle}, and the corresponding POSG in subsection \ref{app:rt:mg}.

\subsubsection{Market model}
\label{app:rt:clearing}
Let $\tau$ index the 30-minute real-time intervals and let $h(\tau)$ denote the corresponding hourly period in the day-ahead market. The day-ahead outcome enters the real-time market as fixed data through the commitment $u^{\mathrm{da}}_{i,h(\tau)}$, scheduled generation $q^{\mathrm{da}}_{i,h(\tau)}$, and day-ahead price $\mathrm{LMP}^{\mathrm{da}}_{n,h(\tau)}$. The realized net demand at bus $n$ is $\tilde{d}_{n,\tau}$. Generators submit price--quantity bids with the same structure as in the day-ahead market, and one market clearing problem is solved for each real-time interval:
\allowdisplaybreaks
\begin{subequations}
\label{eq:rt}
\begin{equation}
    \min_{\{g,\,s\}} \; \sum_{i,k} \pi_{i,k,\tau} \, g_{i,k,\tau} + \mathrm{VOLL} \sum_n s_{n,\tau},
    \label{eq:rt-obj}
\end{equation}
subject to
\begin{gather}
    p_{i,\tau} = \underline p_i\,u^{\mathrm{da}}_{i,h(\tau)} + \sum_k g_{i,k,\tau}, \label{eq:rt-output} \\
    0 \le g_{i,k,\tau} \le G_{i,k} \, u^{\mathrm{da}}_{i,h(\tau)}, \label{eq:rt-segment} \\
    P_{n,\tau} = \sum_{i:\mathrm{bus}(i)=n}p_{i,\tau} + s_{n,\tau} - \tilde d_{n,\tau}, \label{eq:rt-injection} \\
    \sum_n P_{n,\tau} = 0 \qquad (: \lambda_\tau), \label{eq:rt-balance} \\
    -F \le \mathrm{PTDF} \cdot P_\tau \le F \qquad (:\mu_\tau^+,\mu_\tau^-), \label{eq:rt-flow} \\
    0 \le s_{n,\tau} \le \max(\tilde d_{n,\tau},0) \quad (: \rho_{n,\tau}), \label{eq:rt-shed} \\
    -R_i^{\mathrm{dn}} \Delta^{\mathrm{rt}} - \bar p_i w^{\mathrm{da}}_{i,\tau} \le p_{i,\tau} - p_{i,\tau-1} \le R_i^{\mathrm{up}} \Delta^{\mathrm{rt}} +\underline{p}_i v^{\mathrm{da}}_{i,\tau}.
    \label{eq:rt-ramp}
\end{gather}
\end{subequations}

The objective minimizes the as-bid redispatch cost and load-shedding penalty for each real-time interval $\Delta^{\mathrm{rt}}$. Unlike the day-ahead market, unit commitment is not re-optimized: $u^{\mathrm{da}}_{i,h(\tau)}$ is fixed by the day-ahead outcome, and the real-time market only adjusts the dispatch of committed units. The nodal balance and transmission constraints ensure that the resulting dispatch remains physically feasible under realized net demand. Consecutive real-time intervals are coupled through the ramping constraint, where $p_{i,\tau-1}$ is the realized dispatch from the previous interval. The start-up and shut-down indicators $v^{\mathrm{da}}_{i,\tau}$ and $w^{\mathrm{da}}_{i,\tau}$ are determined by changes in the mapped day-ahead commitment and are non-zero only when the commitment status changes between consecutive real-time intervals.

Constraint (\ref{eq:rt-output}) defines the real-time output of generator $i$. Because the commitment $u^{\mathrm{da}}_{i,h(\tau)}$ is fixed by the day-ahead market, only generators committed in the corresponding day-ahead period can produce in real time. Their output equals the minimum generation level plus the accepted quantities from the submitted bid segments. Constraint (\ref{eq:rt-segment}) limits each accepted segment $g_{i,k,\tau}$ to its available quantity $G_{i,k}$. Constraint (\ref{eq:rt-injection}) defines the net injection $P_{n,\tau}$ at bus $n$ as local generation plus load shedding minus realized net demand. Constraint (\ref{eq:rt-balance}) enforces system-wide supply--demand balance, while constraint (\ref{eq:rt-flow}) keeps transmission flows within the line ratings. The associated dual variables $\lambda_\tau$, $\mu_\tau^+$, and $\mu_\tau^-$ are used to determine the real-time LMPs. 

Constraint (\ref{eq:rt-shed}) allows emergency load curtailment only up to the positive net demand at each bus. Hence, $s_{n,\tau}=0$ when renewable generation exceeds local demand. The dual variable $\rho_{n,\tau}$ is associated with the upper bound on load shedding. Constraint (\ref{eq:rt-ramp}) limits changes in generator output between consecutive real-time intervals according to the upward and downward ramp rates. Here, $p_{i,\tau-1}$ is the realized dispatch from the previous real-time interval. The fixed indicators $v^{\mathrm{da}}_{i,\tau}$ and $w^{\mathrm{da}}_{i,\tau}$ account for start-up and shut-down transitions implied by the day-ahead commitment.

Unlike the day-ahead market, the real-time market does not re-optimize unit commitment. The day-ahead commitment is fixed, and the real-time clearing adjusts only the dispatch quantities in response to realized system conditions. Consecutive real-time intervals are coupled through generator ramping.

\subsubsection{Pricing and settlement}
\label{app:rt:settle}
Real-time LMPs are computed from the dual variables of the real-time clearing problem using the same energy--congestion decomposition as in equation (\ref{eq:da-lmp}):
\begin{equation}
    \mathrm{LMP}^{\mathrm{rt}}_{n,\tau} = \lambda_\tau - \sum_l \left( \mu^+_{l,\tau} - \mu^-_{l,\tau} \right) \, \mathrm{PTDF}_{l,n} - \rho_{n,\tau}.
    \label{eq:rt-lmp}
\end{equation}
Unlike day-ahead LMPs, the real-time LMPs reflect realized demand and short-term operating constraints. They can therefore be more volatile and may become negative under local oversupply or binding downward-flexibility constraints.

The market follows a two-settlement rule. The day-ahead schedule remains settled at the day-ahead LMP, while only the deviation from that schedule is settled at the real-time LMP. For generator $i$ in real-time interval $\tau$,
\begin{subequations}
\label{eq:rt-settle}
\begin{align}
    \mathrm{revenue}^{\mathrm{da}}_{i,\tau} &= \Delta^{\mathrm{rt}} \, \mathrm{LMP}^{\mathrm{da}}_{\mathrm{bus}(i),h(\tau)} \, q^{\mathrm{da}}_{i,h(\tau)}, \\
    \mathrm{revenue}^{\mathrm{rt}}_{i,\tau} &= \Delta^{\mathrm{rt}} \, \mathrm{LMP}^{\mathrm{rt}}_{\mathrm{bus}(i),\tau} \left( p_{i,\tau} - q^{\mathrm{da}}_{i,h(\tau)} \right), \\
    \mathrm{cost}_{i,\tau} &= \Delta^{\mathrm{rt}} \, \mathrm{TC}_i(p_{i,\tau}) + \Delta^{\mathrm{rt}} \, \mathrm{NL}_i u^{\mathrm{da}}_{i,h(\tau)} + S_i v^{\mathrm{da}}_{i,\tau}, \\
    \mathrm{profit}_{i,\tau} &= \mathrm{revenue}^{\mathrm{da}}_{i,\tau} + \mathrm{revenue}^{\mathrm{rt}}_{i,\tau} - \mathrm{cost}_{i,\tau}.
\end{align}
\end{subequations}

In this context, a generator producing above its day-ahead schedule sells the positive deviation at the real-time LMP, while a generator producing below its schedule effectively buys back the shortfall at the real-time LMP. The day-ahead revenue, commitment status, and associated no-load and start-up costs are fixed by the day-ahead outcome and do not depend on the real-time bid. They are retained in the reward so that the reported profit represents the full economic outcome of delivery.

\textbf{Remark on the day-ahead and real-time coupling.} 
In practice, the two markets operate at different temporal scales: the day-ahead market is cleared on day $D-1$, while the real-time market is cleared at each interval of delivery day $D$. The commitment $u^{\mathrm{da}}_{i,h(\tau)}$, binding schedule $q^{\mathrm{da}}_{i,h(\tau)}$, and day-ahead price $\mathrm{LMP}^{\mathrm{da}}_{n,h(\tau)}$ are fixed parameters of the real-time clearing, while the real-time market prices deviations arising from uncertainty not captured by the day-ahead forecast.

\subsubsection{Partially Observable Stochastic Game}
\label{app:rt:mg}
The real-time balancing market can be also formulated as a general-sum partially observable stochastic game (POSG),
\begin{equation}
    \mathcal{G}^{\mathrm{rt}} = \left\langle \ \mathcal{I}, \, \mathcal{S}, \{\mathcal{O}_i\}_{i\in\mathcal{I}}, \{\mathcal{A}_i\}_{i\in\mathcal{I}}, \, P, \, \{r_i\}_{i\in\mathcal{I}}, \, \gamma \ \right\rangle,
\end{equation}
where $\mathcal{I}$ is the set of strategic generators. One episodic market step $\tau$ corresponds to one 30-minute real-time interval $\Delta^{\mathrm{rt}}$. For a 24-hour delivery day, the real-time market therefore contains $T^{\mathrm{rt}}=48$ sequential clearing steps.

\paragraph{State.}
The state at interval $\tau$ is
\begin{equation}
    s_\tau = \left(\, x^{\mathrm{sys}}_\tau, \, x^{\mathrm{da}}_\tau, \, \{x_{i,\tau}\}_{i\in\mathcal{I}} \, \right) \in \mathcal{S},
    \label{eq:rt-state}
\end{equation}
where $x^{\mathrm{sys}}_\tau$ contains the realized system conditions used for real-time clearing, i.e., the net bus demand $\tilde{d}_{n,\tau}$. The day-ahead component is
\begin{equation}
    x^{\mathrm{da}}_\tau = \left\{\, u^{\mathrm{da}}_{i,h(\tau)}, \, q^{\mathrm{da}}_{i,h(\tau)}, \, \mathrm{LMP}^{\mathrm{da}}_{n,h(\tau)} \, \right\}_{i,n},
    \label{eq:rt-state-da}
\end{equation}
which contains the fixed commitment $u^{\mathrm{da}}_{i,h(\tau)}$, binding generation schedules $q^{\mathrm{da}}_{i,h(\tau)}$, and nodal prices $\mathrm{LMP}^{\mathrm{da}}_{n,h(\tau)}$ from the corresponding day-ahead period. The operating state of generator $i$ is
\begin{equation}
    x_{i,\tau} = \left(\, p_{i,\tau-1},\, v^{\mathrm{da}}_{i,\tau},\, w^{\mathrm{da}}_{i,\tau} \,\right),
    \label{eq:rt-state-gen}
\end{equation}
where $p_{i,\tau-1}$ is the realized dispatch from the previous real-time interval $\tau-1$, and $v^{\mathrm{da}}_{i,\tau}$ and $w^{\mathrm{da}}_{i,\tau}$ are the start-up and shut-down indicators implied by the mapped day-ahead commitment, which are used in the ramping constraint (\ref{eq:rt-ramp}) to account for output changes associated with start-up and shut-down transitions. Network parameters, generator operating limits, and cost parameters are fixed environment parameters and are omitted from the dynamic state.

\paragraph{Observation.}
Each generator $i$ observes its own operating information together with public market information. Its observation at real-time interval $\tau$ is
\begin{equation}
    o_{i,\tau} = \left(\, p_{i,\tau-1},\, u^{\mathrm{da}}_{i,h(\tau)},\, q^{\mathrm{da}}_{i,h(\tau)}, \,
        \mathrm{LMP}^{\mathrm{da}}_{\mathrm{bus}(i),h(\tau)},\, x^{\mathrm{pub}}_{\tau} \,\right) = \Omega_i(s_\tau) \in \mathcal{O}_{i},
    \label{eq:rt-observation}
\end{equation}
where $p_{i,\tau-1}$ is the generator's realized dispatch in the previous real-time interval $\tau-1$, $u^{\mathrm{da}}_{i,h(\tau)}$ and $q^{\mathrm{da}}_{i,h(\tau)}$ are its day-ahead commitment and schedule, and $\mathrm{LMP}^{\mathrm{da}}_{\mathrm{bus}(i),h(\tau)}$ is the corresponding day-ahead LMP. 
The public real-time market information is denoted by $x^{\mathrm{pub}}_{\tau}$, which is the realized system conditions of net bus demand $x^{\mathrm{sys}}_\tau = \tilde{d}_{n,\tau}$.
A generator does not observe the private parameters or contemporaneous bids of other generators. The market is therefore partially observable from the perspective of each participant.

\paragraph{Action.}
At each real-time interval, generator $i$ submits a stepwise supply bid $\boldsymbol{\pi}_{i,\tau}$, obtained from a single markup $\alpha_{i,\tau}$ on its truthful offer $\pi^{0}_{i,1} \le \cdots \le \pi^{0}_{i,K}$, in which each segment is priced at the unit's true marginal cost $\mathrm{MC}_i$,
\begin{equation}
    a_{i,\tau} = \alpha_{i,\tau} \in \mathcal{A}_i = [1, \bar{\alpha}], \qquad \pi_{i,k,\tau} = \alpha_{i,\tau}\, \pi^{0}_{i,k}, \quad k = 1, \ldots, K,
    \label{eq:rt-action}
\end{equation}
where $\bar{\alpha}$ is the largest admissible markup, subject to the admissible bid-price bounds and the monotonicity condition
\begin{equation}
    \pi_{i,1,\tau} \le \pi_{i,2,\tau} \le \cdots \le \pi_{i,K,\tau}.
\end{equation}
The segment quantities $G_{i,k}$ are fixed, so the strategic action determines the bid prices associated with these quantities. The joint action is $a_\tau=(a_{i,\tau})_{i\in\mathcal{I}}$.

\paragraph{Market clearing and reward.}
Given the current state and joint bids, the market-clearing mechanism
\begin{equation}
    z_\tau = \mathcal{C}^{\mathrm{rt}}(s_\tau,a_\tau)
    \label{eq:rt-clearing-map}
\end{equation}
solves optimization (\ref{eq:rt}) and returns the realized dispatch $p_{i,\tau}$, accepted bid quantities $g_{i,k,\tau}$, and real-time LMPs $\mathrm{LMP}^{\mathrm{rt}}_{n,\tau}$. Each generator receives its realized profit as the stage reward,
\begin{align}
    r_{i,\tau} &= \mathrm{profit}_{i,\tau} = \mathrm{revenue}^{\mathrm{da}}_{i,\tau} + \mathrm{revenue}^{\mathrm{rt}}_{i,\tau} - \mathrm{cost}_{i,\tau} \nonumber \\ &= \Delta^{\mathrm{rt}} \, \mathrm{LMP}^{\mathrm{da}}_{\mathrm{bus}(i),h(\tau)} \, q^{\mathrm{da}}_{i,h(\tau)} + \Delta^{\mathrm{rt}} \, \mathrm{LMP}^{\mathrm{rt}}_{\mathrm{bus}(i),\tau} \left( p_{i,\tau} - q^{\mathrm{da}}_{i,h(\tau)} \right) \nonumber \\ &- \big(\, \Delta^{\mathrm{rt}} \, \mathrm{TC}_i(p_{i,\tau}) + \Delta^{\mathrm{rt}} \, \mathrm{NL}_i u^{\mathrm{da}}_{i,h(\tau)} + S_i v^{\mathrm{da}}_{i,\tau} \,\big).
    \label{eq:rt-reward}
\end{align}
where $\mathrm{profit}_{i,\tau}$ is defined by the two-settlement rule in equation (\ref{eq:rt-settle}). The reward therefore accounts for the fixed day-ahead position, the real-time settlement of deviations, and the realized generation cost.

\paragraph{State transition.}
The clearing result determines the operating state entering the next real-time interval. In particular, the realized dispatch $p_{i,\tau}$ becomes the previous-period dispatch in interval $\tau+1$, so that
\begin{equation}
    x_{i,\tau+1} = \left( p_{i,\tau}, v^{\mathrm{da}}_{i,\tau+1}, w^{\mathrm{da}}_{i,\tau+1} \right).
    \label{eq:rt-transition-gen}
\end{equation}
Thus, the dispatch determined at interval $\tau$ enters the ramping constraint of interval $\tau+1$. The corresponding day-ahead commitment, schedule, and price are updated deterministically according to $h(\tau+1)$, while realized net demand evolves according to the exogenous data process. The overall state transition is therefore
\begin{equation}
    s_{\tau+1} \sim P \left( \cdot \mid s_\tau, a_\tau \right).
    \label{eq:rt-transition}
\end{equation}

\paragraph{Agent objective.}
Each generator uses a policy $\sigma_i(a_{i,\tau}\mid o_{i,\tau})$ to select its real-time bid. Over a
horizon of $H$ real-time intervals, generator $i$ seeks to maximize its expected discounted profit,
\begin{equation}
    J_i(\sigma_i,\sigma_{-i}) = \mathbb{E} \left[ \sum_{\tau=1}^{H} \gamma^{\tau-1} r_{i,\tau} \right],
    \label{eq:rt-objective}
\end{equation}
where $\sigma_{-i}$ denotes the joint policies of the other generators. Because both dispatch and real-time prices are determined jointly by all submitted bids, the return of each generator depends on both its own bidding policy and those of the other participants. Agents therefore learn their real-time bidding policies through repeated market interactions without observing competitors' private information or policies.

\subsection{M3: Ancillary Services Market}
\label{app:as}
The ancillary services market procures operating reserve, where generation capacity is held available to respond to system contingencies. Unlike the energy market, it primarily trades availability: a provider is paid for maintaining reserve capacity that can be activated when needed but may never be called. In some market designs, operating reserve is co-optimized with real-time energy, so energy dispatch and reserve procurement are determined jointly for the same resources subject to shared operating constraints. This section presents the ancillary services market clearing model in subsection \ref{app:as:clearing}, the pricing and settlement rules in subsection \ref{app:as:settle}, and the corresponding POSG in subsection \ref{app:as:mg}.

\subsubsection{Market model}
\label{app:as:clearing}
The market contains $N^{\mathrm{prod}}=2$ reserve products distinguished by their response times: a fast product with $\theta_1 = 10~\mathrm{mins}$ and a slow product with $\theta_2 = 30~\mathrm{mins}$. Before bids are submitted, the system operator publishes the reserve requirement for each product, which can be modeled as a fraction of forecast system demand,
\begin{equation}
    d^{\mathrm{res}}_{j,\tau} = \beta_j \, \sum_n d^{\mathrm{forecast}}_{n,\tau},
    \label{eq:as-req}
\end{equation}
where $d^{\mathrm{forecast}}_{n,\tau}$ is the published demand forecast at bus $n$ and $\beta_j$ is the requirement fraction for reserve product $j$.

For each real-time interval, generator $i$ submits the energy offer described in section~\ref{app:rt:clearing}, together with a reserve offer price $\pi^{\mathrm{res}}_{i,j,\tau} \ge 0$ for each product $j$, expressed in \$/MW-h. The energy and reserve markets are cleared jointly as
\allowdisplaybreaks
\begin{subequations}
\label{eq:as}
\begin{equation}
    \min_{\{g,\,r,\,s,\,s^{\mathrm{res}}\}} \;
    \sum_{i,k}\pi_{i,k,\tau}g_{i,k,\tau} +
    \sum_{i,j}\pi^{\mathrm{res}}_{i,j,\tau}r_{i,j,\tau} +
    \mathrm{VOLL}\sum_n s_{n,\tau} +
    \mathrm{VOLR}\sum_j s^{\mathrm{res}}_{j,\tau},
    \label{eq:as-obj}
\end{equation}
subject to
\begin{gather}
    p_{i,\tau} = \underline p_i\,u^{\mathrm{da}}_{i,h(\tau)} + \sum_k g_{i,k,\tau}, \label{eq:as-output} \\
    0 \le g_{i,k,\tau} \le G_{i,k}\,u^{\mathrm{da}}_{i,h(\tau)}, \label{eq:as-segment} \\
    P_{n,\tau} = \sum_{i:\mathrm{bus}(i)=n}p_{i,\tau} + s_{n,\tau} - \tilde d_{n,\tau}, \label{eq:as-injection} \\
    \sum_n P_{n,\tau} = 0 \qquad (: \lambda_\tau), \label{eq:as-balance} \\
    -F \le \mathrm{PTDF}\cdot P_\tau \le F \qquad (: \mu_\tau^+,\,\mu_\tau^-), \label{eq:as-flow} \\
    0 \le s_{n,\tau} \le \max(\tilde d_{n,\tau},0) \qquad (: \rho_{n,\tau}), \label{eq:as-shed} \\
    -R_i^{\mathrm{dn}}\Delta^{\mathrm{rt}} - \bar p_i w^{\mathrm{da}}_{i,\tau} \le p_{i,\tau}-p_{i,\tau-1} \le R_i^{\mathrm{up}}\Delta^{\mathrm{rt}} + \underline p_i v^{\mathrm{da}}_{i,\tau}, \label{eq:as-ramp} \\
    0 \le r_{i,j,\tau} \le R_i^{\mathrm{up}}\theta_j, \label{eq:as-response} \\
    p_{i,\tau} + \sum_j r_{i,j,\tau} \le \bar p_i\,u^{\mathrm{da}}_{i,h(\tau)} \qquad (: \nu^{\mathrm{cap}}_{i,\tau}), \label{eq:as-capacity} \\
    d^{\mathrm{res}}_{j,\tau} - \sum_i r_{i,j,\tau} - s^{\mathrm{res}}_{j,\tau} \le 0 \qquad (: \lambda^{\mathrm{res}}_{j,\tau}), \label{eq:as-reserve} \\
    0 \le s^{\mathrm{res}}_{j,\tau} \le d^{\mathrm{res}}_{j,\tau}. \label{eq:as-reserve-shortage}
\end{gather}
\end{subequations}

The objective in equation (\ref{eq:as-obj}) minimizes the joint as-bid cost of energy dispatch and reserve procurement for each real-time interval $\Delta^{\mathrm{rt}}$, together with penalties for unserved energy and unmet reserve requirements. The first term represents the accepted energy offers, and the second term represents payments associated with reserve capacity held available for contingency response. The penalties $\mathrm{VOLL}$ and $\mathrm{VOLR}$ are assigned to load shedding and reserve shortage, respectively. We set $\mathrm{VOLR} < \mathrm{VOLL}$ so that, when available capacity is insufficient to satisfy both requirements, the model permits a reserve shortage before shedding firm load.

Constraints (\ref{eq:as-output})--(\ref{eq:as-ramp}) describe the energy dispatch and follow the real-time balancing model in section~\ref{app:rt:clearing}. Constraint (\ref{eq:as-output}) defines the output of generator $i$ as its committed minimum output plus the accepted quantities from its energy-offer segments. Constraint (\ref{eq:as-segment}) bounds each accepted segment by its available quantity. Constraints (\ref{eq:as-injection}) and (\ref{eq:as-balance}) define nodal injections and enforce system-wide supply--demand balance, respectively. Constraint (\ref{eq:as-flow}) keeps transmission flows within the line ratings, and (\ref{eq:as-shed}) allows emergency load shedding when available generation is insufficient. Constraint (\ref{eq:as-ramp}) limits the change in generator output between consecutive real-time intervals.

Constraints (\ref{eq:as-response})--(\ref{eq:as-reserve-shortage}) introduce the reserve products. Constraint (\ref{eq:as-response}) limits the reserve award $r_{i,j,\tau}$ to the amount of additional output that generator $i$ can deliver within the required response time $\theta_j$. Therefore, a product with a longer response time allows a generator to provide more reserve capacity for a given ramp rate. Constraint (\ref{eq:as-capacity}) couples energy and reserve procurement. The scheduled energy output and the total reserve capacity awarded to a generator must jointly remain below its committed capacity. Consequently, capacity used to produce energy cannot simultaneously be sold as reserve, creating an opportunity-cost trade-off between the two products. The dual variable $\nu^{\mathrm{cap}}_{i,\tau}$ represents the marginal value of additional available capacity of generator $i$. Finally, constraint (\ref{eq:as-reserve}) requires the total procured capacity for reserve product $j$ to satisfy the published requirement $d^{\mathrm{res}}_{j,\tau}$. The slack variable $s^{\mathrm{res}}_{j,\tau}$ represents any unmet reserve requirement and is bounded by constraint (\ref{eq:as-reserve-shortage}). The corresponding dual variable $\lambda^{\mathrm{res}}_{j,\tau}$ gives the marginal clearing price of reserve product $j$.

\subsubsection{Pricing and settlement}
\label{app:as:settle}
Energy prices follow the same LMP formulation as in equation (\ref{eq:rt-lmp}), evaluated using the dual variables of the joint energy--reserve clearing problem (\ref{eq:as}). The reserve price of product $j$ is determined by the dual variable $\lambda^{\mathrm{res}}_{j,\tau}$ associated with the reserve requirement constraint (\ref{eq:as-reserve}). It represents the marginal cost of increasing the system-wide requirement for reserve product $j$ by one additional unit of capacity.

A key feature of co-optimization is that the opportunity cost of reserving generation capacity is determined within the clearing problem. When the response-time constraint is non-binding, the marginal cost of providing reserve consists of two components: the submitted reserve offer and the opportunity cost of using generation capacity for reserve rather than energy. The latter is represented by the shadow price $\nu^{\mathrm{cap}}_{i,\tau}$ of the shared capacity constraint. Hence, for a marginal reserve provider,
\begin{equation}
    \lambda^{\mathrm{res}}_{j,\tau} = \pi^{\mathrm{res}}_{i,j,\tau} + \nu^{\mathrm{cap}}_{i,\tau}.
    \label{eq:as-price-cap}
\end{equation}

If the same generator is also marginal in the energy market and no additional operating constraint is binding, the capacity shadow value corresponds to the energy margin,
\begin{equation}
    \nu^{\mathrm{cap}}_{i,\tau} = \mathrm{LMP}^{\mathrm{rt}}_{\mathrm{bus}(i),\tau} - \pi_{i,k,\tau},
\end{equation}
and therefore
\begin{equation}
    \lambda^{\mathrm{res}}_{j,\tau} = \pi^{\mathrm{res}}_{i,j,\tau} + \left( \mathrm{LMP}^{\mathrm{rt}}_{\mathrm{bus}(i),\tau} - \pi_{i,k,\tau} \right).
    \label{eq:as-price}
\end{equation}
Thus, the reserve price accounts not only for the submitted reserve offer but also for the value of energy production that the marginal provider gives up when capacity is allocated to reserve.

Energy is settled according to the two-settlement rule in equation (\ref{eq:rt-settle}). Reserve capacity is settled separately at the clearing price of each product. For generator $i$,
\begin{equation}
    \mathrm{revenue}^{\mathrm{res}}_{i,\tau} = \Delta^{\mathrm{rt}} \sum_j \lambda^{\mathrm{res}}_{j,\tau} \ r_{i,j,\tau},
    \label{eq:as-res-revenue}
\end{equation}
and its total profit for the interval is
\begin{equation}
    \mathrm{profit}_{i,\tau} = \mathrm{revenue}^{\mathrm{da}}_{i,\tau} + \mathrm{revenue}^{\mathrm{rt}}_{i,\tau} + \mathrm{revenue}^{\mathrm{res}}_{i,\tau} - \mathrm{cost}_{i,\tau}.
    \label{eq:as-settle}
\end{equation}
where
\begin{align}
    \mathrm{revenue}^{\mathrm{da}}_{i,\tau} &= \Delta^{\mathrm{rt}} \, \mathrm{LMP}^{\mathrm{da}}_{\mathrm{bus}(i),h(\tau)} \, q^{\mathrm{da}}_{i,h(\tau)}, \\
    \mathrm{revenue}^{\mathrm{rt}}_{i,\tau} &= \Delta^{\mathrm{rt}} \, \mathrm{LMP}^{\mathrm{rt}}_{\mathrm{bus}(i),\tau} \left( p_{i,\tau} - q^{\mathrm{da}}_{i,h(\tau)} \right), \\
    \mathrm{cost}_{i,\tau} &= \Delta^{\mathrm{rt}} \, \mathrm{TC}_i(p_{i,\tau}) + \Delta^{\mathrm{rt}} \, \mathrm{NL}_i u^{\mathrm{da}}_{i,h(\tau)} + S_i v^{\mathrm{da}}_{i,\tau}.
\end{align}
Reserve availability itself does not consume fuel, so no additional generation cost is assigned to a reserve award unless the reserve is activated. Every cleared MW of the same reserve product is paid the same product price, so a provider whose reserve offer is below the clearing price earns an inframarginal margin.

\textbf{Remark on energy--reserve co-optimization.}
Constraint (\ref{eq:as-capacity}) couples energy dispatch and reserve procurement because both compete for the same generation capacity. Increasing the reserve award can reduce the capacity available for energy production, while increasing energy output reduces the headroom available for reserve. The joint clearing therefore allows an agent's energy and reserve offers to interact through this shared capacity constraint.

The ancillary services market is implemented as a separate benchmark task that extends the real-time balancing market with reserve products. The day-ahead commitment, schedule, and prices enter as fixed inputs, while real-time energy and reserve are cleared jointly. Setting $d^{\mathrm{res}}_{j,\tau}=0$ for all reserve products removes the reserve layer and recovers the real-time balancing model of section~\ref{app:rt}.

\subsubsection{Partially Observable Stochastic Game}
\label{app:as:mg}
The ancillary services market is formulated as a general-sum POSG,
\begin{equation}
    \mathcal{G}^{\mathrm{as}} = \left\langle \ \mathcal{I}, \, \mathcal{S}, \, \{\mathcal{O}_i\}_{i\in\mathcal{I}}, \{\mathcal{A}_i\}_{i\in\mathcal{I}}, \, P, \, \{r_i\}_{i\in\mathcal{I}}, \, \gamma \ \right\rangle,
\end{equation}
where $\mathcal{I}$ denotes the set of strategic generators. One episodic market step corresponds to one 30-minute real-time interval $\tau$, during which energy and reserve are cleared jointly.

\paragraph{State.}
The state at interval $\tau$ is
\begin{equation}
    s_\tau = \left(\, x^{\mathrm{sys}}_\tau, \, x^{\mathrm{da}}_\tau, \, x^{\mathrm{res}}_\tau, \, \{x_{i,\tau}\}_{i\in\mathcal{I}} \, \right) \in \mathcal{S},
    \label{eq:as-state}
\end{equation}
where $x^{\mathrm{sys}}_\tau$ contains the realized system conditions of demand $\tilde{d}_{n,\tau}$ used in the joint energy--reserve clearing. The day-ahead component is
\begin{equation}
    x^{\mathrm{da}}_\tau = \left\{\, u^{\mathrm{da}}_{i,h(\tau)}, \, q^{\mathrm{da}}_{i,h(\tau)}, \, \mathrm{LMP}^{\mathrm{da}}_{n,h(\tau)} \, \right\}_{i,n},
    \label{eq:as-state-da}
\end{equation}
which contains the fixed commitment, binding generation schedules, and nodal prices from the corresponding day-ahead period. The reserve component is
\begin{equation}
    x^{\mathrm{res}}_\tau = \left(\, d^{\mathrm{res}}_{1,\tau}, \ldots, d^{\mathrm{res}}_{N^{\mathrm{prod}},\tau} \,\right),
    \label{eq:as-state-res}
\end{equation}
where $d^{\mathrm{res}}_{j,\tau}$ is the system-wide requirement for reserve product $j$ published before bidding. The operating state of generator $i$ is
\begin{equation}
    x_{i,\tau} = \left(\, p_{i,\tau-1}, \, v^{\mathrm{da}}_{i,\tau}, \, w^{\mathrm{da}}_{i,\tau} \,\right),
    \label{eq:as-state-gen}
\end{equation}
where $p_{i,\tau-1}$ is the realized dispatch from the previous real-time interval, and $v^{\mathrm{da}}_{i,\tau}$ and $w^{\mathrm{da}}_{i,\tau}$ are the start-up and shut-down indicators implied by the mapped day-ahead commitment. Network parameters, reserve-product response times, generator operating limits, and cost parameters are fixed environment parameters and are omitted from the dynamic state.

\paragraph{Observation.}
Each generator observes its own operating information together with the public energy and reserve market information available before bidding. Its observation at interval $\tau$ is
\begin{equation}
    o_{i,\tau} = \left(\, p_{i,\tau-1}, \, u^{\mathrm{da}}_{i,h(\tau)}, \, q^{\mathrm{da}}_{i,h(\tau)}, \, \mathrm{LMP}^{\mathrm{da}}_{\mathrm{bus}(i),h(\tau)}, \, x^{\mathrm{sys}}_\tau, \, x^{\mathrm{res}}_\tau \,\right) = \Omega_i(s_\tau) \in \mathcal{O}_i.
    \label{eq:as-observation}
\end{equation}
Thus, the generator observes the realized net-demand conditions and the published reserve requirements in addition to its own day-ahead position and previous dispatch. It does not observe competitors' private cost information or their contemporaneous energy and reserve offers. The market is therefore partially observable from the perspective of each participant.

\paragraph{Action.}
At each interval, generator $i$ jointly submits an energy offer and one reserve offer price for each reserve product. Its action is
\begin{equation}
    a_{i,\tau} = \left(\, \boldsymbol{\pi}^{\mathrm{en}}_{i,\tau}, \, \boldsymbol{\pi}^{\mathrm{res}}_{i,\tau} \right) \in \mathcal{A}_i,
    \label{eq:as-action}
\end{equation}
where
\begin{equation}
    \boldsymbol{\pi}^{\mathrm{en}}_{i,\tau} = \left( \pi_{i,1,\tau}, \ldots, \pi_{i,K,\tau} \right), \qquad \pi_{i,k,\tau} = \alpha_{i,\tau}\, \pi^{0}_{i,k},
\end{equation}
is the stepwise energy-offer price vector, obtained from a single markup $\alpha_{i,\tau} \in [1, \bar{\alpha}]$ on the truthful offer $\pi^{0}_{i,1} \le \cdots \le \pi^{0}_{i,K}$ as in the real-time balancing market, and
\begin{equation}
    \boldsymbol{\pi}^{\mathrm{res}}_{i,\tau} = \left( \pi^{\mathrm{res}}_{i,1,\tau}, \ldots, \pi^{\mathrm{res}}_{i,N^{\mathrm{prod}},\tau} \right)
\end{equation}
contains the reserve offer prices. The energy offer satisfies
\begin{equation}
    \pi_{i,1,\tau} \le \pi_{i,2,\tau} \le \cdots \le \pi_{i,K,\tau},
\end{equation}
while
$0 \le \pi^{\mathrm{res}}_{i,j,\tau} \le \mathrm{VOLR}$ for each reserve product $j$. The energy segment quantities $G_{i,k}$ are fixed, and no reserve quantity is submitted by the agent; the reserve awards are determined by the co-optimization. The joint action is $a_\tau=(a_{i,\tau})_{i\in\mathcal I}$.

\paragraph{Market clearing and reward.}
Given the current state and joint energy--reserve offers, the market-clearing mechanism
\begin{equation}
    z_\tau = \mathcal{C}^{\mathrm{as}}(s_\tau,a_\tau)
    \label{eq:as-clearing-map}
\end{equation}
solves the co-optimization problem (\ref{eq:as}). The clearing outcome includes the energy dispatch $p_{i,\tau}$, reserve awards $r_{i,j,\tau}$, real-time LMPs $\mathrm{LMP}^{\mathrm{rt}}_{n,\tau}$, and reserve prices $\lambda^{\mathrm{res}}_{j,\tau}$.

Each generator receives its total realized profit as the stage reward,
\begin{align}
    r_{i,\tau} &= \mathrm{profit}_{i,\tau} = \mathrm{revenue}^{\mathrm{da}}_{i,\tau} + \mathrm{revenue}^{\mathrm{rt}}_{i,\tau} + \mathrm{revenue}^{\mathrm{res}}_{i,\tau} - \mathrm{cost}_{i,\tau} \nonumber \\ &= \Delta^{\mathrm{rt}} \, \mathrm{LMP}^{\mathrm{da}}_{\mathrm{bus}(i),h(\tau)} \, q^{\mathrm{da}}_{i,h(\tau)} + \Delta^{\mathrm{rt}} \, \mathrm{LMP}^{\mathrm{rt}}_{\mathrm{bus}(i),\tau} \left( p_{i,\tau} - q^{\mathrm{da}}_{i,h(\tau)} \right) \nonumber \\ &+ \Delta^{\mathrm{rt}} \sum_j \lambda^{\mathrm{res}}_{j,\tau} \ r_{i,j,\tau} - \big(\, \Delta^{\mathrm{rt}} \, \mathrm{TC}_i(p_{i,\tau}) + \Delta^{\mathrm{rt}} \, \mathrm{NL}_i u^{\mathrm{da}}_{i,h(\tau)} + S_i v^{\mathrm{da}}_{i,\tau} \,\big).
    \label{eq:as-reward}
\end{align}
as defined in section~\ref{app:as:settle}. The reward therefore captures both energy-market profit and reserve-capacity revenue.

\paragraph{State transition.}
The clearing outcome determines the operating state entering the next real-time interval. In particular,
\begin{equation}
    x_{i,\tau+1} = \left(\, p_{i,\tau}, v^{\mathrm{da}}_{i,\tau+1}, w^{\mathrm{da}}_{i,\tau+1} \,\right),
    \label{eq:as-transition-gen}
\end{equation}
so the realized dispatch $p_{i,\tau}$ enters the ramping constraint of the next interval. The corresponding day-ahead commitment, schedule, and price are updated deterministically according to $h(\tau+1)$, while the realized net demand and reserve requirements evolve according to their exogenous data processes. The overall transition is therefore
\begin{equation}
    s_{\tau+1} \sim P \! \left( \cdot \mid s_\tau, a_\tau \right).
    \label{eq:as-transition}
\end{equation}

\paragraph{Agent objective.}
Each generator uses a policy $\sigma_i(a_{i,\tau}\mid o_{i,\tau})$ to jointly determine its energy and reserve offers. Over a horizon of $H$ real-time intervals, generator $i$ maximizes its expected discounted profit,
\begin{equation}
    J_i(\sigma_i,\sigma_{-i}) = \mathbb{E} \left[ \sum_{\tau=1}^{H} \gamma^{\tau-1} r_{i,\tau} \right],
    \label{eq:as-objective}
\end{equation}
where $\sigma_{-i}$ denotes the joint policies of the other generators. Because energy dispatch, reserve awards, and both energy and reserve prices are jointly determined by all submitted offers, the return of each generator depends on both its own bidding policy and those of other participants. Agents therefore learn how to allocate their available capacity between energy and reserve through repeated market interactions without observing competitors' private information or policies.

\subsection{M4: Peer-to-Peer Local Energy Market}
\label{app:p2p}
Rooftop photovoltaic generation and behind-the-meter storage have enabled many consumers to become prosumers that can both consume and supply electricity. When surplus energy is exported directly to the grid, it is typically settled at an export price such as feed-in-tariff, which is below the retail tariff paid for grid consumption. A peer-to-peer (P2P) local energy market allows neighboring participants to trade their surplus and deficit locally at prices between these two outside options, creating the possibility of mutual benefits for buyers and sellers. Unlike the day-ahead and real-time markets in sections~\ref{app:da} and~\ref{app:rt}, which operate at the transmission level, the P2P market considered here operates at the end-use level and facilitates local energy exchange among prosumers.

The P2P market uses 15-minute trading intervals. Before each auction, the net energy position of every participant is determined from its local generation, consumption, and storage operation. 
Hence, a participant knows the quantity it has available to buy or sell before submitting its bid, and its strategic decisions are its battery power and the submitted price.
A local market operator collects the bids and offers, clears the auction, and settles the resulting transactions. The market does not issue physical dispatch instructions: the participants' net quantities are fixed before clearing, so the auction determines which quantities are matched and at what price rather than how the underlying devices are operated.

In our model, the P2P local energy market is implemented as a periodic double auction among a fixed population of $N^{\mathrm{agent}}$ prosumers, each equipped with rooftop photovoltaic generation, an inelastic local load, and a battery. In each trading interval, a participant appears on one side of the market according to its net energy position: a surplus is offered for sale, whereas a deficit is submitted as demand. The local market operator collects one price--quantity pair from each participant and constructs aggregate supply and demand curves. The market clears at their intersection, with all matched buyers and sellers settling at a common market clearing price. Any unmatched deficit is supplied by the upstream grid at the retail tariff, while any unmatched surplus is exported to the grid at the export price. This section presents the market clearing model in subsection \ref{app:p2p:clearing}, the pricing and settlement rules in subsection  \ref{app:p2p:settle}, and the corresponding POSG in subsection \ref{app:p2p:mg}.

\subsubsection{Market model}
\label{app:p2p:clearing}
Let $\tau$ index the 15-minute P2P trading intervals, with $\Delta^{\mathrm{p2p}} = 0.25~\mathrm{h}$. Before each auction, the local generation, consumption, and battery operation of participant $i$ determine its net power position,
\begin{subequations}
\label{eq:p2p-pos}
\begin{equation}
    p^{\mathrm{net}}_{i,\tau} = - p^{\mathrm{pv}}_{i,\tau} + p^{\mathrm{ch}}_{i,\tau} + d_{i,\tau} - p^{\mathrm{dis}}_{i,\tau},
    \label{eq:p2p-net-power}
\end{equation}
and the corresponding quantities available for P2P trading are
\begin{equation}
    q^{\mathrm{sell}}_{i,\tau} = \Delta^{\mathrm{p2p}} \max\!\left( -p^{\mathrm{net}}_{i,\tau},0 \right),
    \qquad
    q^{\mathrm{buy}}_{i,\tau} = \Delta^{\mathrm{p2p}} \max\!\left( p^{\mathrm{net}}_{i,\tau},0 \right).
    \label{eq:p2p-net}
\end{equation}
\end{subequations}
Here, $p^{\mathrm{pv}}_{i,\tau}$ and $d_{i,\tau}$ denote photovoltaic generation and local demand, respectively. $p^{\mathrm{dis}}_{i,\tau}$ and $p^{\mathrm{ch}}_{i,\tau}$ denote battery discharging and charging power, respectively. Only one of $q^{\mathrm{sell}}_{i,\tau}$ and $q^{\mathrm{buy}}_{i,\tau}$ can be positive. A participant with surplus energy is therefore a seller, while one with an energy deficit is a buyer.

The battery is the only local device that carries a physical state between trading intervals. Its state of charge evolves according to
\begin{equation}
    \mathrm{soc}_{i,\tau+1} = \mathrm{soc}_{i,\tau} + \frac{\Delta^{\mathrm{p2p}}}{E_i} \left( \eta_i^{\mathrm{ch}} p^{\mathrm{ch}}_{i,\tau} - \frac{p^{\mathrm{dis}}_{i,\tau}} {\eta_i^{\mathrm{dis}}} \right),
    \label{eq:p2p-soc}
\end{equation}
where $E_i$ is the battery energy capacity, and $\eta_i^{\mathrm{ch}}$ and $\eta_i^{\mathrm{dis}}$ are the charging and discharging loss efficiencies. The feasible charging and discharging powers are bounded by
\begin{equation}
    0 \le p^{\mathrm{dis}}_{i,\tau} \le \min\!\left( \bar p_i^{\mathrm{bat}}, \frac{ \left( \mathrm{soc}_{i,\tau} - \underline{\mathrm{soc}}_i \right) E_i \eta_i^{\mathrm{dis}} }{ \Delta^{\mathrm{p2p}} } \right),
    \label{eq:p2p-dis}
\end{equation}
and
\begin{equation}
    0 \le p^{\mathrm{ch}}_{i,\tau} \le \min\!\left( \bar p_i^{\mathrm{bat}}, \frac{ \left( \overline{\mathrm{soc}}_i - \mathrm{soc}_{i,\tau} \right) E_i }{ \eta_i^{\mathrm{ch}} \Delta^{\mathrm{p2p}} } \right),
    \label{eq:p2p-ch}
\end{equation}
where $\bar p_i^{\mathrm{bat}}$ is the battery power rating and $[\underline{\mathrm{soc}}_i,\overline{\mathrm{soc}}_i]$ is the admissible state-of-charge range. Charging and discharging are mutually exclusive within each trading interval.

Battery operation is chosen by the participant before the P2P auction. Hence, $p^{\mathrm{net}}_{i,\tau}$ and the corresponding trading quantity are fixed when the participant submits its bid, and within the auction only the submitted price is a strategic bidding decision.

The seller and buyer sets in interval $\tau$ are
\begin{equation}
    \mathcal{S}_\tau = \left\{ i : q^{\mathrm{sell}}_{i,\tau}>0 \right\},
    \qquad
    \mathcal{B}_\tau = \left\{ i : q^{\mathrm{buy}}_{i,\tau}>0 \right\}.
    \label{eq:p2p-sets}
\end{equation}
Each participant submits one price $\pi_{i,\tau}$: an ask if $i\in\mathcal{S}_\tau$ and a bid if $i\in\mathcal{B}_\tau$. Submitted prices are restricted by the outside options provided by the upstream grid,
\begin{equation}
    \pi^{\mathrm{exp}} \le \pi_{i,\tau} \le \pi^{\mathrm{ret}},
    \label{eq:p2p-price-bound}
\end{equation}
where $\pi^{\mathrm{ret}}$ is the retail tariff for grid imports and $\pi^{\mathrm{exp}}$ is the export price for grid injections, with $\pi^{\mathrm{exp}}<\pi^{\mathrm{ret}}$. The submitted quantity is fixed by the participant's net position, so $\pi_{i,\tau}$ is the only strategic variable in the auction.

The local market operator clears the market through a periodic double auction. Seller asks are sorted in ascending order of price, while buyer bids are sorted in descending order of price, with ties broken by participant index. Let $\sigma^{\mathrm{s}}_\tau(j)$ denote the participant occupying position $j$ in the sorted seller order and $\sigma^{\mathrm{b}}_\tau(j)$ the participant occupying position $j$ in the sorted buyer order. The associated cumulative quantities are
\begin{equation}
    Q^{\mathrm{sell}}_{j,\tau} = \sum_{\ell=1}^{j} q^{\mathrm{sell}}_{\sigma^{\mathrm{s}}_\tau(\ell),\tau},
    \qquad
    Q^{\mathrm{buy}}_{j,\tau} = \sum_{\ell=1}^{j} q^{\mathrm{buy}}_{\sigma^{\mathrm{b}}_\tau(\ell),\tau}.
    \label{eq:p2p-cumulative}
\end{equation}
The total quantities available on the two sides are
\begin{equation}
    Q^{\mathrm{sell}}_\tau = \sum_{i\in\mathcal{S}_\tau} q^{\mathrm{sell}}_{i,\tau},
    \qquad
    Q^{\mathrm{buy}}_\tau = \sum_{i\in\mathcal{B}_\tau} q^{\mathrm{buy}}_{i,\tau}.
    \label{eq:p2p-total}
\end{equation}

The sorted submissions define the aggregate supply curve $S_\tau(x)$ and aggregate demand curve $D_\tau(x)$. The supply curve returns the ask price associated with the $x$-th unit of cumulative supply, while the demand curve returns the bid price associated with the $x$-th unit of cumulative demand. Both are step functions whose breakpoints are given by the cumulative
quantities in equation (\ref{eq:p2p-cumulative}).

Let $\mathcal{X}_\tau$ denote the union of the supply and demand breakpoints, together with zero. The cleared P2P trading volume is
\begin{equation}
    Q^\star_\tau = \max \left\{ x\in\mathcal{X}_\tau : x \le \min\!\left( Q^{\mathrm{sell}}_\tau, Q^{\mathrm{buy}}_\tau \right), \; D_\tau(x) \ge S_\tau(x) \right\}.
    \label{eq:p2p-volume}
\end{equation}
Thus, trading continues while the marginal buyer's bid is no lower than the marginal seller's ask. If no bid meets an ask, $Q^\star_\tau=0$.

Individual awards follow the sorted orders. For the seller occupying position $j$,
\begin{equation}
    q^{\mathrm{sell}\star}_{\sigma^{\mathrm{s}}_\tau(j),\tau} = \min \left[ \max\!\left( Q^\star_\tau - Q^{\mathrm{sell}}_{j-1,\tau}, 0 \right), q^{\mathrm{sell}}_{\sigma^{\mathrm{s}}_\tau(j),\tau} \right],
    \label{eq:p2p-sell-award}
\end{equation}
with $Q^{\mathrm{sell}}_{0,\tau}=0$. Similarly, the award to the buyer occupying position $j$ is
\begin{equation}
    q^{\mathrm{buy}\star}_{\sigma^{\mathrm{b}}_\tau(j),\tau} = \min \left[ \max\!\left( Q^\star_\tau - Q^{\mathrm{buy}}_{j-1,\tau}, 0 \right), q^{\mathrm{buy}}_{\sigma^{\mathrm{b}}_\tau(j),\tau} \right],
    \label{eq:p2p-buy-award}
\end{equation}
with $Q^{\mathrm{buy}}_{0,\tau}=0$.

Equations (\ref{eq:p2p-sell-award}) and (\ref{eq:p2p-buy-award}) fill the most competitive submissions first: lower-priced sellers and higher-priced buyers receive priority. Since $Q^\star_\tau$ coincides with a cumulative-quantity breakpoint on at least one side of the market, all accepted participants on that side are either fully cleared or rejected. On the opposite side, only the marginal participant may be partially cleared. 

\subsubsection{Pricing and settlement}
\label{app:p2p:settle}
Let $S^{\star}_\tau$ and $D^{\star}_\tau$ denote the marginal accepted ask and bid, respectively, and let $S^{+}_\tau$ and $D^{+}_\tau$ denote the first rejected ask and bid beyond the cleared volume $Q^{\star}_\tau$. A uniform clearing price must be high enough for all accepted sellers and low enough for all accepted buyers, while remaining unattractive to rejected submissions. This defines the feasible price interval
\begin{equation}
    \underline{\lambda}_\tau = \max\!\left( S^{\star}_\tau, D^{+}_\tau \right),
    \qquad
    \bar{\lambda}_\tau = \min\!\left( D^{\star}_\tau, S^{+}_\tau \right),
    \qquad
    \lambda^{\mathrm{loc}}_\tau = \frac{1}{2} \left( \underline{\lambda}_\tau + \bar{\lambda}_\tau \right).
    \label{eq:p2p-price}
\end{equation}
The local market therefore publishes the midpoint of the admissible price interval. When an adjacent accepted or rejected submission does not exist because one side of the market is exhausted, the export price $\pi^{\mathrm{exp}}$ and retail tariff $\pi^{\mathrm{ret}}$ provide the corresponding lower and upper price bounds. Consequently,
\begin{equation}
    \pi^{\mathrm{exp}} \le \lambda^{\mathrm{loc}}_\tau \le \pi^{\mathrm{ret}}.
\end{equation}
Thus, a matched seller receives no less than the grid export price and a matched buyer pays no more than the retail tariff. If a marginal participant is only partially cleared, the admissible interval collapses to that participant's submitted price.

P2P awards are financially binding, while any unmatched quantity is settled with the upstream grid. Define the residual export and import quantities as
\begin{equation}
    q^{\mathrm{ex}}_{i,\tau} = q^{\mathrm{sell}}_{i,\tau} - q^{\mathrm{sell}\star}_{i,\tau},
    \qquad
    q^{\mathrm{im}}_{i,\tau} = q^{\mathrm{buy}}_{i,\tau} - q^{\mathrm{buy}\star}_{i,\tau}.
    \label{eq:p2p-grid-residual}
\end{equation}
The settlement of participant $i$ is then
\begin{subequations}
\label{eq:p2p-settle}
\begin{align}
    \mathrm{revenue}_{i,\tau} &= \lambda^{\mathrm{loc}}_\tau q^{\mathrm{sell}\star}_{i,\tau} + \pi^{\mathrm{exp}} q^{\mathrm{ex}}_{i,\tau}, \\
    \mathrm{cost}_{i,\tau} &= \lambda^{\mathrm{loc}}_\tau q^{\mathrm{buy}\star}_{i,\tau} + \pi^{\mathrm{ret}} q^{\mathrm{im}}_{i,\tau} + c^{\mathrm{deg}}_i \Delta^{\mathrm{p2p}} \left( p^{\mathrm{ch}}_{i,\tau} + p^{\mathrm{dis}}_{i,\tau} \right), \\
    \mathrm{profit}_{i,\tau} &= \mathrm{revenue}_{i,\tau} - \mathrm{cost}_{i,\tau}.
\end{align}
\end{subequations}
Here, $c^{\mathrm{deg}}_i$ is the battery degradation cost in EUR/MWh of throughput. Photovoltaic generation is assumed to have zero marginal cost, while local demand is inelastic. A participant with a net energy deficit may therefore have negative profit, corresponding to a net electricity bill; maximizing profit in this case is equivalent to minimizing that bill. Because all matched buyers and sellers settle at the same local price, P2P payments are budget balanced within the local market, excluding transactions with the upstream grid and any network charges not modeled here.

\textbf{Remark on the P2P local market.}
The P2P market provides an intermediate trading opportunity between the two grid settlement prices. Sellers can receive more than the export price, while buyers can pay less than the retail tariff when mutually acceptable bids are matched. The market can also increase the share of locally produced renewable energy consumed within the local network. It therefore complements, rather than replaces, upstream electricity markets and grid transactions.

\subsubsection{Partially Observable Stochastic Game}
\label{app:p2p:mg}
The P2P local energy market is formulated as a general-sum POSG,
\begin{equation}
    \mathcal{G}^{\mathrm{p2p}} = \left\langle \ \mathcal{I}, \, \mathcal{S}, \{\mathcal{O}_i\}_{i\in\mathcal{I}}, \{\mathcal{A}_i\}_{i\in\mathcal{I}}, \, P, \, \{r_i\}_{i\in\mathcal{I}}, \, \gamma \ \right\rangle,
\end{equation}
where $\mathcal{I}$ is the fixed set of $N^{\mathrm{agent}}$ prosumers. One game step corresponds to one 15-minute P2P trading interval $\tau$.

\paragraph{State.}
The state at interval $\tau$ is
\begin{equation}
    s_\tau = \left(\, \{x_{i,\tau}\}_{i\in\mathcal{I}}, \pi^{\mathrm{exp}}, \pi^{\mathrm{ret}} \, \right) \in \mathcal{S},
    \label{eq:p2p-state}
\end{equation}
where the local physical state of prosumer $i$ is
\begin{equation}
    x_{i,\tau} = \left(\, \mathrm{soc}_{i,\tau}, p^{\mathrm{pv}}_{i,\tau}, d_{i,\tau}, p^{\mathrm{ch}}_{i,\tau}, p^{\mathrm{dis}}_{i,\tau} \,\right).
    \label{eq:p2p-state-agent}
\end{equation}
These variables determine the participant's net power position $p^{\mathrm{net}}_{i,\tau}$ and therefore its available selling or buying quantity through \eqref{eq:p2p-net}. The export price $\pi^{\mathrm{exp}}$ and retail tariff $\pi^{\mathrm{ret}}$ define the outside settlement options with the upstream grid.

\paragraph{Observation.}
Each participant observes its own physical state and net trading position, together with the public grid settlement prices. Its observation is
\begin{equation}
    o_{i,\tau} = \left( \mathrm{soc}_{i,\tau}, p^{\mathrm{net}}_{i,\tau}, q^{\mathrm{sell}}_{i,\tau}, q^{\mathrm{buy}}_{i,\tau}, \pi^{\mathrm{exp}}, \pi^{\mathrm{ret}} \right) = \Omega_i(s_\tau) \in \mathcal{O}_i.
    \label{eq:p2p-observation}
\end{equation}
Thus, each participant knows whether it is a seller or buyer and the quantity it has available to trade before submitting its price. It does not observe the contemporaneous bids, asks, or private local states of other participants. The market is therefore partially observable from the perspective of each individual prosumer.

\paragraph{Action.}
The trading quantity is fixed by the battery power before market clearing, so the strategic action of participant $i$ is its battery power together with its submitted price,
\begin{equation}
    a_{i,\tau} = \left( p^{\mathrm{bat}}_{i,\tau}, \pi_{i,\tau} \right), \qquad \pi_{i,\tau} \in \left[ \pi^{\mathrm{exp}}, \pi^{\mathrm{ret}} \right],
    \label{eq:p2p-action}
\end{equation}
where $p^{\mathrm{bat}}_{i,\tau} = p^{\mathrm{dis}}_{i,\tau} - p^{\mathrm{ch}}_{i,\tau}$ is the signed battery power within the feasible range (\ref{eq:p2p-dis})--(\ref{eq:p2p-ch}). If $q^{\mathrm{sell}}_{i,\tau}>0$, the submitted price is an ask; if $q^{\mathrm{buy}}_{i,\tau}>0$, it is a bid. A participant with zero net position does not actively participate in the auction. The joint action is $a_\tau=(a_{i,\tau})_{i\in\mathcal{I}}$.

\paragraph{Market clearing and reward.}
Given the participants' net positions and joint submitted prices, the local market-clearing mechanism
\begin{equation}
    z_\tau = \mathcal{C}^{\mathrm{p2p}}(s_\tau,a_\tau)
    \label{eq:p2p-clearing-map}
\end{equation}
sorts the asks and bids and applies the double-auction clearing rules in Section~\ref{app:p2p:clearing}. The clearing outcome contains the total traded volume $Q^\star_\tau$, the uniform local clearing price $\lambda^{\mathrm{loc}}_\tau$, and the individual awards $q^{\mathrm{sell}\star}_{i,\tau}$ and $q^{\mathrm{buy}\star}_{i,\tau}$. Any unmatched quantity is settled with the upstream grid.

The stage reward of participant $i$ is its realized profit,
\begin{align}
    r_{i,\tau} &= \mathrm{profit}_{i,\tau} = \mathrm{revenue}_{i,\tau} - \mathrm{cost}_{i,\tau} \nonumber \\ &= \left( \lambda^{\mathrm{loc}}_\tau q^{\mathrm{sell}\star}_{i,\tau} + \pi^{\mathrm{exp}} q^{\mathrm{ex}}_{i,\tau} \right) - \left( \lambda^{\mathrm{loc}}_\tau q^{\mathrm{buy}\star}_{i,\tau} + \pi^{\mathrm{ret}} q^{\mathrm{im}}_{i,\tau} + c^{\mathrm{deg}}_i \Delta^{\mathrm{p2p}} ( p^{\mathrm{ch}}_{i,\tau} + p^{\mathrm{dis}}_{i,\tau} ) \right)
    \label{eq:p2p-reward}
\end{align}
where $\mathrm{profit}_{i,\tau}$ is defined by \eqref{eq:p2p-settle}. For a seller, the reward reflects revenue from P2P sales and any residual export to the grid. For a buyer, the reward reflects the cost of P2P purchases and any residual import from the grid, together with battery degradation cost. A buyer may therefore receive a negative reward corresponding to its electricity bill, so maximizing reward is equivalent to minimizing that bill.

\paragraph{State transition.}
The battery state of charge provides the physical coupling between consecutive trading intervals. Following \eqref{eq:p2p-soc},
\begin{equation}
    \mathrm{soc}_{i,\tau+1} = \mathrm{soc}_{i,\tau} + \frac{\Delta^{\mathrm{p2p}}}{E_i} \left( \eta_i^{\mathrm{ch}}p^{\mathrm{ch}}_{i,\tau} - \frac{p^{\mathrm{dis}}_{i,\tau}} {\eta_i^{\mathrm{dis}}} \right).
    \label{eq:p2p-transition-soc}
\end{equation}
Photovoltaic generation and local demand evolve according to their exogenous data processes, while battery operation for the next period is determined before the next P2P auction. The next net position and trading quantity are then obtained from \eqref{eq:p2p-net}. The overall state transition is
\begin{equation}
    s_{\tau+1} \sim P\!\left( \cdot \mid s_\tau,a_\tau \right).
    \label{eq:p2p-transition}
\end{equation}
Under the market design considered here, the submitted P2P price affects market matching and settlement but does not directly alter the participant's physical net position or battery dynamics.

\paragraph{Agent objective.}
Each participant uses a policy $\sigma_i(a_{i,\tau}\mid o_{i,\tau})$ to determine its submitted price. Over a horizon of $H$ trading intervals, participant $i$ seeks to maximize its expected discounted profit,
\begin{equation}
    J_i(\sigma_i,\sigma_{-i}) = \mathbb{E} \left[ \sum_{\tau=1}^{H} \gamma^{\tau-1} r_{i,\tau} \right],
    \label{eq:p2p-objective}
\end{equation}
where $\sigma_{-i}$ denotes the joint policies of the other participants. Because the cleared volume, matching outcome, and local clearing price depend on all submitted bids and asks, the return of each participant depends on both its own pricing policy and those of the other prosumers. Participants therefore learn bidding strategies through repeated local-market interactions without observing competitors' contemporaneous submissions or private states.

\subsection{M5: Local Flexibility Market}
\label{app:flex}
Growing electrification and distributed generation can place distribution networks under increasing operational stress, leading to feeder congestion and voltage violations. Reinforcing distribution infrastructure can be costly and slow, particularly when the underlying constraint occurs only during a limited number of periods. A local flexibility market provides an alternative means of managing these constraints by allowing the distribution system operator (DSO) to procure temporary changes in power injection or consumption from resources already connected to the affected network.

The market operates at the distribution level and clears once per delivery period. It addresses local network constraints that are not represented in the transmission-level wholesale markets, while the upstream energy price faced by each aggregator is treated as exogenous. The DSO is the sole buyer, making the market a one-sided procurement auction. It specifies the locations and periods where flexibility is required, and accepts offers that restore a feasible distribution-network operating point at minimum procurement cost. The sellers are aggregators, with one aggregator at each bus, each representing a portfolio of distributed energy resources.

Two features distinguish this market from the preceding settings. First, the traded product is a deviation from a predefined baseline rather than energy itself. Second, settlement follows a pay-as-bid rule: each accepted aggregator is paid its submitted offer price for the flexibility it provides, rather than a common market-clearing price. The settlement therefore does not rely on dual-based marginal prices. This section presents the market-clearing model in subsection \ref{app:flex:clearing}, the pricing and settlement rule in subsection \ref{app:flex:settle}, and the corresponding POSG in subsection \ref{app:flex:mg}.

\subsubsection{Market model}
\label{app:flex:clearing}
We consider a radial distribution network with $N^{\mathrm{bus}}$ buses supplied by a single substation, indexed by $0$, whose voltage is fixed at $v_0=1$ p.u. Let $\tau$ index the local flexibility trading intervals, with duration $\Delta^{\mathrm{flex}} = 1~\text{hour}$.
Before the flexibility market clears, each aggregator has a baseline operating position determined by local photovoltaic generation, demand, and planned battery charging. With no flexibility activated, the baseline active-power injection at bus $n$ is
\begin{equation}
    P^{\mathrm{base,inj}}_{n,\tau} = \sum_{i:\mathrm{bus}(i)=n} \left( p^{\mathrm{pv}}_{i,\tau} - d_{i,\tau} - p^{\mathrm{plan}}_{i,\tau} \right) - d^{\mathrm{bg}}_{n,\tau},
    \label{eq:flex-base-inj}
\end{equation}
where $p^{\mathrm{pv}}_{i,\tau}$ is the photovoltaic output, $d_{i,\tau}$ is the local demand of aggregator $i$, $p^{\mathrm{plan}}_{i,\tau}$ is its planned battery charging power, and $d^{\mathrm{bg}}_{n,\tau}$ denotes background demand not controlled by any aggregator.

Under the radial-network approximation, the corresponding baseline line flows and squared voltages are obtained from
\begin{equation}
    P^{\mathrm{base}}_{l,\tau} = - \sum_{m\in\mathcal D(l)} P^{\mathrm{base,inj}}_{m,\tau},
    \label{eq:flex-base-flow}
\end{equation}
and
\begin{equation}
    \left(v^{\mathrm{base}}_{n,\tau}\right)^2 = v_0^2 + 2\sum_m \left( R_{n,m}P^{\mathrm{base,inj}}_{m,\tau} + X_{n,m}Q^{\mathrm{base,inj}}_{m,\tau} \right),
    \label{eq:flex-base-voltage}
\end{equation}
where $\mathcal D(l)$ denotes the set of buses downstream of line $l$. The quantities $R_{n,m}$ and $X_{n,m}$ are the sums of resistance and reactance over the network path shared by buses $n$ and $m$, respectively. Reactive injections are determined from the active loads using a fixed power-factor assumption.

Before offers are submitted, the DSO publishes indicators of the local flexibility requirement. For an undervoltage at bus $n$, the indicator is
\begin{equation}
    \mathrm{req}^{\mathrm v}_{n,\tau} = \frac{ \max\!\left[ \left( \underline v_n+\gamma^{\mathrm v} \right)^2 - \left( v^{\mathrm{base}}_{n,\tau} \right)^2, 0 \right] }{ 2R_{n,n} },
    \label{eq:flex-req-v}
\end{equation}
while the requirement associated with forward-flow congestion on line $l$ is
\begin{equation}
    \mathrm{req}^{\mathrm{th}}_{l,\tau} = \max\!\left[ P^{\mathrm{base}}_{l,\tau} - \left( 1-\gamma^{\mathrm{th}} \right)\bar P_l, 0 \right].
    \label{eq:flex-req-th}
\end{equation}
These quantities are published as market information for the aggregators; the market clearing itself enforces the full network constraints below.

The traded product is upward flexibility, defined as an increase in net active
power injection relative to the baseline. Aggregator $i$ submits one
price--quantity offer
$(\pi_{i,\tau},\bar q_{i,\tau})$, where $\pi_{i,\tau}$ is its offer price and
$\bar q_{i,\tau}$ is the maximum flexibility quantity offered. Any quantity
$q_{i,\tau}\in[0,\bar q_{i,\tau}]$ may be accepted by the DSO. Unlike the
stepwise offers used in the wholesale markets, each aggregator submits a
single price--quantity pair in each flexibility trading interval.

Upward flexibility can be delivered by reducing planned battery charging or, once planned charging has been fully removed, by discharging the battery. The offered quantity must therefore satisfy
\begin{equation}
    0 \le \bar q_{i,\tau} \le q^{\mathrm{phys}}_{i,\tau},
    \label{eq:flex-offer-cap}
\end{equation}
where
\begin{equation} 
    q^{\mathrm{phys}}_{i,\tau} = p^{\mathrm{plan}}_{i,\tau} + \min\!\left( \bar p^{\mathrm{dis}}_i,\, \frac{ \eta^{\mathrm{dis}}_i E_i \left( \mathrm{soc}_{i,\tau} - \underline{\mathrm{soc}}_i \right) }{ \Delta^{\mathrm{flex}} } \right).
    \label{eq:flex-cap}
\end{equation}
Here, $\bar p^{\mathrm{dis}}_i$ is the battery discharge-power limit, $E_i$ is its energy capacity, $\eta^{\mathrm{dis}}_i$ is its discharge efficiency, and $\underline{\mathrm{soc}}_i$ is the minimum admissible state of charge. Because $q^{\mathrm{phys}}_{i,\tau}$ depends on the current state of charge, the amount of flexibility available from an aggregator varies across trading intervals.

Given the submitted offers, the DSO selects flexibility awards by solving
\allowdisplaybreaks
\begin{subequations}
\label{eq:flex-clear}
\begin{equation}
    \min_{\{q,s,P^{\mathrm{inj}},P,v\}} \; \Delta^{\mathrm{flex}} \left[ \sum_i \pi_{i,\tau}q_{i,\tau} + \mathrm{VOLL}\sum_n s_{n,\tau} \right],
    \label{eq:flex-obj}
\end{equation}
subject to
\begin{gather}
    P^{\mathrm{inj}}_{n,\tau} = \sum_{i:\mathrm{bus}(i)=n} \left( p^{\mathrm{pv}}_{i,\tau} - d_{i,\tau} + q_{i,\tau} - p^{\mathrm{plan}}_{i,\tau} \right) - d^{\mathrm{bg}}_{n,\tau} + s_{n,\tau}, \label{eq:flex-injection} \\
    P_{l,\tau} = - \sum_{m\in\mathcal D(l)} P^{\mathrm{inj}}_{m,\tau}, \label{eq:flex-lineflow} \\
    v_{n,\tau}^{2} = v_0^{2} + 2\sum_m \left( R_{n,m}P^{\mathrm{inj}}_{m,\tau} + X_{n,m}Q^{\mathrm{inj}}_{m,\tau} \right), \label{eq:flex-voltage} \\
    \left( \underline v_n+\gamma^{\mathrm v} \right)^2 \le v_{n,\tau}^{2} \le \left( \bar v_n-\gamma^{\mathrm v} \right)^2, \qquad n\neq0, \label{eq:flex-vlim} \\
    - \left( 1-\gamma^{\mathrm{th}} \right)\bar P_l \le P_{l,\tau} \le \left( 1-\gamma^{\mathrm{th}} \right)\bar P_l, \label{eq:flex-thermal} \\
    0 \le q_{i,\tau} \le \bar q_{i,\tau}, \label{eq:flex-award} \\
    0 \le s_{n,\tau} \le d^{\mathrm{bg}}_{n,\tau} + \sum_{i:\mathrm{bus}(i)=n} d_{i,\tau}. \label{eq:flex-shed}
\end{gather}
\end{subequations}

The objective (\ref{eq:flex-obj}) minimizes the total procurement cost of local flexibility in each trading interval. The first term represents the pay-as-bid cost of accepted flexibility offers, while the second penalizes emergency load curtailment. The value of lost load $\mathrm{VOLL}$ is set above the admissible flexibility offer prices so that curtailment is used only when the available flexibility cannot restore a feasible network operating point.

Constraint (\ref{eq:flex-injection}) defines the active-power injection at each distribution bus. Without flexibility, the baseline injection consists of photovoltaic generation minus local demand and planned battery charging. An accepted flexibility quantity $q_{i,\tau}$ increases the bus injection relative to this baseline, either by reducing charging or by discharging the battery. Background demand $d^{\mathrm{bg}}_{n,\tau}$ represents demand not controlled by the aggregators, while $s_{n,\tau}$ denotes emergency load curtailment. 

Constraint (\ref{eq:flex-lineflow}) computes the active-power flow on each line as the total downstream withdrawal under the radial-network structure. Here, $\mathcal D(l)$ denotes the set of buses downstream of line $l$. Constraint (\ref{eq:flex-voltage}) uses a linearized radial distribution power-flow model. The quantities $R_{n,m}$ and $X_{n,m}$ represent the resistance and reactance shared by the paths from the substation to buses $n$ and $m$, respectively. The reactive injections $Q^{\mathrm{inj}}_{m,\tau}$ are determined from the fixed power-factor assumptions of the underlying loads and are not strategic decision variables. 

Constraints (\ref{eq:flex-vlim}) and (\ref{eq:flex-thermal}) enforce the distribution-network operating limits. The parameters $\gamma^{\mathrm v}$ and $\gamma^{\mathrm{th}}$ introduce safety margins by tightening the admissible voltage range and thermal line ratings used during clearing. Constraint (\ref{eq:flex-award}) allows the DSO to accept any quantity between zero and the quantity offered by aggregator $i$. The submitted quantity $\bar q_{i,\tau}$ must itself satisfy the physical deliverability condition (\ref{eq:flex-cap}). Consequently, the clearing may partially accept an offer when the full quantity is not required to restore network feasibility. Finally, constraint (\ref{eq:flex-shed}) bounds emergency curtailment by the total active demand at each bus. Curtailment acts as a feasibility backstop when the submitted flexibility offers are insufficient to satisfy the tightened network constraints.

Since the network equations (\ref{eq:flex-injection})--(\ref{eq:flex-voltage}) are affine in the clearing variables under the linearized radial power-flow model, the resulting flexibility-market clearing problem is a linear program.

\subsubsection{Payment and settlement}
\label{app:flex:settle}
Settlement follows a pay-as-bid rule. Each accepted aggregator is paid its submitted offer price for the quantity cleared by the DSO,
\begin{equation}
    \mathrm{revenue}_{i,\tau} = \Delta^{\mathrm{flex}} \pi_{i,\tau} q_{i,\tau}, \qquad \mathrm{profit}_{i,\tau} = \mathrm{revenue}_{i,\tau} - \mathrm{cost}_{i,\tau},
    \label{eq:flex-pay}
\end{equation}
where $\mathrm{cost}_{i,\tau}$ is the true delivery cost defined in Section~\ref{app:flex:mg}. Because settlement is pay-as-bid, two aggregators providing flexibility at the same bus and in the same interval may receive different payments if they submit different offer prices.

Although the market imposes no uniform clearing price, the curtailment backstop creates an implicit upper value for flexibility. Emergency curtailment can relieve the same network constraints at a marginal cost of $\mathrm{VOLL}$, so a flexibility offer is selected only when its network benefit justifies its submitted cost relative to this alternative. This value is inherently locational because the effect of an injection depends on where it enters the network. For a binding forward-flow constraint, an additional MW of injection downstream of the constrained line reduces the line flow by one MW under \eqref{eq:flex-lineflow}, whereas an injection outside that downstream region does not relieve that constraint. For a binding voltage constraint, the value of flexibility depends on the voltage sensitivity of the participant's bus, so injections at electrically more effective locations can support higher accepted offer prices than injections with weaker voltage impact.

The flexibility product is defined as a deviation from the pre-market baseline. For aggregator $i$, this baseline is determined from its photovoltaic generation, local demand, and planned battery charging,
\begin{equation}
    p^{\mathrm{base}}_{i,\tau} = p^{\mathrm{pv}}_{i,\tau} - d_{i,\tau} - p^{\mathrm{plan}}_{i,\tau}.
\end{equation}
An accepted flexibility award $q_{i,\tau}$ therefore requires the aggregator to increase its net injection by exactly that amount relative to the baseline. In the benchmark, delivery is assumed to equal the cleared award, so no payment is made for flexibility that is not delivered.

\subsubsection{Partially Observable Stochastic Game}
\label{app:flex:mg}
The local flexibility market is formulated as a general-sum partially observable stochastic game (POSG),
\begin{equation}
    \mathcal{G}^{\mathrm{flex}} = \left\langle \mathcal{I}, \mathcal{S}, \{\mathcal{O}_i\}_{i\in\mathcal{I}}, \{\mathcal{A}_i\}_{i\in\mathcal{I}}, P, \{r_i\}_{i\in\mathcal{I}}, \gamma \right\rangle,
\end{equation}
where $\mathcal{I}$ denotes the set of strategic aggregators. One game step corresponds to one local flexibility trading interval $\tau$.

\paragraph{State.}
The state at interval $\tau$ is
\begin{equation}
    s_\tau = \left( x^{\mathrm{sys}}_\tau, \{x_{i,\tau}\}_{i\in\mathcal{I}} \right) \in \mathcal{S},
    \label{eq:flex-state}
\end{equation}
where $x^{\mathrm{sys}}_\tau$ describes the distribution-network operating condition before flexibility procurement. It contains the baseline bus injections, line flows, voltages, and the flexibility-need indicators published by the DSO,
\begin{equation}
    x^{\mathrm{sys}}_\tau = \left( \mathbf P^{\mathrm{base,inj}}_\tau, \mathbf P^{\mathrm{base}}_\tau, \mathbf v^{\mathrm{base}}_\tau, \mathbf{req}^{\mathrm v}_\tau, \mathbf{req}^{\mathrm{th}}_\tau \right).
    \label{eq:flex-state-system}
\end{equation}
The local physical state of aggregator $i$ is
\begin{equation}
    x_{i,\tau} = \left( \mathrm{soc}_{i,\tau}, p^{\mathrm{pv}}_{i,\tau}, d_{i,\tau}, p^{\mathrm{plan}}_{i,\tau} \right),
    \label{eq:flex-state-agent}
\end{equation}
which, together with the planned charging in its action, determines its baseline injection and the maximum physically deliverable flexibility $q^{\mathrm{phys}}_{i,\tau}$ through \eqref{eq:flex-cap}.
Network parameters, battery ratings, and other fixed technical parameters are omitted from the dynamic state.

\paragraph{Observation.}
Each aggregator observes its own operating state together with the public network information released by the DSO. Its observation is
\begin{equation}
    o_{i,\tau} = \left( \mathrm{soc}_{i,\tau}, p^{\mathrm{pv}}_{i,\tau}, d_{i,\tau}, \mathbf{req}^{\mathrm v}_\tau, \mathbf{req}^{\mathrm{th}}_\tau \right) = \Omega_i(s_\tau) \in\mathcal{O}_i.
    \label{eq:flex-observation}
\end{equation}
Thus, each aggregator knows its own available flexibility and the locations and magnitudes of the network needs published before bidding. It does not observe the private operating states, delivery costs, or contemporaneous offers of other aggregators. The market is therefore partially observable from the perspective of each participant.

\paragraph{Action.}
At each trading interval, aggregator $i$ submits one price--quantity offer together with its planned battery charging,
\begin{equation}
    a_{i,\tau} = \left( \pi_{i,\tau}, \bar q_{i,\tau}, p^{\mathrm{plan}}_{i,\tau} \right) \in \mathcal{A}_i,
    \label{eq:flex-action}
\end{equation}
subject to
\begin{equation}
    c^{\mathrm{rep}} \le \pi_{i,\tau} \le \overline{\pi}, \qquad 0 \le \bar q_{i,\tau} \le q^{\mathrm{phys}}_{i,\tau}, \qquad 0 \le p^{\mathrm{plan}}_{i,\tau} \le \bar p^{\mathrm{ch}}_{i,\tau}.
    \label{eq:flex-action-bound}
\end{equation}
The offer price $\pi_{i,\tau}$ determines the payment requested per unit of flexibility, $\bar q_{i,\tau}$ specifies the maximum quantity the aggregator is willing to provide, and $p^{\mathrm{plan}}_{i,\tau}$ enters the baseline injection (\ref{eq:flex-base-inj}). Here, $c^{\mathrm{rep}}$ is the replacement cost of one delivered MWh, i.e., the energy price and the degradation cost corrected for the round-trip efficiency, $\overline{\pi} < \mathrm{VOLL}$ is the market price cap, and $\bar p^{\mathrm{ch}}_{i,\tau}$ is the charging power the battery can absorb in the interval given its state of charge. The DSO may accept any quantity $q_{i,\tau}\in[0,\bar q_{i,\tau}]$. The joint action is $a_\tau=(a_{i,\tau})_{i\in\mathcal I}$.

\paragraph{Market clearing and reward.}
Given the current state and joint offers, the DSO applies the network-constrained clearing mechanism
\begin{equation}
    z_\tau = \mathcal{C}^{\mathrm{flex}}(s_\tau,a_\tau),
    \label{eq:flex-clearing-map}
\end{equation}
which solves \eqref{eq:flex-clear}. The clearing outcome contains the accepted flexibility quantities $q_{i,\tau}$, resulting bus injections, line flows and voltages, together with any emergency load curtailment.

Under the pay-as-bid settlement rule, the revenue of aggregator $i$ is
\begin{equation}
    \mathrm{revenue}_{i,\tau} = \Delta^{\mathrm{flex}} \pi_{i,\tau}q_{i,\tau}.
    \label{eq:flex-posg-revenue}
\end{equation}
Its stage reward is the realized profit,
\begin{equation}
    r_{i,\tau} = \mathrm{profit}_{i,\tau} = \mathrm{revenue}_{i,\tau} - C^{\mathrm{flex}}_i \left( q_{i,\tau}; x_{i,\tau} \right),
    \label{eq:flex-reward}
\end{equation}
where $C^{\mathrm{flex}}_i = \Delta^{\mathrm{flex}} \left[ \lambda_\tau\, p^{\mathrm{ch}}_{i,\tau} + c^{\mathrm{cyc}}_i \left( p^{\mathrm{ch}}_{i,\tau} + p^{\mathrm{dis}}_{i,\tau} \right) \right]$ is the true cost of the interval, the energy bought for charging at the exogenous price $\lambda_\tau$ plus the degradation cost $c^{\mathrm{cyc}}_i$ per MWh of battery throughput, with $p^{\mathrm{ch}}$ and $p^{\mathrm{dis}}$ given by the state transition below. This cost depends on the aggregator's physical operating state and the quantity actually delivered rather than on its submitted offer price.

\paragraph{State transition.}
An accepted flexibility award changes the battery operating trajectory and therefore affects the flexibility available in subsequent intervals. Let
\begin{equation}
    q^{\mathrm{red}}_{i,\tau} = \min\!\left( q_{i,\tau}, p^{\mathrm{plan}}_{i,\tau} \right)
\end{equation}
denote the part of the flexibility award delivered by reducing planned charging, and let
\begin{equation}
    q^{\mathrm{dis}}_{i,\tau} = \max\!\left( q_{i,\tau} - p^{\mathrm{plan}}_{i,\tau}, 0 \right)
\end{equation}
denote the remaining part delivered through battery discharge. The realized battery charging and discharging powers are therefore
\begin{equation}
    p^{\mathrm{ch}}_{i,\tau} = p^{\mathrm{plan}}_{i,\tau} - q^{\mathrm{red}}_{i,\tau}, \qquad p^{\mathrm{dis}}_{i,\tau} = q^{\mathrm{dis}}_{i,\tau}.
\end{equation}
The state of charge evolves according to
\begin{equation}
    \mathrm{soc}_{i,\tau+1} = \mathrm{soc}_{i,\tau} + \frac{\Delta^{\mathrm{flex}}}{E_i} \left( \eta_i^{\mathrm{ch}} p^{\mathrm{ch}}_{i,\tau} - \frac{ p^{\mathrm{dis}}_{i,\tau} }{ \eta_i^{\mathrm{dis}} } \right).
    \label{eq:flex-transition-soc}
\end{equation}
Hence, providing flexibility in the current interval can reduce the battery energy available for future flexibility provision. 
Photovoltaic generation, local demand, planned battery operation, and the resulting network operating conditions evolve according to their exogenous data processes. 
The overall
state transition is
\begin{equation}
    s_{\tau+1} \sim P\!\left( \cdot \mid s_\tau, a_\tau \right).
    \label{eq:flex-transition}
\end{equation}

\paragraph{Agent objective.}
Each aggregator uses a policy $\sigma_i(a_{i,\tau}\mid o_{i,\tau})$ to determine its flexibility offer. Over a horizon of $H$ trading intervals, aggregator $i$ seeks to maximize its expected discounted profit,
\begin{equation}
    J_i(\sigma_i,\sigma_{-i}) = \mathbb{E} \left[ \sum_{\tau=1}^{H} \gamma^{\tau-1} r_{i,\tau} \right],
    \label{eq:flex-objective}
\end{equation}
where $\sigma_{-i}$ denotes the joint policies of the other aggregators. Because the DSO jointly selects offers according to their submitted prices, available quantities, locations, and effects on network constraints, the return of each aggregator depends on both its own bidding policy and those of the other participants. Aggregators therefore learn price--quantity offering strategies through repeated market interactions without observing competitors' private states, costs, or contemporaneous offers.
\clearpage

\section{Computation Performance}
\label{app:speed}

This section compares the throughput of the same environment code under three layouts: all on the GPU; environment on the host CPU, policy on the GPU; all on the host CPU. With 64 parallel environments, the throughput of all on the GPU is 2.3 to 22 times that of all on the host CPU and 2.3 to 33 times that of environment on the host CPU, policy on the GPU, in every one of the five markets, and it does not change with the number of host threads allotted. Figure~\ref{fig.appendix.speed.layouts} compares the three layouts in the market with the most participants, the M4 peer-to-peer local energy market with 1\,200 households; Figure~\ref{fig.appendix.speed.markets} covers all five markets, and Figure~\ref{fig.appendix.speed.tools} open-source tools. All experiments run under Python 3.11.15 with JAX and jaxlib 0.10.2, CUDA support added as the same-version plugin rather than by changing either, and double precision and highest matmul precision enabled, on a single shared workstation with three NVIDIA RTX~4500 Ada (24~GB) GPUs and an AMD Ryzen Threadripper PRO 7985WX host (64 cores, 128 threads); each measurement point is run three times and compilation time is excluded. Throughput is the number of environment steps advanced per second, summed over the parallel environments, where one step covers the market clearing and one forward pass of the policy, without gradient updates.

\subsection{Three layouts on the peer-to-peer market}
\label{app:speed:layouts}

\begin{figure}[h!]
\centering
\includegraphics[width=\linewidth]{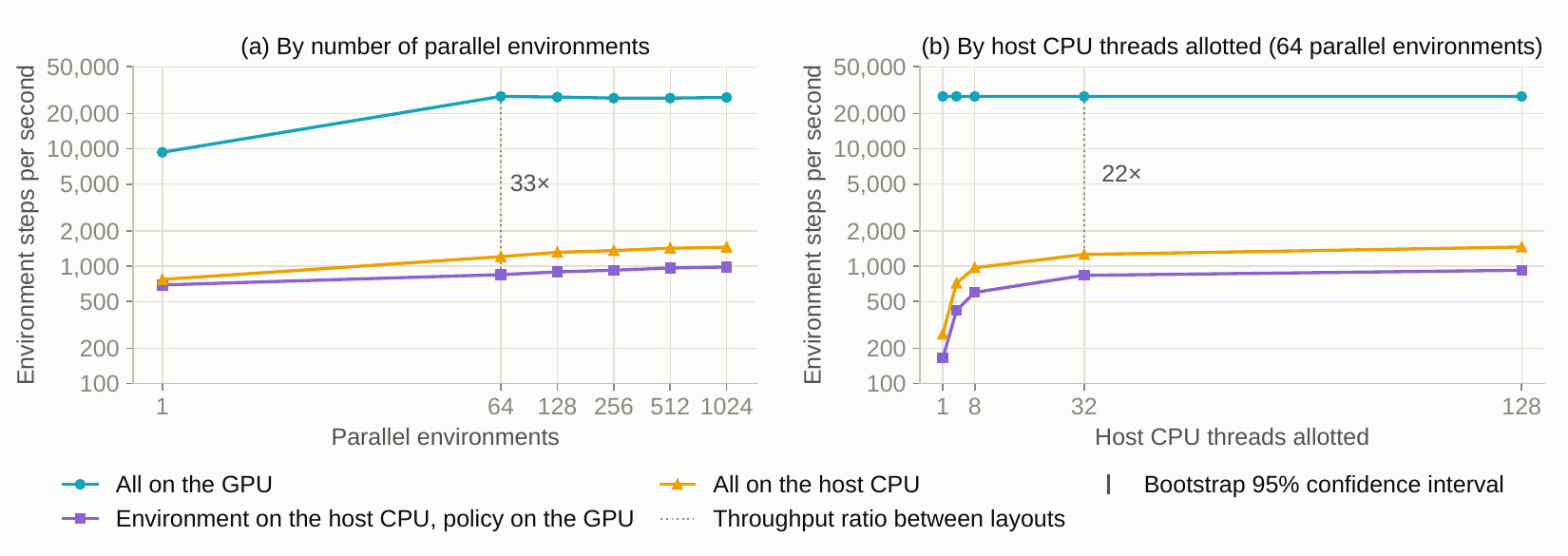}
\caption{Throughput of the peer-to-peer local energy market environment (1\,200 households, 96 periods per episode) under three layouts: (a) against the number of parallel environments; (b) against the number of host CPU threads allotted, with 64 parallel environments. Each point is the mean of three runs with a bootstrap 95\% confidence interval; each dotted line joins two points and the number beside it is their throughput ratio.}
\label{fig.appendix.speed.layouts}
\end{figure}

In the peer-to-peer local energy market with 64 parallel environments, all on the GPU advances 27\,929 environment steps per second, 22 times the throughput of all on the host CPU and 33 times that of environment on the host CPU, policy on the GPU (Figure~\ref{fig.appendix.speed.layouts}a). Raising the number of parallel environments from 64 to 1\,024 leaves the throughput unchanged, and so does the number of host threads allotted (Figure~\ref{fig.appendix.speed.layouts}b).

\subsection{Across the five markets}
\label{app:speed:markets}

\begin{figure}[h!]
\centering
\includegraphics[width=\linewidth]{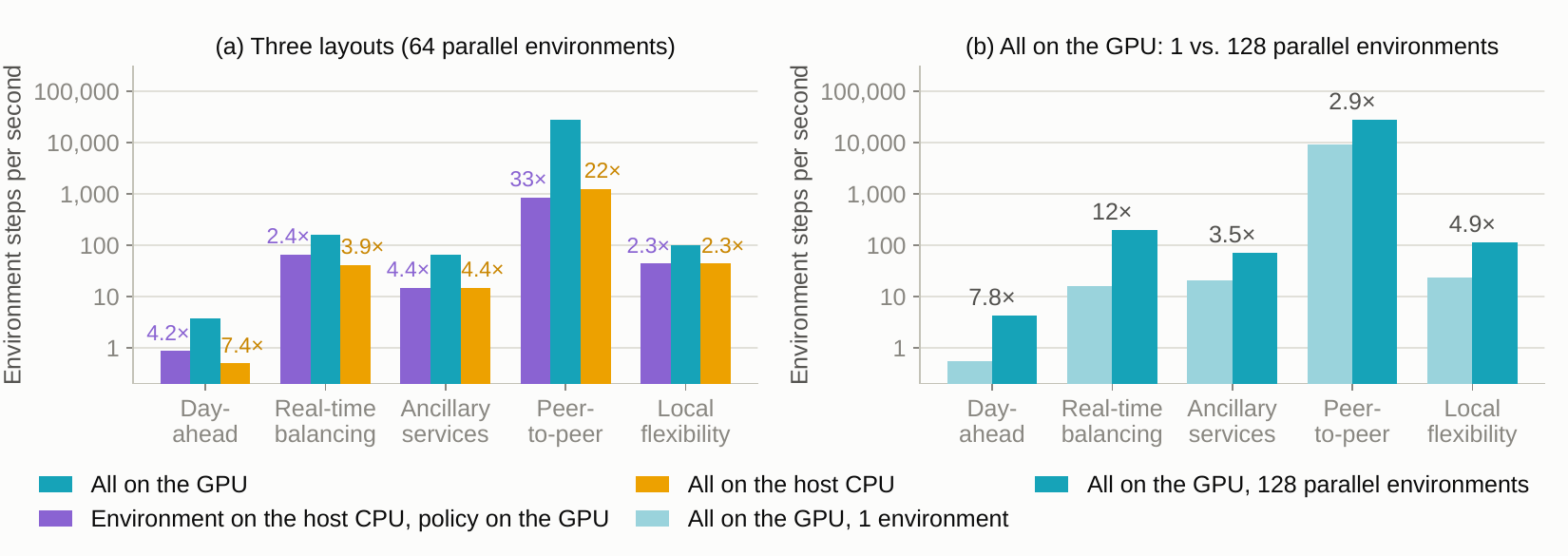}
\caption{Throughput of the five market environments: (a) the three layouts with 64 parallel environments, the number above each host bar being the ratio of all on the GPU to that layout; (b) all on the GPU with 1 and with 128 parallel environments, the number above each pair being the ratio of the two bars. Each bar is the mean of three runs.}
\label{fig.appendix.speed.markets}
\end{figure}

All five markets show the speed and throughput advantage of PowerMarketJax: with 64 parallel environments, all on the GPU gives 2.3 to 22 times the throughput of all on the host CPU and 2.3 to 33 times that of environment on the host CPU, policy on the GPU (Figure~\ref{fig.appendix.speed.markets}a); all on the GPU, raising the number of parallel environments from 1 to 128 multiplies the throughput by 2.9 to 12 (Figure~\ref{fig.appendix.speed.markets}b).

\subsection{Against open-source tools}
\label{app:speed:tools}

\begin{figure}[h!]
\centering
\includegraphics[width=\linewidth]{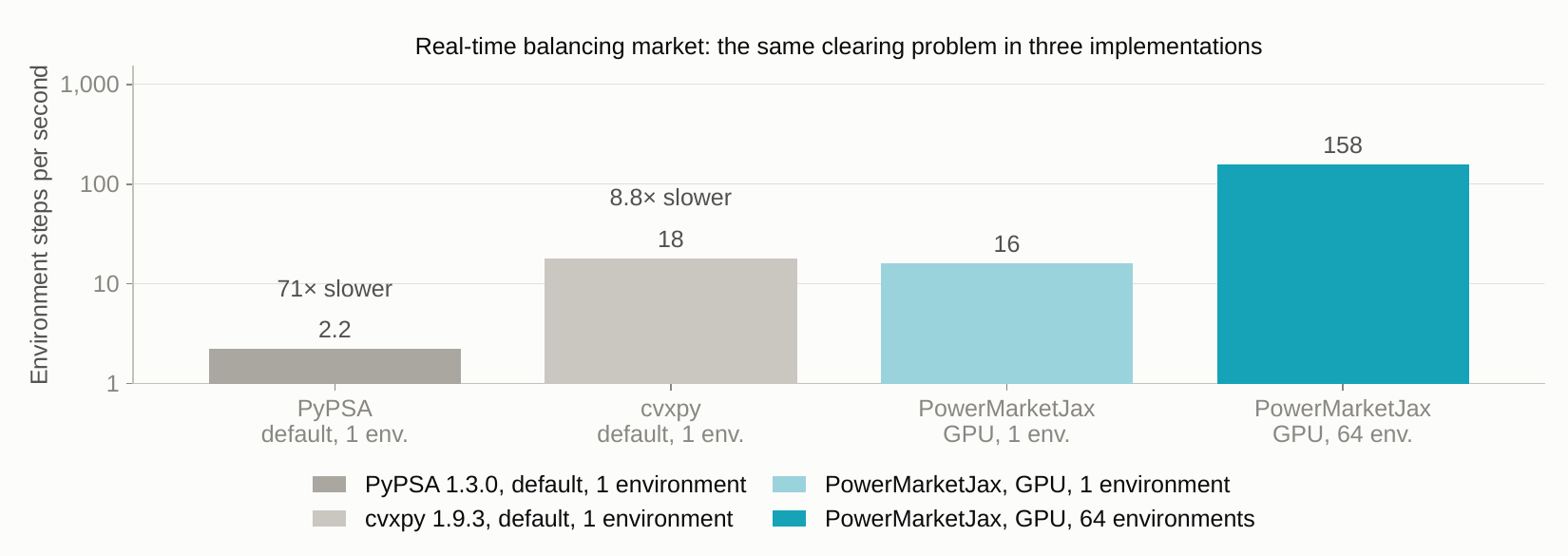}
\caption{Throughput of the real-time balancing market environment in three implementations: PyPSA 1.3.0 and cvxpy 1.9.3 in their default usage, one environment at a time, and PowerMarketJax all on the GPU with 1 and with 64 parallel environments. Each bar is the mean of three runs; the label above each grey bar is how many times slower it is than PowerMarketJax with 64 parallel environments.}
\label{fig.appendix.speed.tools}
\end{figure}

The clearing problem of the M2 real-time balancing market was also implemented in PyPSA~\citep{brown2018-pypsa} and in cvxpy~\citep{diamond2016-cvxpy,agrawal2018-rewriting}, each in its default usage (default formulation and solver), stepping one environment at a time. All on the GPU with 64 parallel environments gives 71 times the throughput of PyPSA and 8.8 times that of cvxpy (Figure~\ref{fig.appendix.speed.tools}). Training PPO on the real-time balancing market for the same number of environment steps, PowerMarketJax all on the GPU is 3.8 times faster than Stable-Baselines3~\citep{raffin2021-sb3} with its default settings (wall-clock time, including gradient updates).

\subsection{Implementation validation}
\label{app:speed:validation}

We validate the implementation in three ways: market clearing and settlement against a NumPy reference implementation; the dual prices of the JAX linear-program solver against HiGHS~\citep{huangfu2018-highs}, an independent solver; and the three-stage clearing of the M1 day-ahead market against the exact solution of the same commitment problem, computed by HiGHS with the commitment decisions kept binary.

\paragraph{Against a NumPy reference implementation.}
Each market has a separate NumPy reference implementation that rewrites the same clearing and settlement rules with explicit matrices and loops. We run both on every period of the evaluation days of each market under truthful offers and compare each quantity (Table~\ref{tab:speed-numpy}). In M3, load shedding and reserve shortfall are compared as totals per period, because the optimum does not fix how they are split across buses and reserve products.

\begin{table}[h]
  \caption{Largest difference between the JAX implementation and the NumPy reference. The relative difference divides the absolute difference by the largest absolute reference value.}
  \label{tab:speed-numpy}
  \centering
  \footnotesize
  \begin{tabular}{lllrr}
    \toprule
    Market & Quantity & Data & Absolute difference & Relative difference \\
    \midrule
    \multirow{4}{*}{M1} & LMP & \multirow{4}{0.22\linewidth}{\raggedright 36 days $\times$ 24 h, case29gb} & $2.2\times10^{-8}$~\$/MWh & $1.4\times10^{-10}$ \\
     & Award &  & $1.3\times10^{-11}$~MW & $1.7\times10^{-15}$ \\
     & Objective value &  & $9.3\times10^{-10}$~\$ & $4.9\times10^{-16}$ \\
     & Settlement (profit) &  & $2.8\times10^{-9}$~\$ & $6.3\times10^{-16}$ \\
    \midrule
    \multirow{3}{*}{M2} & LMP & \multirow{3}{0.22\linewidth}{\raggedright 36 days $\times$ 48 half-hours, case29gb} & $9.1\times10^{-8}$~\$/MWh & $9.1\times10^{-12}$ \\
     & Award &  & $1.6\times10^{-11}$~MW & $2.1\times10^{-15}$ \\
     & Settlement (profit) &  & $1.5\times10^{-10}$~\$ & $1.7\times10^{-17}$ \\
    \midrule
    \multirow{7}{*}{M3} & LMP & \multirow{7}{0.22\linewidth}{\raggedright 36 days $\times$ 48 half-hours, case29gb} & $7.3\times10^{-6}$~\$/MWh & $7.3\times10^{-10}$ \\
     & Reserve price &  & $1.2\times10^{-4}$~\$/MWh & $4.7\times10^{-7}$ \\
     & Award &  & $5.1\times10^{-9}$~MW & $6.5\times10^{-13}$ \\
     & Load shed, total &  & $1.4\times10^{-11}$~MW & $7.7\times10^{-15}$ \\
     & Reserve shortfall, total &  & $9.3\times10^{-12}$~MW & $4.0\times10^{-15}$ \\
     & Objective value &  & $3.0\times10^{-4}$~\$ & $3.0\times10^{-11}$ \\
     & Settlement (profit) &  & $4.7\times10^{-2}$~\$ & $5.5\times10^{-9}$ \\
    \midrule
    \multirow{4}{*}{M4} & Clearing price & \multirow{4}{0.22\linewidth}{\raggedright 64 episodes $\times$ 96 quarter-hours, 1\,200 households} & $6.1\times10^{-6}$~EUR/MWh & $1.8\times10^{-8}$ \\
     & Traded volume &  & $1.4\times10^{-8}$~MWh & $1.0\times10^{-7}$ \\
     & Award &  & $1.9\times10^{-8}$~MWh & $3.9\times10^{-6}$ \\
     & Settlement (profit) &  & $1.8\times10^{-6}$~EUR & $1.0\times10^{-6}$ \\
    \midrule
    \multirow{4}{*}{M5} & Award & \multirow{4}{0.22\linewidth}{\raggedright 36 days $\times$ 24 h, feeder 459\_0} & $3.4\times10^{-9}$~MW & $1.3\times10^{-8}$ \\
     & Load shed &  & $1.5\times10^{-9}$~MW & $5.0\times10^{-7}$ \\
     & Objective value &  & $5.2\times10^{-7}$~CHF & $1.1\times10^{-9}$ \\
     & Settlement (profit) &  & $5.2\times10^{-7}$~CHF & $1.5\times10^{-8}$ \\
    \bottomrule
  \end{tabular}
\end{table}

\paragraph{Dual prices against HiGHS.}
The prices of M1, M2 and M3 are duals of a linear program. On every period of the same evaluation days, we solve the same linear program with HiGHS and compare its prices with those of the JAX solver (Table~\ref{tab:speed-highs}). M4 and M5 set prices without a linear program.

\begin{table}[h]
  \caption{Largest difference between the JAX solver and HiGHS.}
  \label{tab:speed-highs}
  \centering
  \footnotesize
  \begin{tabular}{lllrrr}
    \toprule
    Market & Test system & Price & Periods & Price difference (\$/MWh) & Relative objective difference \\
    \midrule
    M1 & case29gb & LMP & 864 & $2.6\times10^{-8}$ & $3.1\times10^{-16}$ \\
    M1 & case73rts & LMP & 864 & $4.1\times10^{-8}$ & $6.2\times10^{-8}$ \\
    M1 & case813nem & LMP & 864 & $3.5\times10^{-7}$ & $1.2\times10^{-7}$ \\
    M2 & case29gb & LMP & 1\,728 & $1.1\times10^{-7}$ & $1.8\times10^{-12}$ \\
    M3 & case29gb & LMP & 1\,728 & $7.3\times10^{-6}$ & $3.1\times10^{-11}$ \\
    M3 & case29gb & Reserve price & 1\,728 & $2.1\times10^{-7}$ & $3.1\times10^{-11}$ \\
    \bottomrule
  \end{tabular}
\end{table}

\paragraph{Gap to the exact optimum.}
\label{app:speed:gap}
The M1 day-ahead clearing has binary commitment variables, and no mixed-integer solver runs inside the compiled graph, so the market is cleared by a three-stage procedure: a linear program with the on/off decisions relaxed to values between 0 and 1, rounding of the relaxed commitment, and a re-solve of the economic dispatch with the commitment fixed. On the 36 evaluation days of each of the three M1 test systems, under truthful offers, we solve the same unit commitment problem exactly and offline with HiGHS (commitment variables 0 or 1) and compare its production cost with that of the three-stage procedure.

\begin{figure}[h!]
\centering
\includegraphics[width=\linewidth]{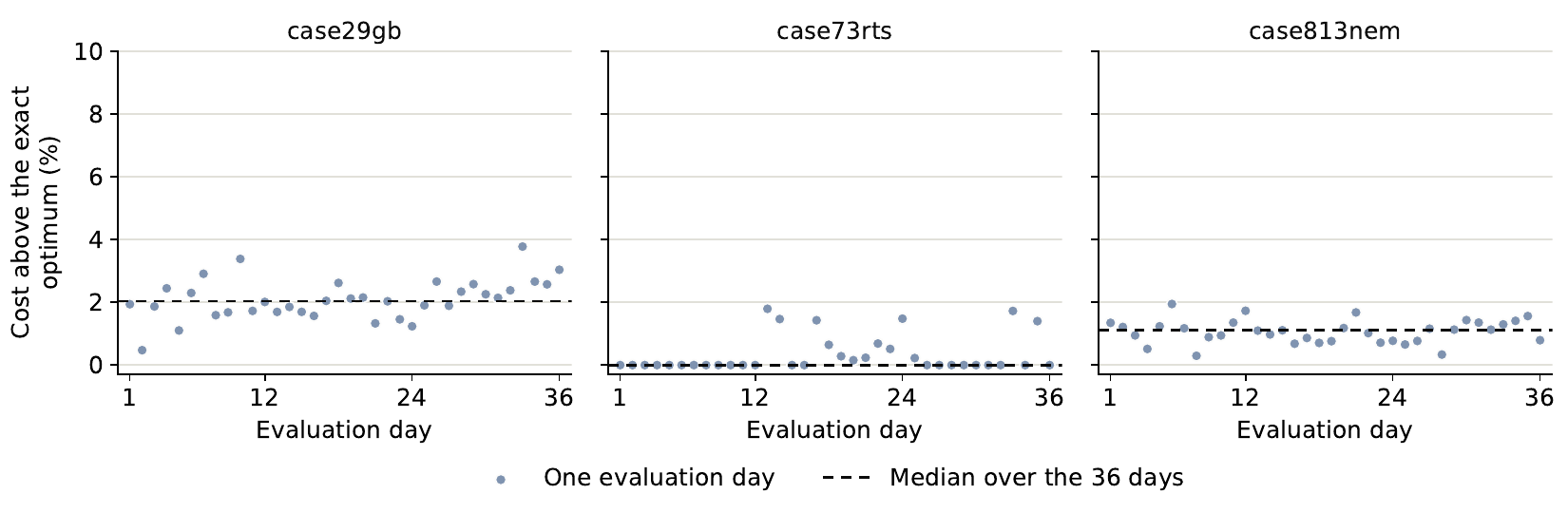}
\caption{Production cost of the three-stage clearing above the exact optimum on the evaluation days of the three M1 test systems, under truthful offers. Each point is one evaluation day, in date order; the dashed line is the median.}
\label{fig.appendix.speed.gap}
\end{figure}

Over the 36 evaluation days, the median excess production cost of the three-stage procedure over the exact optimum is 2.04\% on case29gb, 0 on case73rts and 1.10\% on case813nem, and the maximum is 3.78\%, 1.79\% and 1.95\% (Figure~\ref{fig.appendix.speed.gap}). On 23 of the 36 evaluation days of case73rts, the three-stage commitment is the same as the exact optimum and the gap is 0.

\FloatBarrier
\clearpage

\section{Quickstart and API}
\label{app:api}

This section is a walk-through of how to install PowerMarketJax, run a minimal rollout, read the interface contract that the five markets share, construct each of them, find where an experimental configuration is written down, and reproduce a result from the command line. Listing~C.2 is a complete runnable program, Listing~C.3 a rollout function that takes an environment built by any of the five constructors, Listing~C.5 is abridged from a released source file, and the bash blocks in Sections~\ref{app:api:install} and~\ref{app:api:cli} cover installation and experiment script usage. The market models themselves, including the observation, action, clearing and settlement of each market, are in Appendix~\ref{app:market}; this section covers only the software interface to them.

\subsection{Installation}
\label{app:api:install}

PowerMarketJax requires Python 3.10--3.12 and has no build step. The base install ships a CPU-only JAX; CUDA~12 support is opt-in via the \texttt{cuda} extra, which adds the CUDA plugin alongside the already resolved \texttt{jaxlib} at the same version and therefore does not move the JAX version the suite is verified on. GPU support follows JAX's own platform support: the CUDA build is available on Linux only, so on Windows the package runs on the CPU natively and on the GPU through WSL2. All reported results and test runs were obtained on Linux. The learning stack of \texttt{rlax} and \texttt{distrax} used by the P2P trading experiment script lives behind \texttt{rl}, and the figure scripts behind \texttt{figs}. Nothing under \texttt{powermarketjax/} imports \texttt{matplotlib}, \texttt{rlax} or \texttt{distrax}, so the five markets and both learners are available on a \texttt{dev} install alone.

\begin{lstbox}{Listing C.1: PowerMarketJax Installation commands}
\begin{lstlisting}[style=pmjbash, label={lst:installation}]
git clone https://github.com/powermarketjax/PowerMarketJax
conda create -y -n powermarketjax python=3.11 && conda activate powermarketjax

pip install -e ".[dev]"                   # CPU JAX, the five markets, both learners
pip install -e ".[dev,cuda,rl,figs]"      # Linux + CUDA 12, plus scripts and figures

python -c "import powermarketjax, jax; print(powermarketjax.__version__, jax.__version__, jax.devices())"

\end{lstlisting}
\end{lstbox}

\subsection{Minimal rollout}
\label{app:api:rollout}

Listing~C.2 opens the day-ahead wholesale market on the 29-bus GB case, resets it, and runs a three-step rollout under \texttt{jax.lax.scan} with every agent bidding at twice its cost. One step is one market day, so the rollout clears three consecutive days of a multi-period security-constrained unit commitment; every transition, the clearing included, stays inside \texttt{jit}. 
\begin{lstbox}{Listing C.2: Minimal rollout on \texttt{case29gb} (day-ahead wholesale)}
\begin{lstlisting}[style=pmjpython,label={lst:rollout}]
import jax, jax.numpy as jnp
jax.config.update("jax_enable_x64", True) # M1/M2/M3/M5 refuse to build without it

from powermarketjax.case import load_case
from powermarketjax.envs.day_ahead import make_env, load_commitment, load_gb_demand
from powermarketjax.utils.jax_utils import scan_rollout

env, spec = make_env(load_case("29gb"),
                     load_commitment(n_periods=24), load_gb_demand(),
                     kind="markup", markup_max=2.0,
                     cap_scale=0.60, ramp_scale=1.00)
params = env.make_params(episode_len=3)

obs, state = env.reset(jax.random.PRNGKey(0), params)
actions = jnp.full((3, spec["n_agents"]), 2.0)          # (n_steps, n_agents)
final_state, obs_traj, reward_traj, cost_traj, done_traj, info_traj = jax.jit(lambda k, s: scan_rollout(env, k, s, params, actions)
)(jax.random.PRNGKey(1), state)
\end{lstlisting}
\end{lstbox}
At this configuration the market has 66 agents, one per thermal unit: \texttt{obs\_traj} is \texttt{(3, 66, 112)}, \texttt{reward\_traj} is \texttt{(3, 66)} and \texttt{cost\_traj} is \texttt{(3, 66, 2)}, and \texttt{info\_traj["converged"]} reports per step whether the clearing met its tolerance. \texttt{scan\_rollout} drives \texttt{step\_auto\_reset}, so episodes restart at their boundary without a Python conditional and the trajectory has a fixed length.

\emph{The scenario parameters have no defaults.} \texttt{cap\_scale} and \texttt{ramp\_scale} in Listing~C.2 are required: omitting either raises \texttt{ValueError}. The two scale the line ratings and the ramp limits the case ships with, so \texttt{cap\_scale} 1.00 means every line keeps the capacity written in the case file and in that network no line ever reaches its limit. With no line at its limit there is no congestion, and with no congestion every bus clears at the same price.

Listing~C.3 is the same contract on any of the five markets, vectorised over a batch of environments. \texttt{unpack\_env} accepts whichever shape a market's constructor returns (Section~\ref{app:api:markets}) and yields the same four objects, so the rollout below is market-agnostic.
\begin{lstbox}{Listing C.3: One rollout function for five markets: \texttt{vmap} over environments inside \texttt{scan} over time}
\begin{lstlisting}[style=pmjpython,label={lst:batched}]
import math, jax, jax.numpy as jnp, numpy as np
from functools import partial
from powermarketjax.learning import unpack_env
from powermarketjax.resources.battery import make_battery_bundle
from powermarketjax.envs.p2p import (make_p2p_env, make_p2p_params, load_fluvius_households)
jax.config.update("jax_enable_x64", True) # M1/M2/M3/M5 refuse to build without it
# `built` is whatever a market's constructor returned: the (env, spec) pair  for M1/M2, or the (reset, step, step_auto_reset, spec) tuple for M3-M5.  Here it is M4, the P2P trading market, on 12 metered houses.
N, DT, ETA = 12, 0.25, math.sqrt(0.85)
series = load_fluvius_households(n_households=N)
battery = make_battery_bundle(n_devices=N, dt_hours=DT,
                              capacity_mwh=[0.01] * N,
                              power_mw=[0.005] * N,
                              eta_charge=ETA, eta_discharge=ETA,
                              soc_min=0.15, soc_max=1.0,
                              initial_soc=0.5,
                              cycle_cost_per_mwh=0.0)
params = make_p2p_params(p_pv=series.injection, load=series.offtake,
                         battery=battery, 
                         kappa=np.full(N, 13.88, np.float32),
                         learner_mask=np.ones(N, bool), 
                         episode_len=96)
built = make_p2p_env(N, 73.0, 333.4, DT)   # n_agents, export/retail, dt
reset, step, step_auto_reset, spec = unpack_env(built)
# unpack_env takes either shape and gives the four names below. 

def rollout(key, params, policy, n_envs, n_steps):
    k_reset, k_roll = jax.random.split(key)
    obs, state = jax.vmap(reset, in_axes=(0, None))(
        jax.random.split(k_reset, n_envs), params)

    def one_step(carry, k):
        obs, state = carry
        action = policy(obs)                     # (n_envs, *action_shape)
        obs, state, reward, costs, done, info = jax.vmap(
            step_auto_reset, in_axes=(0, 0, 0, None))(
                jax.random.split(k, n_envs), state, action, params)
        return (obs, state), (reward, costs, done)

    return jax.lax.scan(one_step, (obs, state),
                        jax.random.split(k_roll, n_steps))

# The action shape comes from the market, not from the learner.
zero = lambda obs: jnp.zeros((obs.shape[0], *spec["action_shape"]),
                             jnp.float32)
(obs, state), (reward, costs, done) = jax.jit(
    partial(rollout, policy=zero, n_envs=8, n_steps=6)
)(jax.random.PRNGKey(0), params)
\end{lstlisting}
\end{lstbox}
The returned \texttt{reward} has shape \texttt{(n\_steps, n\_envs, N)} and \texttt{costs} has shape \texttt{(n\_steps, n\_envs, N, C)}: time, parallel environments and market participants are three separate axes, and the second and third are the two the accelerator exploits. \texttt{params} is closed over as a broadcast argument, so a sweep over anything carried there is a loop over calls and not a recompilation.

\subsection{The environment interface}
\label{app:api:interface}
\emph{Signatures.} Every market exposes the same three pure functions:

\begin{lstbox}{Listing C.4: Market environment signatures}
\begin{lstlisting}[style=pmjpython]
obs, state = reset(key, params)
obs, state, reward, costs, done, info = step(key, state, action, params)
obs, state, reward, costs, done, info = step_auto_reset(key, state, action, params)
\end{lstlisting}
\end{lstbox}
With $N$ agents and $C$ constraint channels, \texttt{obs} is $(N, d)$, \texttt{reward} is $(N,)$, \texttt{costs} is $(N, C)$ and \texttt{done} is a scalar boolean. The agent index is one array axis: observations, actions, rewards and costs are aligned on it, and $N$ is fixed at compile time.

\emph{Reward and constraint costs.} The reward of an agent is the profit it earns in that step, settled on the quantities and prices produced by the market's own clearing (Appendix~\ref{app:market}). Violations of physical or market constraints, such as shed load or an unmet reserve requirement, are returned separately in \texttt{costs}, one column per name in \texttt{spec["cost\_names"]}, and are never subtracted from the reward. A user decides how to handle them, for example by adding a penalty to the reward or by using a constrained learning method.

\emph{Episode boundary.} \texttt{spec["termination"]} tells a learner what the end of an episode means. In four markets it is \texttt{"truncation"}: the episode is cut off in time, the market would continue, and the value of the last state should be bootstrapped. In the P2P market it is \texttt{"terminal"}: the last reward already pays each household for the energy left in its battery, so bootstrapping would count that energy twice.

\subsection{Constructing the five markets}
\label{app:api:markets}
Each market has its own constructor, since the scenarios differ: the wholesale markets take a network case, a precomputed day-ahead commitment and schedule, and a demand series; the ancillary market adds the reserve products; the P2P trading market takes a population size and the export and retail tariffs; the local flexibility market takes a feeder and the placement of the aggregators. The ancillary, P2P and local flexibility constructors return \texttt{(reset, step, step\_auto\_reset, spec)}, and the day-ahead and real-time constructors return \texttt{(env, spec)}; \texttt{powermarketjax.learning.unpack\_env} converts either to the first form, as in Listing~C.3. Anything that changes the size of the clearing problem, such as the network or the placement of agents, is fixed when the environment is built, and changing it requires recompiling; anything that varies within a market, such as the demand series or the episode length, is passed in \texttt{params} on every call.

\subsection{Scenario and learner configuration}
\label{app:api:runpoint}
There is no configuration file format: an experimental configuration is Python. The learner configuration is \texttt{tools/benchmark/hyperparams.py}, abridged in Listing~C.5. The scenario parameters are module-level constants of each market's experiment script, each also exposed as a command-line flag.
\begin{lstbox}{Listing C.5: The shared learner configuration }
\begin{lstlisting}[style=pmjpython,label={lst:hp}]
N_ENVS, MINIBATCHES = 64, 32
HORIZON = {"01": 4, "02": 48, "03": 48}   # steps per update, per market
for _m, _h in HORIZON.items():            # the batch must divide into minibatches
    assert (N_ENVS * _h) % MINIBATCHES == 0

SHARED = IPPOConfig(
    n_envs=N_ENVS,           
    horizon=HORIZON["02"],  # ours, per market via shared_for()
    epochs=10,              # update_epochs
    minibatches=MINIBATCHES,# num_minibatches
    lr=3e-4,                # learning_rate (no annealing here)
    clip_eps=0.2,           # clip_coef
    gamma=0.99,             # gamma
    gae_lambda=0.95,        # gae_lambda
    vf_coef=0.5,            # vf_coef
    ent_coef=0.0,           # ent_coef
    max_grad_norm=0.5,      # max_grad_norm
    hidden=(64, 64),        # Agent: two tanh layers of 64
    weight_decay=0.0,       # zero reproduces optax.adam exactly
    init_scale=1.0,         
)

def shared_for(market):
    return dataclasses.replace(SHARED, horizon=HORIZON[market])
\end{lstlisting}
\end{lstbox}

\subsection{Command-line usage}
\label{app:api:cli}
Each market has one experiment script for its learning runs. The non-learning baselines are run by a separate script for the day-ahead, real-time and ancillary markets (\texttt{tools/benchmark/run\_eval\_0*.py}), by the learning script itself for the P2P trading market (\texttt{--arm truthful}) and for the local flexibility market, where they run alongside the learners. The learning script takes the learner (\texttt{--algo ippo|sac}), parameter sharing or not (\texttt{--per-agent-params} for one network per agent), the seed, the iteration count, the scenario parameters and an output directory. It trains on the training days only and then evaluates the mean action of the policy on the held-out days, so the reported number excludes exploration noise.

\begin{lstbox}{Listing C.6: Command line for RL training}
\begin{lstlisting}[style=pmjbash]
# One-year inputs for M1-M3 (2023-07-10 to 2024-07-08), built once.
python tools/commitment/precommit.py --mode relax --start-day 5 --days 365 --cap-scale 0.6 --ramp-scale 1.0 --out commit_y1.npz
python tools/commitment/da_position.py --case 29gb --fixture commit_y1.npz --out position_y1.npz

# M1 day-ahead: parameter sharing, then one network per agent.
python tools/benchmark/run_rl_01.py --fixture commit_y1.npz --algo ippo --seed 0 --iterations 400 --out-dir runs/m1-ippo-ps
python tools/benchmark/run_rl_01.py --fixture commit_y1.npz --algo ippo --seed 0 --iterations 400 --per-agent-params --out-dir runs/m1-ippo-nops

# M2 real-time and M3 ancillary settle against the day-ahead position.
python tools/benchmark/run_rl_02.py --position position_y1.npz --seed 0 --out-dir runs/m2-ippo-ps
python tools/benchmark/run_rl_03.py --position position_y1.npz --seed 0 --out-dir runs/m3-ippo-ps

# M4 P2P, 1200 households.
python tools/benchmark/run_rl_04.py --agents 1200 --seed 0 --out-dir runs/m4-ippo-ps

# M5 local flexibility.
PYTHONPATH=tools/flex_experiment python -m concentration_baseline --algo ippo --seeds 3 --out runs/m5-ippo-ps

# Non-learning baselines for M1 on the held-out days.
python tools/benchmark/run_eval_01.py --fixture commit_y1.npz --days eval --out-dir results/m1-baselines
\end{lstlisting}
\end{lstbox}

\subsection{Cases and data}
\label{app:api:data}

\texttt{powermarketjax.case.list\_cases()} returns the 18 built-in network cases, 10 transmission and 8 distribution, sorted by bus count and with aliases removed; \texttt{load\_case} accepts a canonical identifier or an alias. The demand, generation, price and smart-meter series are carried in the repository as parquet files, so every listing above runs on a fresh checkout with no separate download step.

\subsection{Clearing and learning in a single computation graph}
\label{app:api:graph}

\begin{figure}[h!]
\centering
\includegraphics[width=\linewidth]{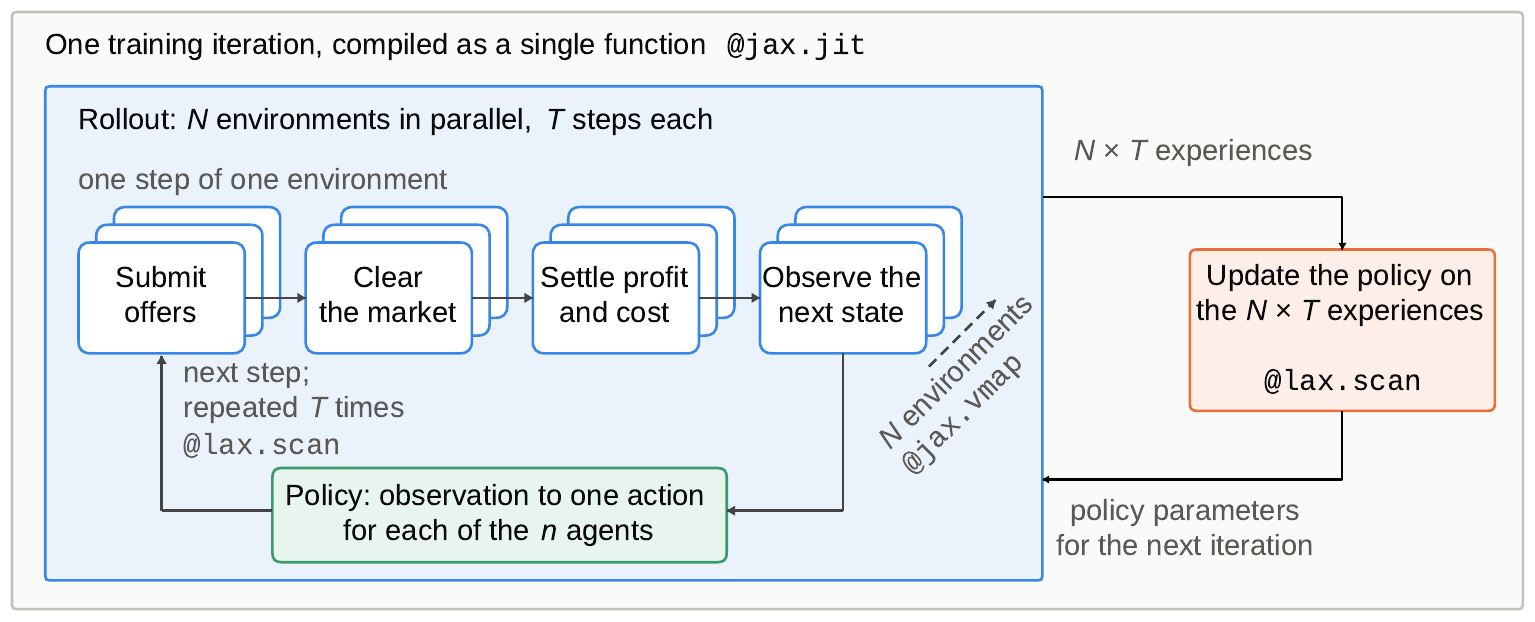}
\caption{One training iteration of IPPO, compiled as a single function. Rollout: $N$ environments advance in parallel for $T$ steps; in each step the $n$ agents submit offers, the market clears, the awards are settled and the next observation is constructed, and the policy maps the observation to the next action inside the same function. Update: the policy is updated on the $N \times T$ experiences, with a second scan over epochs and minibatches (Table~\ref{tab:train-hp}). The values of $N$ and $T$ for each market are in Table~\ref{tab:train-scale}.}
\label{fig.appendix.graph}
\end{figure}

A training iteration has two parts, a rollout and an update. 
In the rollout, $N$ environments advance together for $T$ steps. In each step the participants submit offers, the market clears, the awards are settled, and the next observation is constructed; the policy maps the observation to the action at every step, and the observation, action, reward and constraint cost of one step form one experience. 
The update takes gradient steps on the $N \times T$ experiences of the rollout, split by epoch and minibatch. The whole iteration, including the clearing problem of every step, is compiled as one function on the GPU: the host makes one call and receives the training metrics, and every clearing, every policy forward pass and every gradient step in between runs on the GPU in sequence. Figure~\ref{fig.appendix.graph} shows the data flow of one iteration, and Listing~C.7 writes the same iteration as loops, with the JAX primitive that implements each loop in the comment above it. The values of $N$ and $T$ for each market are in Table~\ref{tab:train-scale}, the epochs and minibatches in Table~\ref{tab:train-hp}, and $n$, the number of agents, in Appendix~\ref{app:market}.

The loop over the $N$ environments is \texttt{vmap(step)}. The function \texttt{step} advances one environment by one step and is written for one environment only; \texttt{vmap} returns the function that advances $N$ environments at once, whose input and output gain one axis of length $N$. The clearing, settlement and observation of the $N$ environments are one batched computation over $N$ sets of data, not $N$ calls one after another. The forward pass of the policy goes through \texttt{vmap} in the same way, so the actions are produced inside the same function and passed directly to \texttt{step}, and the observations never reach the host.

\FloatBarrier
\begin{lstbox}[unbreakable]{Listing C.7: One training iteration as loops, with the primitive that implements each loop}
\begin{lstlisting}[style=pmjyaml,label={lst:graph}]
# iteration = jit(iteration): the whole body below is one compiled
#   function on the GPU, called once per training iteration
iteration(params, states, observations):
    # rollout
    # scan(one_step, carry, length=T): carry = (states, observations);
    #   returns the final carry and the T records stacked
    for t = 1 .. T:
        # policy_N = vmap(policy), step_N = vmap(step):
        #   each takes N inputs and returns N outputs at once
        for each of the N environments:
            for each of the n agents: action <- policy(observation)
            # step(state, actions), one environment
            offers          <- submit(actions)
            awards, prices  <- clear(offers)
            reward, cost    <- settle(awards, prices)
            observation     <- observe(the state after clearing)
            if the episode ended: reset this environment in place
        # one record per step, holding the N environments
        record (observations, actions, rewards, costs)
    # N x T experiences, each with n agents
    experiences <- the T records
    # update
    # scan(gradient_step, ...): the passes over the experiences,
    #   by epoch and minibatch
    for each pass over the experiences:
        params <- optimiser step on the gradient of the loss
    return params, states, observations, training metrics
\end{lstlisting}
\end{lstbox}

The loop over the $T$ steps is \texttt{lax.scan(one\_step, carry, length=T)}. It takes the loop body \texttt{one\_step}, the initial carry, which is the $N$ states and the $N$ observations, and the number of steps $T$; runs the body $T$ times, feeding the carry returned by one run into the next; stacks the record of each run into an array of length $T$; and returns the final carry together with this array. The horizon $T$ is fixed, and an environment whose episode ends before step $T$ is reset in place inside the body, so the input and output of the body have the same shape at every step, which is the condition under which \texttt{lax.scan} applies. The gradient steps of the update are a second \texttt{lax.scan}. None of these loops runs on the host, and the host does not see the boundary between two consecutive steps.

The outer function is \texttt{jit(iteration)}: the rollout followed by the update, compiled for the GPU at the first call and called once per iteration. The host passes in the parameters, the environment states and the observations, and receives the updated parameters, states, observations and training metrics; no intermediate result returns to the host, and the only loop left on the host is the loop over iterations. The iteration of SAC has the same form: the rollout is the same, and the update first places the experiences of the rollout into a replay buffer of fixed size, which also lives inside this function and is carried from one iteration to the next, then samples from it for the gradient steps, which are again a \texttt{lax.scan}.

\clearpage

\section{Training Setup and Hyperparameters}
\label{app:train}

The results of Section~\ref{sec.results} and Appendix~\ref{app:results} share four learners, one set of hyperparameters and one evaluation protocol; where a market departs from them, its subsection of Appendix~\ref{app:results} gives the difference. Appendix~\ref{app:api:runpoint} shows where these values are set in the code.

\subsection{Four learners}
\label{app:train:learners}

Each market is run with four learners: IPPO-PS, IPPO-NoPS, SAC-PS and SAC-NoPS. All four are independent learners: every agent sees only its own observation and optimises only its own profit, with no centralised value network. PS and NoPS stand for parameter sharing and no parameter sharing. Under PS all agents share one set of networks and pool their samples for training; under NoPS each agent has its own set and trains only on its own samples. For IPPO the set is one actor and one value network that share two $\tanh$ layers, and the log standard deviation of the actor does not depend on the observation; for SAC it is one stochastic actor and a pair of Q-networks with their target networks, with an automatically tuned temperature.

\subsection{Hyperparameters}
\label{app:train:hp}

The hyperparameters build on the defaults of the continuous-action PPO and SAC implementations in CleanRL; the five markets use one set, listed in Table~\ref{tab:train-hp}, without tuning for each market. The profit of one step in M1 is of the order of $10^5$ to $10^6$ dollars and the Q regression of SAC is sensitive to this magnitude, so SAC divides the reward by a constant before the Q regression: the standard deviation of the rewards under truthful offers, computed once before training. SAC runs the gradient steps of an iteration as one block after the rollout. In all four learners each component of the observation is normalised with constants fixed before training.

\begin{table}[h]
  \caption{Hyperparameters of the two IPPO and the two SAC learners, shared by the five markets.}
  \label{tab:train-hp}
  \centering
  \footnotesize
  \begin{tabularx}{\linewidth}{>{\raggedright\arraybackslash}X>{\raggedright\arraybackslash}p{0.26\linewidth}>{\raggedright\arraybackslash}p{0.34\linewidth}}
    \toprule
    Hyperparameter & IPPO & SAC \\
    \midrule
    Optimiser & Adam & Adam \\
    Learning rate & $3 \times 10^{-4}$, constant & actor $3 \times 10^{-4}$, Q-networks and temperature $10^{-3}$ \\
    Discount factor $\gamma$ & 0.99 & 0.99 \\
    Hidden layers & two layers of 64 units, $\tanh$ & two layers of 256 units, ReLU \\
    GAE parameter $\lambda$ & 0.95 & --- \\
    Clipping range & 0.2 & --- \\
    Update epochs per iteration & 10 & --- \\
    Minibatches per epoch & 32 & --- \\
    Value loss coefficient & 0.5 & --- \\
    Entropy coefficient & 0 & --- \\
    Maximum gradient norm & 0.5 & --- \\
    Initialisation scale of the actor output layer & 1.0 & --- \\
    Replay buffer & --- & 32\,768 environment steps \\
    Batch size & --- & 256 environment steps, each carrying the transitions of all agents \\
    Gradient steps per environment step & --- & 1 \\
    Random-action steps before training & --- & 0 \\
    Actor update interval & --- & every gradient step \\
    Target smoothing coefficient $\tau$ & --- & 0.005 \\
    Temperature & --- & initial value 0.2, tuned automatically \\
    Target entropy & --- & minus the action dimension \\
    Range of the log standard deviation & --- & $[-5, 2]$ \\
    \bottomrule
  \end{tabularx}
\end{table}

The two SAC learners of M4 and the two IPPO learners of M5 differ from Table~\ref{tab:train-hp} in the settings given in Sections~\ref{app:results:p2p:training} and~\ref{app:results:flex:setting}.

\subsection{Training scale}
\label{app:train:scale}

Table~\ref{tab:train-scale} gives the training scale of each market.

\begin{table}[h]
  \caption{Training scale.}
  \label{tab:train-scale}
  \centering
  \footnotesize
  \begin{tabularx}{\linewidth}{>{\raggedright\arraybackslash}p{0.17\linewidth}>{\raggedright\arraybackslash}p{0.08\linewidth}>{\raggedright\arraybackslash}p{0.12\linewidth}>{\raggedright\arraybackslash}X>{\raggedright\arraybackslash}p{0.13\linewidth}}
    \toprule
    Market & Seeds per learner & Iterations & Parallel environments and steps per iteration & Environment steps per iteration \\
    \midrule
    M1 day-ahead wholesale & 3 & 400 & 64 environments, 4 steps each & 256 \\
    M2 real-time balancing & 3 & 200 & 64 environments, 48 steps each & 3\,072 \\
    M3 ancillary services & 3 & 200 & 64 environments, 48 steps each & 3\,072 \\
    M4 peer-to-peer local energy & 3 & 400 & 64 environments, 96 steps each & 6\,144 \\
    M5 local flexibility & 3 to 10 & early stopping\textsuperscript{a}, at most 300 & 64 environments, 24 steps each & 1\,536 \\
    \bottomrule
  \end{tabularx}
  \par\smallskip
  \parbox{\linewidth}{\scriptsize\raggedright \textsuperscript{a}\,M5 evaluates a checkpoint on the validation days every 10 iterations and stops early once the mean validation return of the last 5 checkpoints exceeds that of the 5 checkpoints before them by 0 to 5\%.}
\end{table}

\subsection{Evaluation protocol}
\label{app:train:eval}

M1 to M4 evaluate the policy at the last iteration. In M1 to M3, evaluation takes the mean action of the policy and runs day by day on the 36 evaluation days of the one-year window of case29gb, 2023-07-10 to 2024-07-08 (the dates of the other two test systems are in Table~\ref{tab:app-da-data}). In M4, evaluation also takes the mean action and runs on 1\,024 held-out episodes (Table~\ref{tab:app-p2p-data}). In M5, the checkpoint with the highest validation return is evaluated on the 36 test days with sampled actions (Section~\ref{app:results:flex:setting}). The training return is the return under sampled actions and includes exploration noise.

The untrained network is the network of the same parameter layout (PS or NoPS) at iteration 0, evaluated in the same way.

Truthful offers are the reference policy in which every agent offers at its true cost: in M1 to M3 every unit takes the markup $\alpha_i = 1$, with a reserve offer of 0 in M3; in M4 the batteries are kept still, sellers ask the export price and buyers bid the retail tariff; and in M5 every aggregator offers its full deliverable quantity at the replacement cost and plans no charging.
\clearpage

\section{Additional Market Results}
\label{app:results}

\subsection{M1: Day-ahead Wholesale Market}
\label{app:results:da}

This section first presents detailed data, training curves and evaluation results of the three test systems, and then elaborates the finding in Section~\ref{sec.results.da} with the supply curves, the offer distribution and the joint sampling probability on the British test system.

\subsubsection{Market setting}
\label{app:results:da:data}

Each unit $i$ offers a price--quantity curve with a single segment ($K = 1$ in Section~\ref{app:da:clearing}). Its offer price is $\alpha_i m_i$, where $m_i$ is its segment cost (the mean of its true marginal cost over its output range) and $\alpha_i \in [1, \bar{\alpha}]$ is its markup. At $\alpha_i = 1$ the offer is truthful, and $\bar{\alpha} > 1$ is the markup cap. A markup above one represents economic withholding: the unit offers its capacity at a price above its cost.

We consider three test systems. As shown in Table~\ref{tab:app-da-data}, each test system uses one year of data, of which 36 days are used for evaluation and the rest for training. On each test system, three selected days are taken from the 36 evaluation days: the days at the 10th, 50th and 90th percentiles of total daily demand, referred to below as the low-load day, the mid-load day and the high-load day.

\begin{table}[h!]
  \caption{Data of the three test systems.}
  \label{tab:app-da-data}
  \centering
  \footnotesize
  \renewcommand{\tabularxcolumn}[1]{m{#1}}
  \begin{tabularx}{\linewidth}{>{\raggedright\arraybackslash}m{0.12\linewidth}YY>{\centering\arraybackslash}m{0.11\linewidth}>{\centering\arraybackslash}m{0.1\linewidth}>{\centering\arraybackslash}m{0.08\linewidth}}
    \toprule
    Test system & Grid & Demand data & Time span & Training / evaluation days & Markup cap $\bar{\alpha}$ \\
    \midrule
    case29gb & A reduced 29-bus model of the GB transmission network\textsuperscript{a} (network diagram\textsuperscript{b}), 66 units & GB transmission system demand (NESO)\textsuperscript{c}, day-ahead forecast from Elexon\textsuperscript{d} & 2023-07-10 to 2024-07-08 & 329 / 36 & 2 \\
    \midrule
    case73rts & The RTS-GMLC test system\textsuperscript{e}, 73 units & Load, wind, solar and hydro of RTS-GMLC\textsuperscript{f} & 2020-01-01 to 2020-12-31 & 330 / 36 & 1.4 \\
    \midrule
    case813nem & An open grid model of the Australian National Electricity Market\textsuperscript{g}, 151 units & Operational demand and day-ahead forecast of the four mainland regions from the Australian Energy Market Operator (AEMO)\textsuperscript{h} & 2025-02-01 to 2026-01-31 & 329 / 36 & 2 \\
    \bottomrule
  \end{tabularx}
  \par\smallskip
  \parbox{\linewidth}{\scriptsize\raggedright \textsuperscript{a}\,\url{https://webhomes.maths.ed.ac.uk/OptEnergy/NetworkData/reducedGB/}; \textsuperscript{b}\,\url{https://webhomes.maths.ed.ac.uk/OptEnergy/NetworkData/reducedGB/GBreducednetwork.pdf}; \textsuperscript{c}\,\url{https://www.neso.energy/data-portal/historic-demand-data}; \textsuperscript{d}\,\url{https://bmrs.elexon.co.uk/}; \textsuperscript{e}\,\url{https://github.com/GridMod/RTS-GMLC}; \textsuperscript{f}\,\url{https://github.com/GridMod/RTS-GMLC/tree/master/RTS_Data/timeseries_data_files}; \textsuperscript{g}\,\url{https://github.com/akxen/egrimod-nem-dataset}; \textsuperscript{h}\,\url{https://visualisations.aemo.com.au/aemo/nemweb/index.html}.}
\end{table}

\begin{figure}[h!]
\centering
\includegraphics[width=\linewidth]{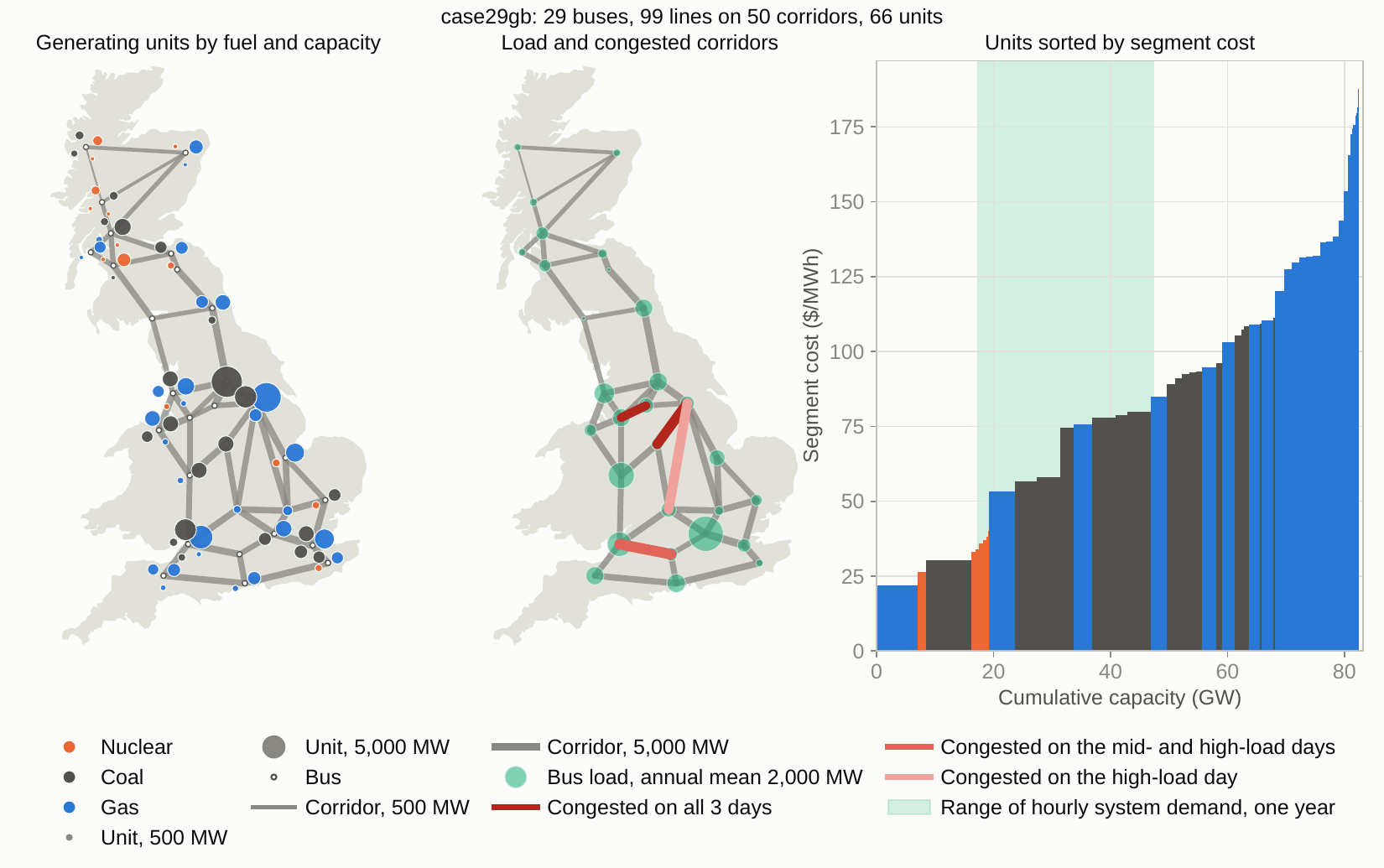}
\caption{The test system case29gb. Left: the 66 units at their buses, coloured by fuel and sized by maximum output; corridor width scales with capacity. Middle: annual mean bus load and the corridors congested under truthful offers on the three example days. Right: the supply curve ordered by segment cost, against the range of hourly system demand over one year. Generation sits mostly in the north and load mostly in the south, so the congested corridors are central and southern. All 14 nuclear units, 4\,359~MW in total, are among the 16 cheapest units, which sum to 19.2~GW; hourly demand ranges from 17.2 to 47.4~GW and exceeds 19.2~GW in almost every hour. The nuclear units therefore sit near the left of the supply curve with demand almost always to their right: a unilateral markup does not move the price, while a joint markup lifts the whole curve and the price with it.}
\label{fig.appendix.da.system}
\end{figure}

In the British test system case29gb, generation lies mostly in the north and load mostly in the south, and power flows from north to south: the north has 63\% of the total capacity but only 43\% of the load. The three example days are 2024-06-23, 2024-03-16 and 2024-01-23. Under truthful offers, 2, 3 and 5 lines are congested on these days, all on central and southern corridors (Figure~\ref{fig.appendix.da.system}). Of the 66 units, 14 are nuclear, 23 coal and 29 gas. In order of segment cost, all 14 nuclear units are among the 16 cheapest units, and their total capacity is only 4.359~GW. These 16 units have a total capacity of 19.2~GW, while hourly system demand over one year ranges from 17.2 to 47.4~GW and is above 19.2~GW in almost every hour. The nuclear units therefore sit near the left of the supply curve with demand almost always to their right, which means a unilateral markup by one nuclear unit does not move the price, and only a joint markup raises it.

\begin{figure}[h!]
\centering
\includegraphics[width=\linewidth]{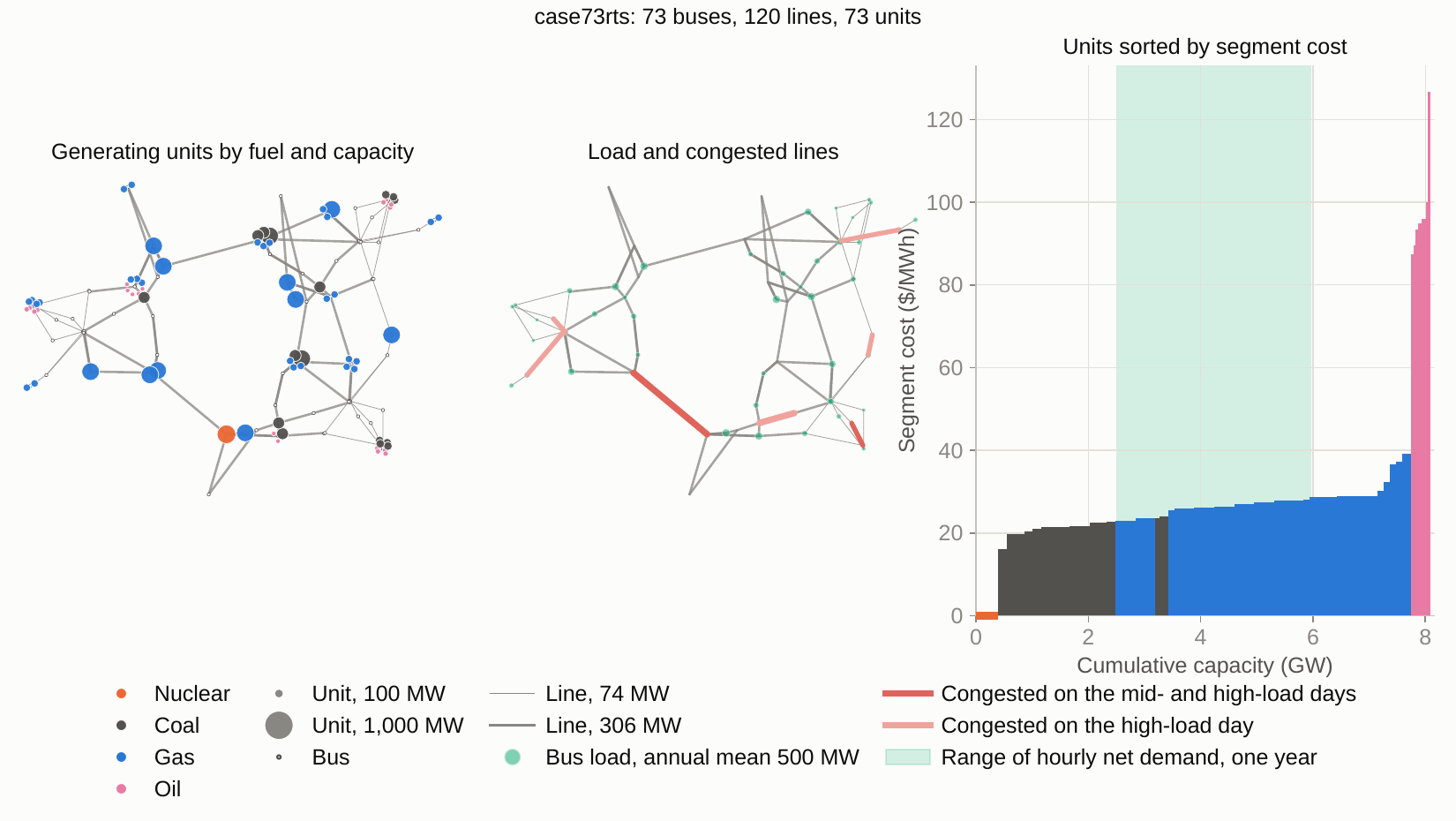}
\caption{The test system case73rts, drawn as Figure~\ref{fig.appendix.da.system}: the three-area RTS-GMLC system with 73 units. Left: units by fuel and maximum output. Middle: annual mean bus load and the lines congested under truthful offers on the three example days. Right: the supply curve ordered by segment cost against the range of hourly net demand over one year. Gas units make up most of the supply curve; no line is congested on the low-load day and seven are on the high-load day.}
\label{fig.appendix.da.rts}
\end{figure}

In the  RTS-GMLC test system, case73rts, wind, solar and hydro (referred to below as renewables) do not submit offers, and the demand cleared by the market is the net demand (total load minus renewable output). The three example days are 2020-02-24, 2020-03-24 and 2020-07-16. Under truthful offers, 0, 2 and 7 lines are congested on these days (Figure~\ref{fig.appendix.da.rts}).

\begin{figure}[h!]
\centering
\includegraphics[width=\linewidth]{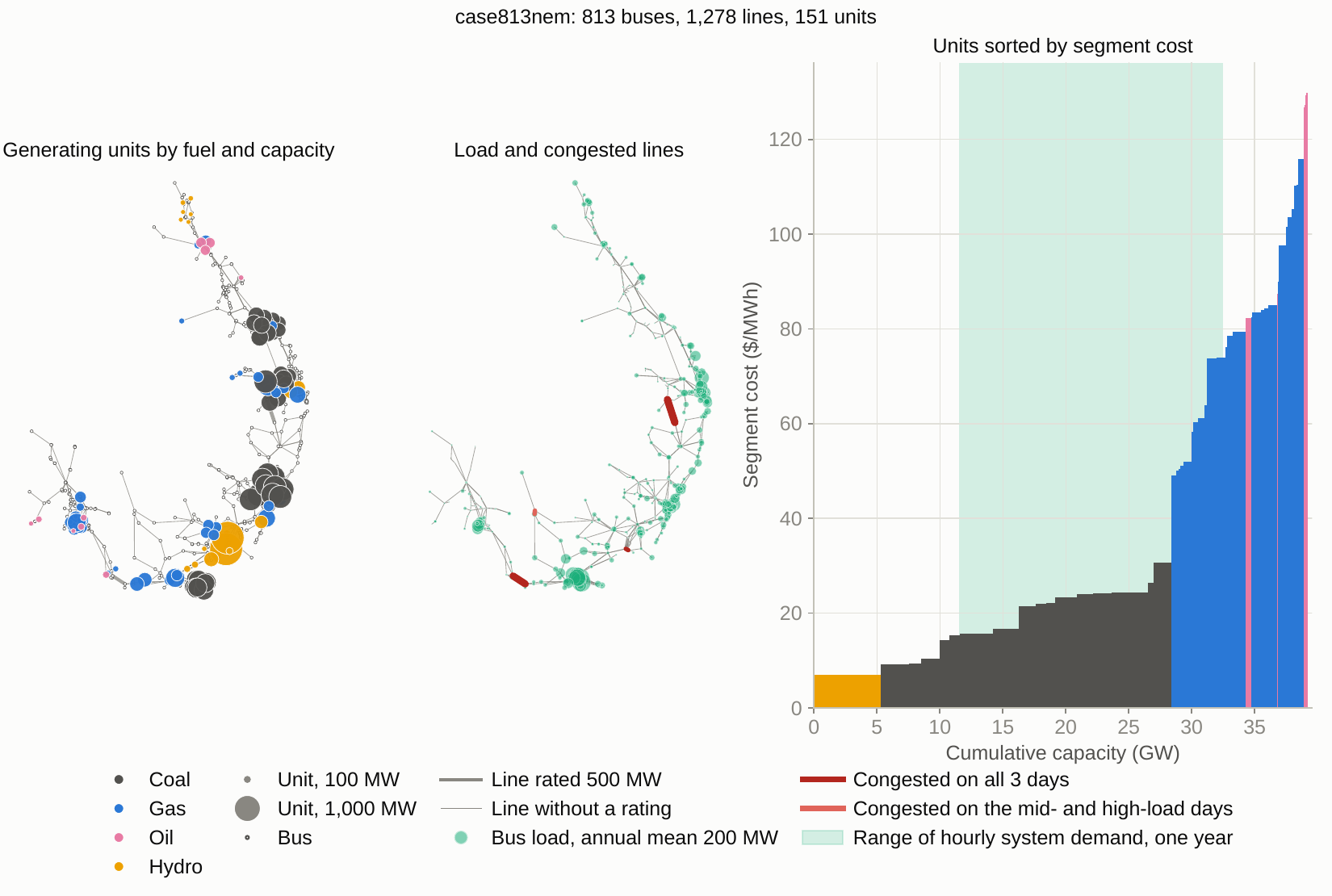}
\caption{The test system case813nem, drawn as Figure~\ref{fig.appendix.da.system}: an open grid model of the mainland Australian National Electricity Market with 151 units. Only 7 of its 1\,278 lines carry a published rating. Left: units by fuel and maximum output. Middle: annual mean bus load and the lines congested under truthful offers on the three example days. Right: the supply curve ordered by segment cost against the range of hourly demand over one year. Hydro and then coal fill the left of the supply curve; three lines are congested on all three days and a fourth on the mid- and high-load days.}
\label{fig.appendix.da.nem}
\end{figure}

Regarding the Australian test system, case813nem, the three example days are 2025-09-20, 2025-12-10 and 2025-06-24. Under truthful offers, 3, 4 and 4 lines are congested on these days (Figure~\ref{fig.appendix.da.nem}).

\begin{figure}[h!]
\centering
\includegraphics[width=\linewidth]{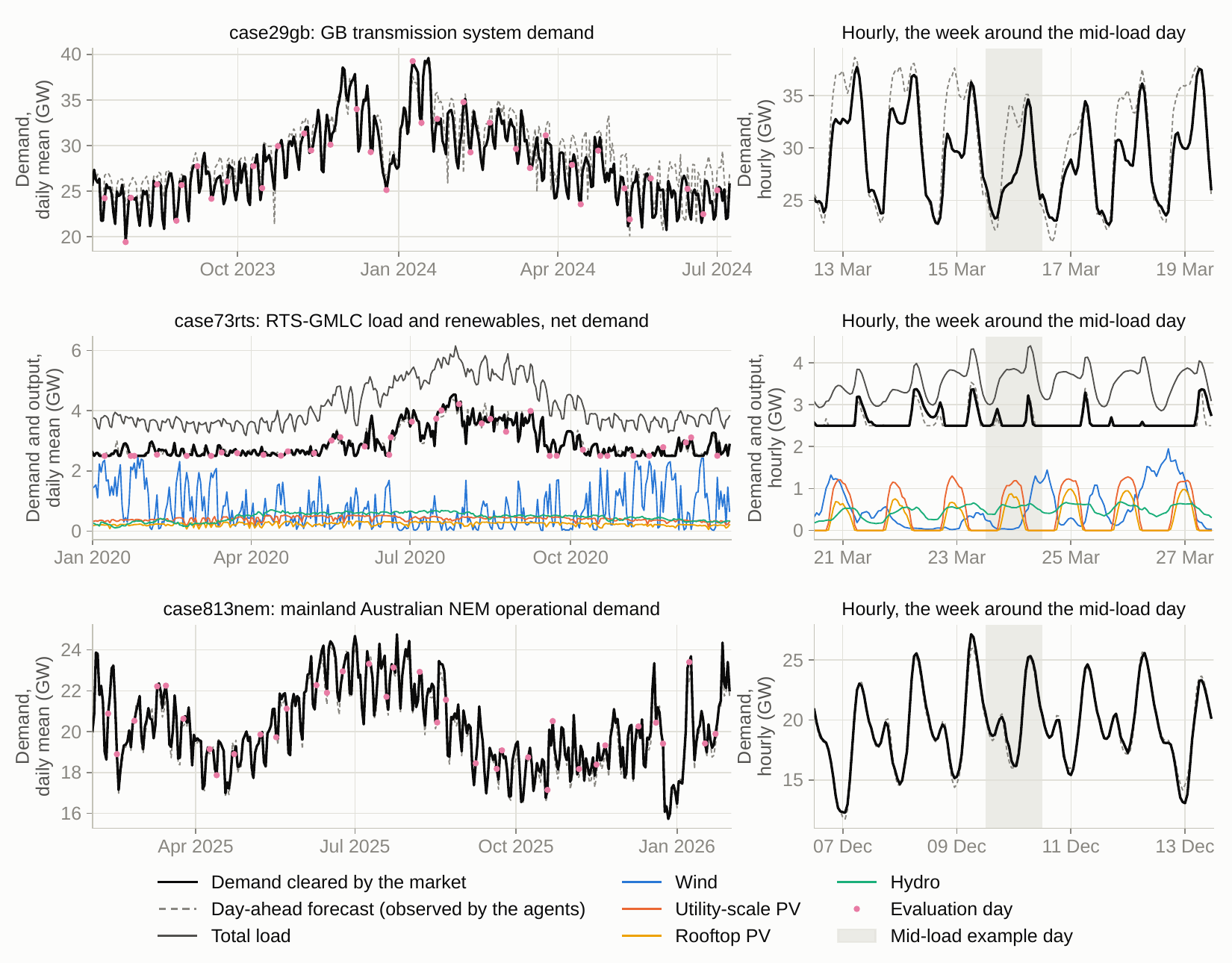}
\caption{Demand on the three test systems, and the renewable output netted out on case73rts. Left: daily means over the one-year window, with the 36 evaluation days marked. Right: hourly values in the week around the mid-load example day. The thick black line is the demand the market clears; the dashed line is the day-ahead forecast the agents observe. On case73rts the market clears net demand, total load minus the renewable output shown, raised to a floor of 2\,500~MW; it sits at that floor in more than half of the hours of the year. Demand on case813nem has a floor of 11\,500~MW, reached in a few hours.}
\label{fig.appendix.da.demand}
\end{figure}

On all three test systems the market clears the demand that actually occurs on the day, while the units observe the day-ahead forecast before they submit offers (Figure~\ref{fig.appendix.da.demand}). On case73rts, net demand has a floor of 2\,500~MW, and it sits at that floor in more than half of the hours of the year. On case813nem, demand has a floor of 11\,500~MW, reached in only a few hours.

\subsubsection{Agent training}
\label{app:results:da:training}

\begin{figure}[h!]
\centering
\includegraphics[width=\linewidth]{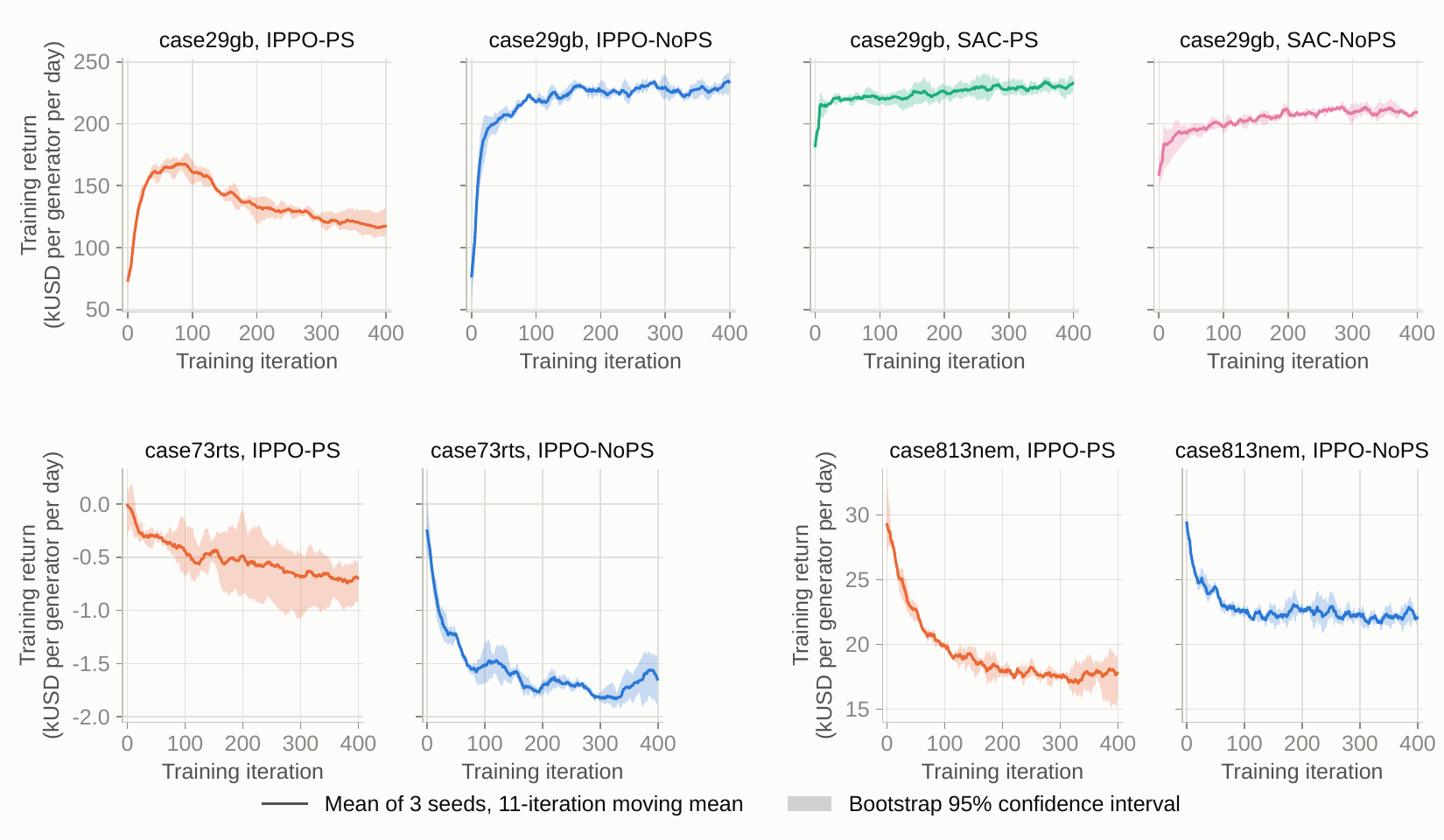}
\caption{Training return on the three test systems. Top: the four learners on case29gb. Bottom: IPPO-PS and IPPO-NoPS on case73rts and case813nem. Each line is the mean of three seeds, each seed first smoothed by an 11-iteration centred moving mean, and the shaded band is the bootstrap 95\% confidence interval. Return is the mean daily profit per generator over all sampled environment steps of an iteration. On case29gb IPPO-PS is the only learner that rises and then falls; on the other two systems the return of both learners falls from its starting level.}
\label{fig.appendix.da.training}
\end{figure}

The training setup of the four learners is that of Appendix~\ref{app:train}. On the British test system, the training return of IPPO-NoPS,  SAC-PS and SAC-NoPS ends close to its highest level (top row of Figure~\ref{fig.appendix.da.training}). IPPO-PS stays below the other three and is the only one whose return rises and then falls; the markup it samples also rises to about 1.7 and then returns to about 1.5 (Figure~\ref{fig.da}a).

On the other two test systems only IPPO-PS and IPPO-NoPS are trained (bottom row of Figure~\ref{fig.appendix.da.training}). The training return of both learners falls from its starting level. The fall comes from competition among the units: at the start of training every unit offers near the middle of its markup range, which amounts to a joint markup; during training the units that set the price learn to offer closer to cost, because a unit that stays high alone loses its output to its rivals, so the price falls while production cost does not, and profit falls with it.

\subsubsection{Discussion of results}
\label{app:results:da:eval}

On each test system, the policies learned by IPPO-PS and IPPO-NoPS are compared with truthful offers on the 36 evaluation days, each learner taken as the mean of three seeds.

\begin{figure}[h!]
\centering
\includegraphics[width=\linewidth]{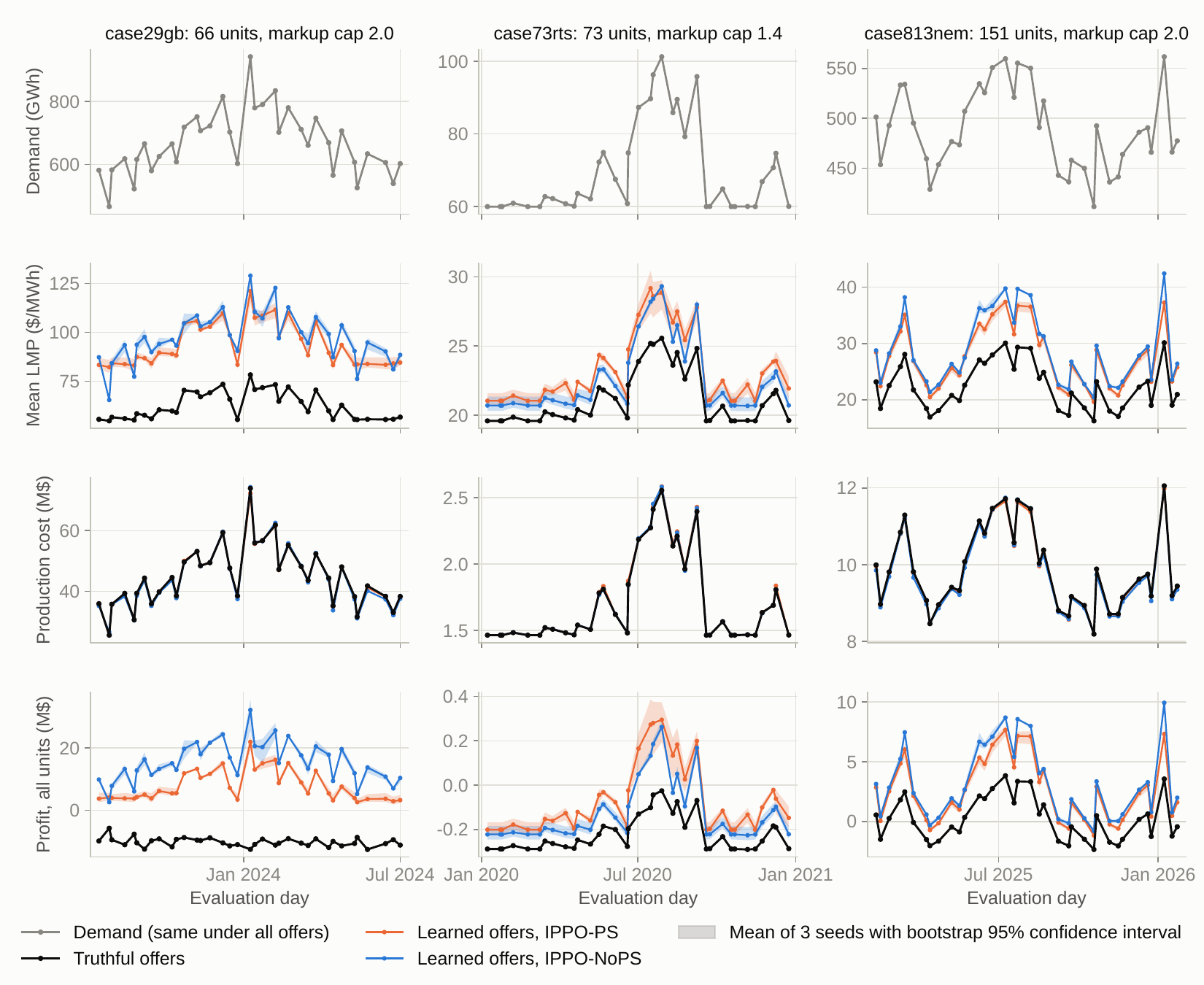}
\caption{Daily system quantities on the 36 evaluation days of each test system, truthful offers (black) against the learned offers of IPPO-PS (orange) and IPPO-NoPS (blue); each learned line is the mean of three seeds, with a bootstrap 95\% confidence interval. Rows: daily demand, the load-weighted mean LMP, production cost, and the total profit of all units. 
}
\label{fig.appendix.da.daily}
\end{figure}

As shown in Figure~\ref{fig.appendix.da.daily}, across all three test systems, the mean LMP under the learned policies (load-weighted over all periods and buses of the day) is higher than under truthful offers on every one of the 36 evaluation days (production cost barely changes) and the profit of all units rises accordingly. Averaged over the 36 days, the mean LMP under the policies learned by IPPO-PS and IPPO-NoPS is 1.50 and 1.56 times, respectively, its value under truthful offers on case29gb, 1.10 and 1.07 times on case73rts, and 1.23 and 1.29 times on case813nem. For both learners, the total production cost over the 36 days differs from that under truthful offers by less than 1\%. The increase in mean LMP is smallest on case73rts, where the mean markups of IPPO-PS and IPPO-NoPS are only 1.19 and 1.21, respectively.

\subsubsection{Details of supply curves}
\label{app:results:da:supply}

\begin{figure}[h!]
\centering
\includegraphics[width=\linewidth]{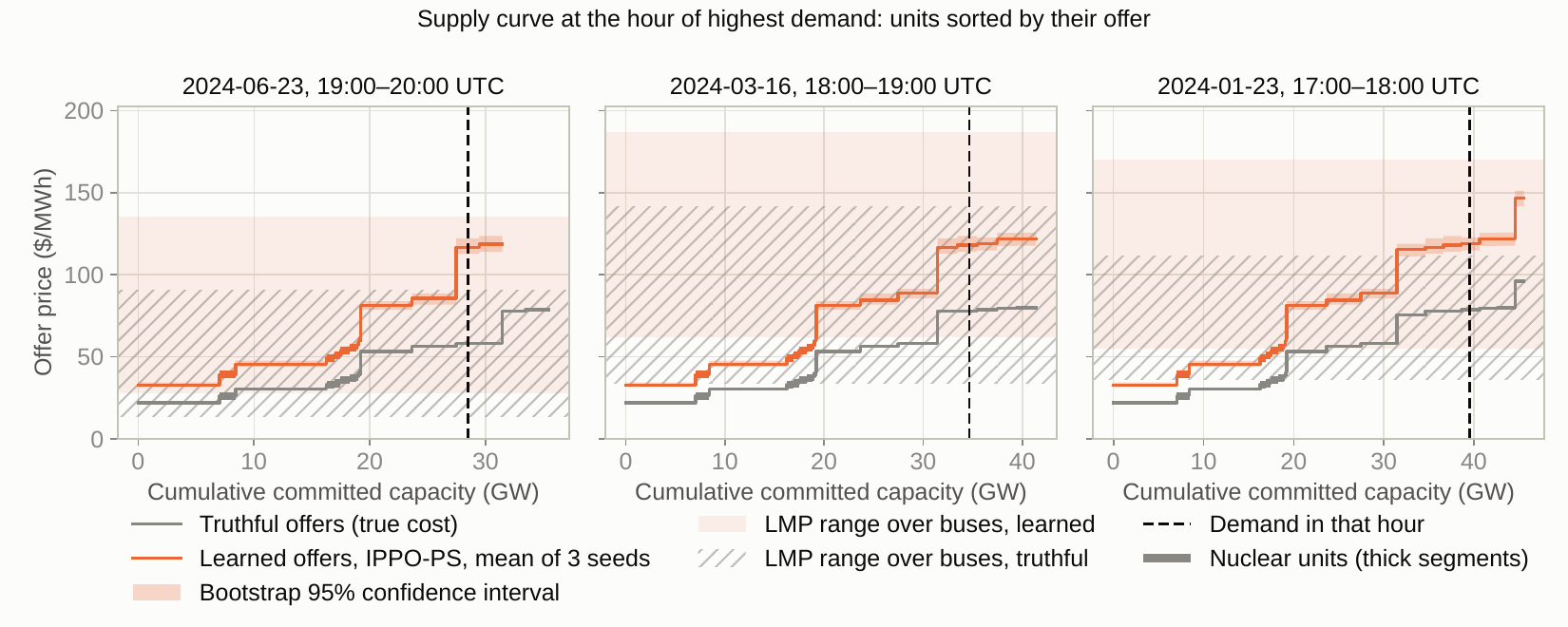}
\caption{Supply curves at the hour of highest demand on each example day of case29gb, the evaluation days at the 10th, 50th and 90th percentiles of daily demand (2024-06-23, 2024-03-16 and 2024-01-23): committed units sorted by offer under truthful (grey) and learned (orange) offers, with the steps of the 14 nuclear units drawn thicker and the demand in that hour as a dashed line; the shaded bands are the range of LMP over the 29 buses under each set of offers. The learned offers are the mean over three IPPO-PS seeds at the end of training, with a bootstrap 95\% confidence interval for each step; the learned LMP band is the range over buses of the seed-mean LMP. The nuclear steps lie far to the left of demand under both sets of offers, and the learned offers raise the whole curve.}
\label{fig.appendix.da.supply}
\end{figure}

At the peak demand periods of the British test system, the committed units are sorted by offer into a stepped supply curve as shown in Figure~\ref{fig.appendix.da.supply}. Under both truthful and learned offers, the steps of the 14 nuclear units lie far to the left of the demand line. The policy learned by IPPO-PS raises the whole supply curve while the demand line stays in place, and the price rises with it.

\subsubsection{Markups of different learners}
\label{app:results:da:offers}

\begin{figure}[h!]
\centering
\includegraphics[width=\linewidth]{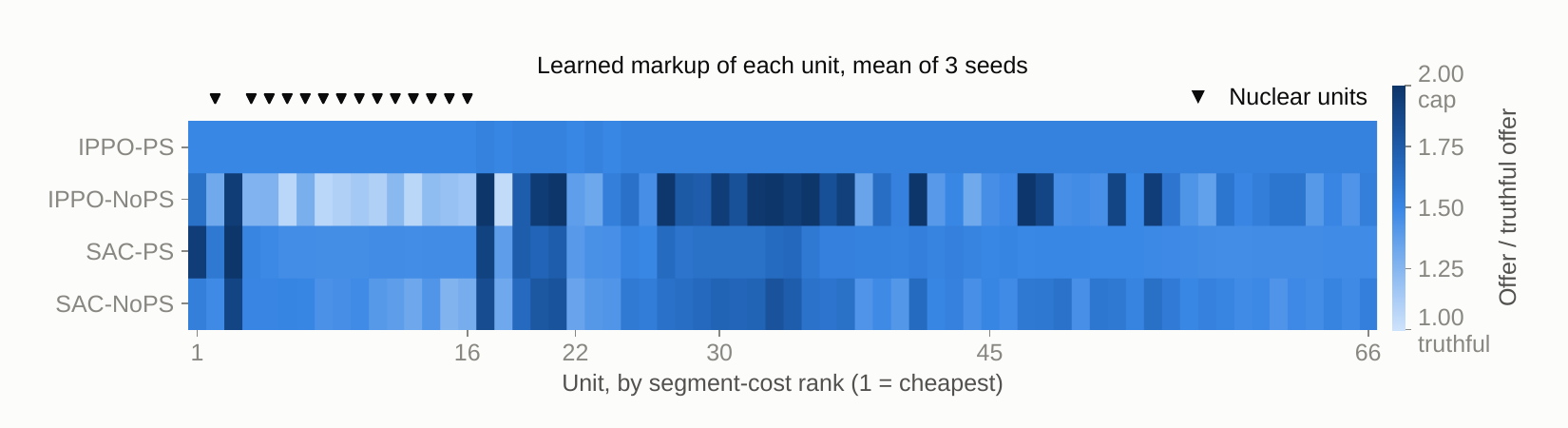}
\caption{Markup of each unit under the four learners on case29gb; rows are learners, each the mean over three seeds, columns the 66 units ordered by segment cost. Colour is the markup of that unit, averaged over the 36 evaluation days and the three seeds, on the full action range from 1.00 (truthful) to 2.00 (cap). Black triangles above the top row mark the 14 nuclear units. 
}
\label{fig.appendix.da.markups}
\end{figure}

For each of the four learners the mean markup over the 66 units is near 1.5, but the spread across units differs. IPPO-PS gives almost the same markup to every unit; IPPO-NoPS has the widest spread, with some units near truthful offers and some near the cap; the two SAC learners fall between the two, as shown in Figure~\ref{fig.appendix.da.markups}. For each algorithm, NoPS has a wider spread than PS. All four learners give the 14 nuclear units a mean markup below their own mean over all 66 units, and IPPO-NoPS gives these units markups close to truthful offers.

\FloatBarrier

\subsection{M2: Real-Time Balancing Market}
\label{app:results:rt}

In this section, we discuss the result of the four learners in the RTM balancing market.

\subsubsection{Market setting}
\label{app:results:rt:data}

The action of unit $i$ is a markup $\alpha_i \in [1, \bar{\alpha}]$ on its generation cost (i.e., true marginal cost over its output range); $\alpha_i = 1$ is a truthful offer, and an offer is submitted once per half-hour period. The test systems of this market are case29gb and case73rts (detailed data in Table~\ref{tab:app-da-data}), with the markup cap $\bar{\alpha}$ set to 2. The grids and units are shown in Figures~\ref{fig.appendix.da.system} and~\ref{fig.appendix.da.rts}.

\begin{figure}[h!]
\centering
\includegraphics[width=\linewidth]{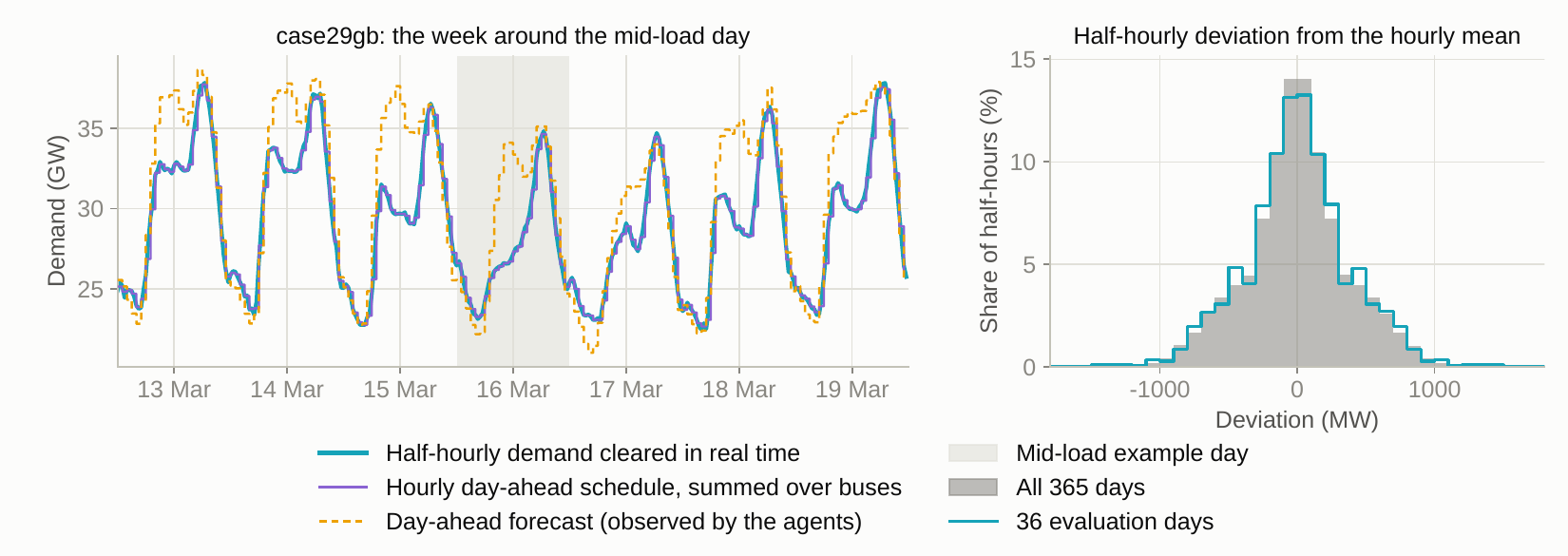}
\caption{
Demand on case29gb. Left: the week around the mid-load day (2024-03-16): the half-hourly demand the real-time market clears (teal), the hourly day-ahead schedule summed over buses, which equals the hourly mean of that demand (purple steps), and the day-ahead forecast the units observe (dashed yellow). Right: the deviation of the demand in each half-hour from the mean of its hour, over all 365 days (grey) and the 36 evaluation days (teal). On the evaluation days the absolute deviation has a median of 214~MW, a 90th percentile of 638~MW and a maximum of 1\,478~MW, against a mean demand of 27\,730~MW. The day-ahead schedule follows the half-hourly demand to within a few hundred MW.
}
\label{fig.appendix.rt.demand}
\end{figure}

Most output is scheduled day ahead, so real-time offers price only small deviations: on the 36 evaluation days, the median absolute half-hourly demand deviation is 214~MW, less than 1\% of mean demand (Figure~\ref{fig.appendix.rt.demand}). This limits the gain from raising an offer alone (Section~\ref{app:results:rt:unilateral}).

\subsubsection{Agent training}
\label{app:results:rt:training}

\begin{figure}[h!]
\centering
\includegraphics[width=\linewidth]{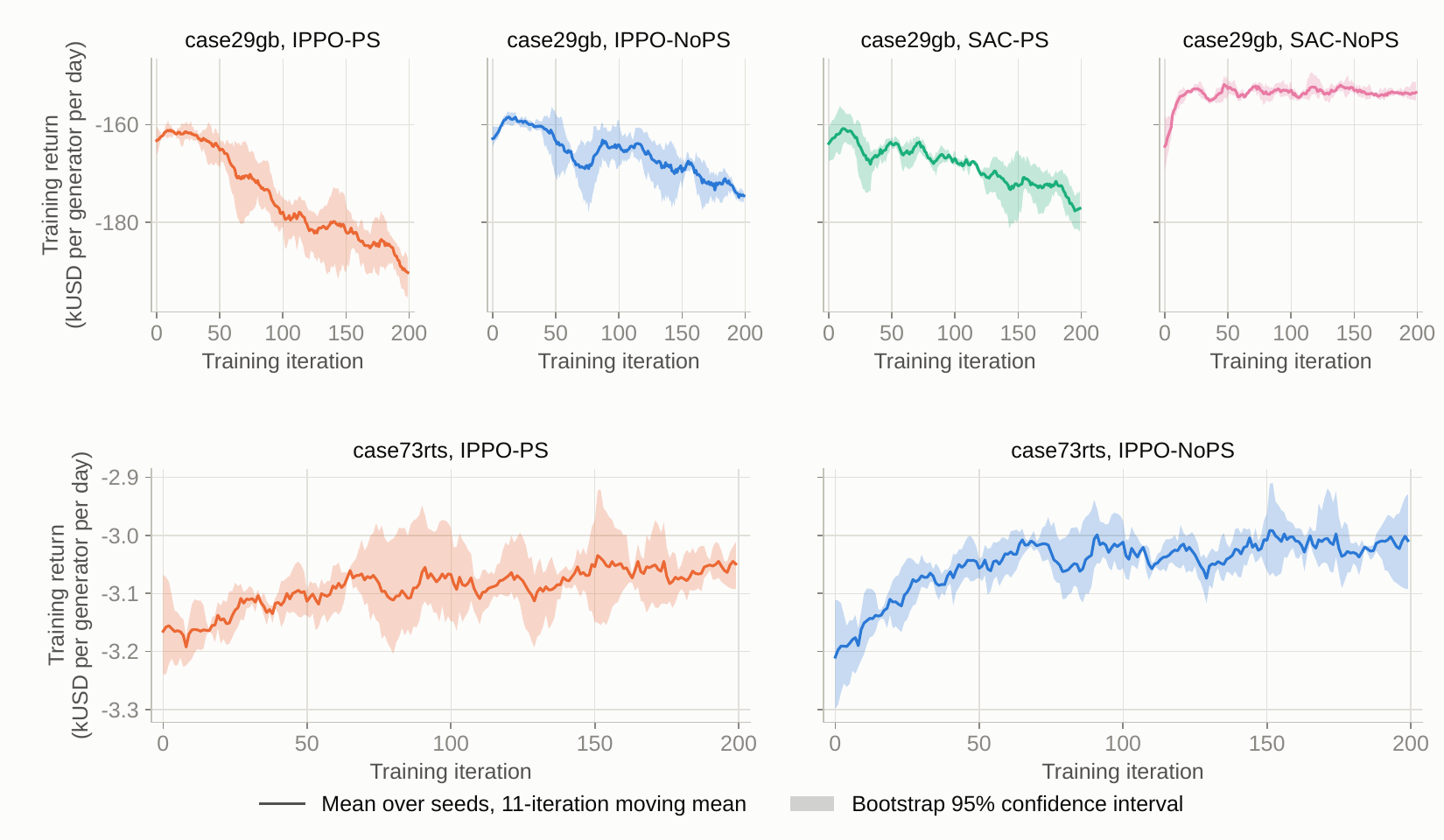}
\caption{Training return on two test systems in the real-time market. Top: the four learners on case29gb. Bottom: IPPO-PS and IPPO-NoPS on case73rts. Each line is the mean over seeds, smoothed by an 11-iteration moving mean, and the shaded band is the bootstrap 95\% confidence interval of the mean. Return is the profit of a unit per day, averaged over units and over the environment steps of an iteration. On case29gb the return of IPPO-PS falls the most, and SAC-NoPS is the only learner whose return rises. On case73rts the return of both learners rises slowly.}
\label{fig.appendix.rt.training}
\end{figure}

The training return is the mean daily profit per unit under sampled actions, which is the objective of the learner itself. Under all four learners the mean markup of the 30 committed units starts between 1.48 and 1.61; at the end of training it is 1.04 under IPPO-PS, and 1.24, 1.40 and 1.39 under IPPO-NoPS, SAC-PS and SAC-NoPS; the markup along training is shown in Figure~\ref{fig.rt}b. On case29gb the training return of IPPO-PS falls throughout and ends lowest of the four learners; IPPO-NoPS and SAC-PS also fall, by less; SAC-NoPS rises first and then stays level (top row of Figure~\ref{fig.appendix.rt.training}). On case73rts IPPO-PS and IPPO-NoPS are trained, and the training return of both rises slowly (bottom row of Figure~\ref{fig.appendix.rt.training}).

\subsubsection{Discussion of results}
\label{app:results:rt:eval}

\begin{figure}[h!]
\centering
\includegraphics[width=\linewidth]{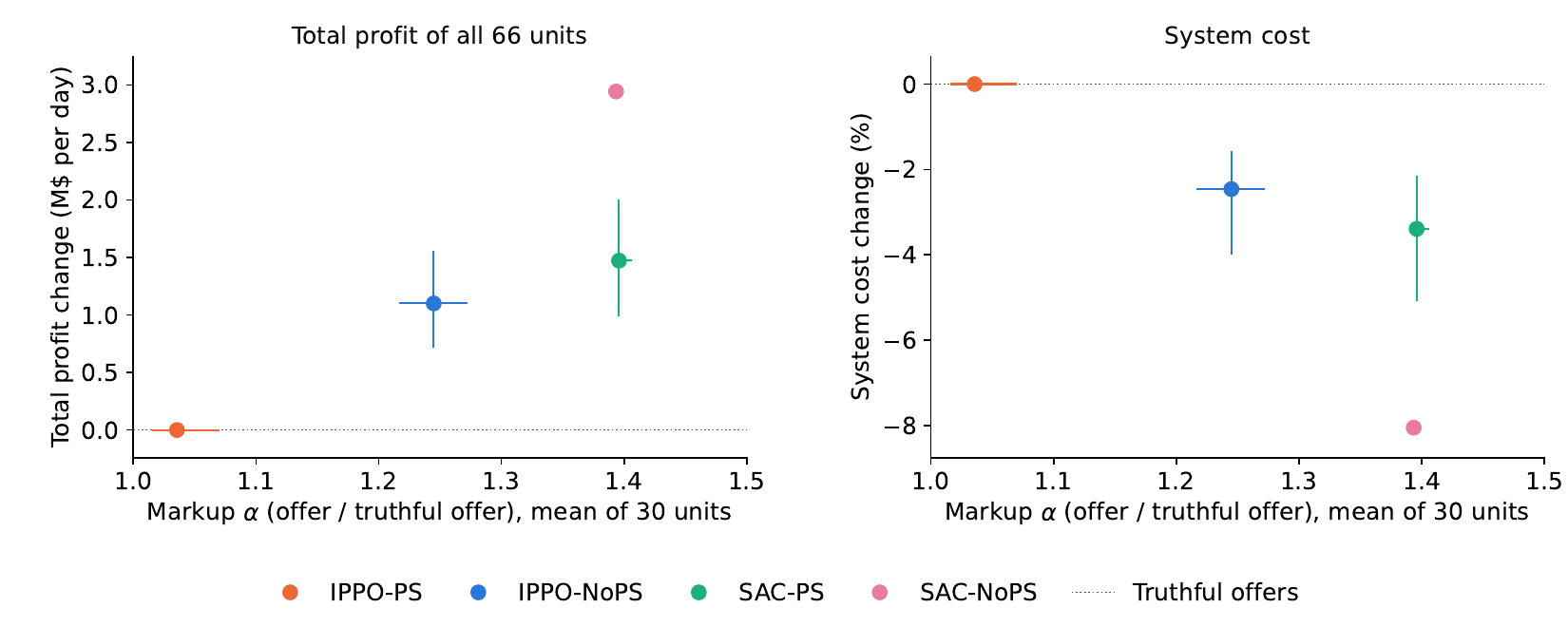}
\caption{Learned offers against truthful offers on case29gb over the 36 evaluation days. Left: change in total profit of all 66 units compared to truthful offers (in million dollars per day); right: change in system cost compared to truthful offers (in percent). Each circle indicates one learner after training, placed at the mean markup of the 30 committed units. The bars are the 95\% confidence interval of the mean over seeds; the dotted line is truthful offers. IPPO-PS returns to truthful offers; the other three learners end with higher total profit and lower system cost.}
\label{fig.appendix.rt.eval}
\end{figure}

On case29gb, IPPO-PS returns to truthful offers, while the other three learners end with higher total profit and lower system cost, as shown in Figure~\ref{fig.appendix.rt.eval}. This position is not learned. The untrained networks already give the units different markups, and training only moves these markups towards truthful offers.
The reason is that an independent learner sees only how its own return responds to its own action. A unit that raises its offer alone only loses output, as shown in section~\ref{app:results:rt:unilateral}. If its output depends on its offer, its gradient points towards truthful offers; if not, its gradient is zero. Without parameter sharing, each unit follows its own gradient: the first kind returns to truthful offers and the second stays where it started. With sharing, the updates from the first kind move all units, so every markup falls together.

There are 36 units that are never committed, which give a clean test. Their offers never enter the clearing, so their own return carries no signal. Without information sharing, training leaves their markups unchanged (IPPO-NoPS from 1.48 to 1.49, SAC-NoPS from 1.50 to 1.51). With sharing, they fall with the committed units (IPPO-PS from 1.52 to 1.03, SAC-PS from 1.57 to 1.47). SAC-PS falls much less, because under SAC the committed units themselves stay far from truthful offers (Section~\ref{sec.results.rt}). These results show an information limit in electricity markets: a unit learns a bidding strategy only when its offer changes its own outcome. Otherwise, its learned offer comes from the initialization (without parameter sharing) or from the other units (with it), and should not be read as a strategy.

\subsubsection{Unilateral and joint markups}
\label{app:results:rt:unilateral}
\label{app:results:rt:joint}

We extend the discussion in Appendix~\ref{app:results:rt:eval} in this section.

\begin{figure}[h!]
\centering
\includegraphics[width=\linewidth]{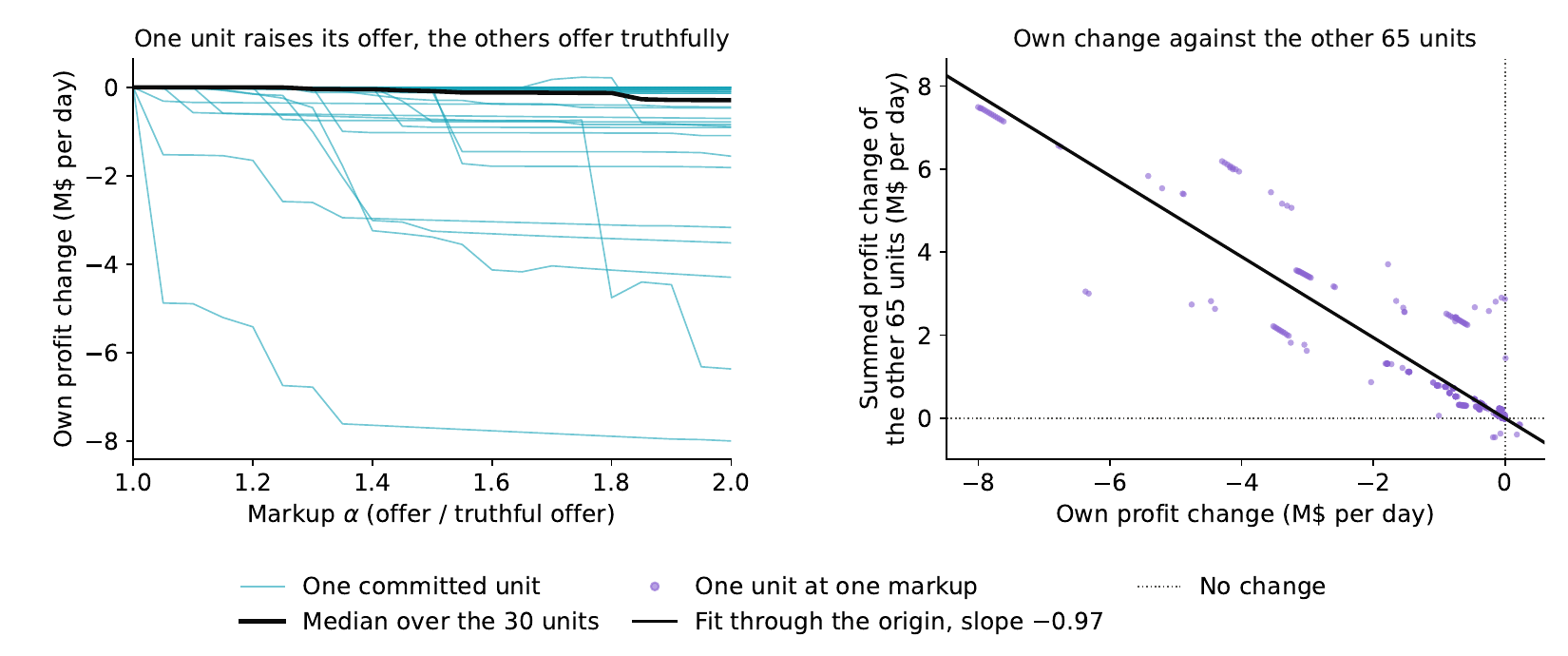}
\caption{
Each of the 30 committed units raises its offer alone while the others offer truthfully. Left: own profit change against markup (one line per unit, median in black). Right: own profit change against the profit change of the other units (with a least-squares fit, near-zero points are omitted). Means over the 36 evaluation days, in million dollars per day. 
}
\label{fig.appendix.rt.unilateral}
\end{figure}

A unit gains little by raising its offer alone. We let each of the 30 committed units raise its offer alone, at 21 markup levels from 1.00 to 2.00, while the other 65 units offer truthfully as shown in the left subfigure of Figure~\ref{fig.appendix.rt.unilateral}. For 19 units, the most profitable level is the truthful offer, and no unit gains more than \$0.23 million per day. At a markup of 2.00, 27 of the 30 units lose profit, by up to \$8.0 million per day. The unit that raises its offer produces less, and the other units take over both its output and its lost profit (Figure~\ref{fig.appendix.rt.unilateral}, right, slope $-0.97$). A unilateral markup therefore moves profit between units but leaves the total almost unchanged.

\begin{figure}[h!]
\centering
\includegraphics[width=\linewidth]{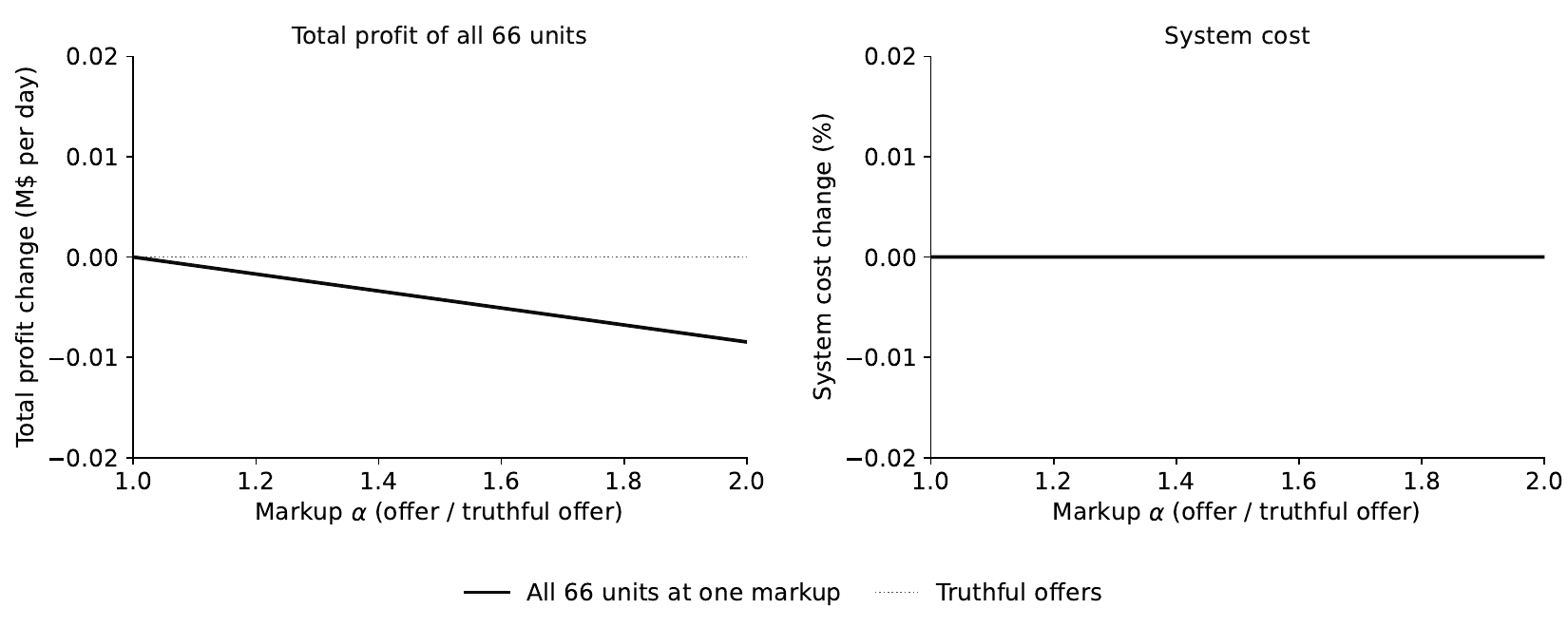}
\caption{
All 66 units at the same markup, from 1.00 to 2.00. Left: change in total profit (million dollars per day). Right: change in system cost (\%). Both are relative to truthful offers (dotted line), averaged over the 36 evaluation days. A common markup leaves both nearly unchanged.
}
\label{fig.appendix.rt.uniform}
\end{figure}

Raising all offers together helps only if the markups differ across units. With the same markup for all 66 units, the order of the offers and the dispatch schedules do not change as shown in Figure~\ref{fig.appendix.rt.uniform}. 
The reason is the two-settlement rule \eqref{eq:rt-settle}. Since the day-ahead market clears the actual hourly demand, output below the schedule in one half-hour is roughly offset by output above it in the other.
A higher real-time price therefore moves money between units without raising the total. Different markups, in contrast, change the dispatch, raise total profit and lower system cost, as under IPPO-NoPS, SAC-PS and SAC-NoPS (as shown in Figure~\ref{fig.appendix.rt.eval}). This joint gain appears only when units offer differently, and no single unit sees it in its own profit.

\FloatBarrier

\subsection{M3: Ancillary Services Market}
\label{app:results:as}

This section first gives the data, training curves and the discussion about results of the ancillary services market.

\subsubsection{Market setting}
\label{app:results:as:data}

We consider the setting in which M2 and M3 are jointly cleared.
Therefore, the action of each unit $i$ has two parts: an energy offer with markup $\alpha_i \in [1, \bar{\alpha}]$, and one reserve offer $\pi^{\mathrm{res}}_{i,j,t}$ per period $t$ on each of the two reserve products $j = 1, 2$, in \pounds/MWh.
The two products require a response within 10 and 30 minutes respectively. The reserve a unit is awarded on a product, called its reserve held, is at most the output it can ramp up within that time.
The reserve requirement of each product is 5\% of the demand forecast of the period. Unmet reserve requirement is priced at $\mathrm{VOLR}$ (as shown in Table~\ref{tab:app-as-data}), and the reserve price $\lambda^{\mathrm{res}}_{j,t}$ does not exceed it.

This market uses the same three test systems as the day-ahead market: case29gb, case73rts and case813nem. Table~\ref{tab:app-da-data} describes their data, and Table~\ref{tab:app-as-data} gives the markup cap and the value of lost reserve for each system. Figures~\ref{fig.appendix.da.system} to~\ref{fig.appendix.da.nem} show the grids and units, and Figure~\ref{fig.appendix.da.demand} shows the demand.

\begin{table}[h!]
  \caption{Markup cap and value of lost reserve of the ancillary services market on the three test systems.}
  \label{tab:app-as-data}
  \centering
  \footnotesize
  \begin{tabular}{lcc}
    \toprule
    Test system & Markup cap $\bar{\alpha}$ & Value of lost reserve $\mathrm{VOLR}$ (\pounds/MWh) \\
    \midrule
    case29gb & 2 & 250 \\
    case73rts & 2 & 136 \\
    case813nem & 2 & 147 \\
    \bottomrule
  \end{tabular}
\end{table}

\subsubsection{Agent training}
\label{app:results:as:training}

The four learners are trained as described in Appendix~\ref{app:train}, and Figure~\ref{fig.appendix.as.training} shows their training return. On the case29gb test system, IPPO-PS is the only learner whose return falls, from $-106$~k \pounds per generator per day over the first 5 iterations to $-126$ over the last 20. The other three rise slightly and stay close to where they started. In test systems case73rts and case813nem, the return of both IPPO learners rises.

\begin{figure}[h!]
\centering
\includegraphics[width=\linewidth]{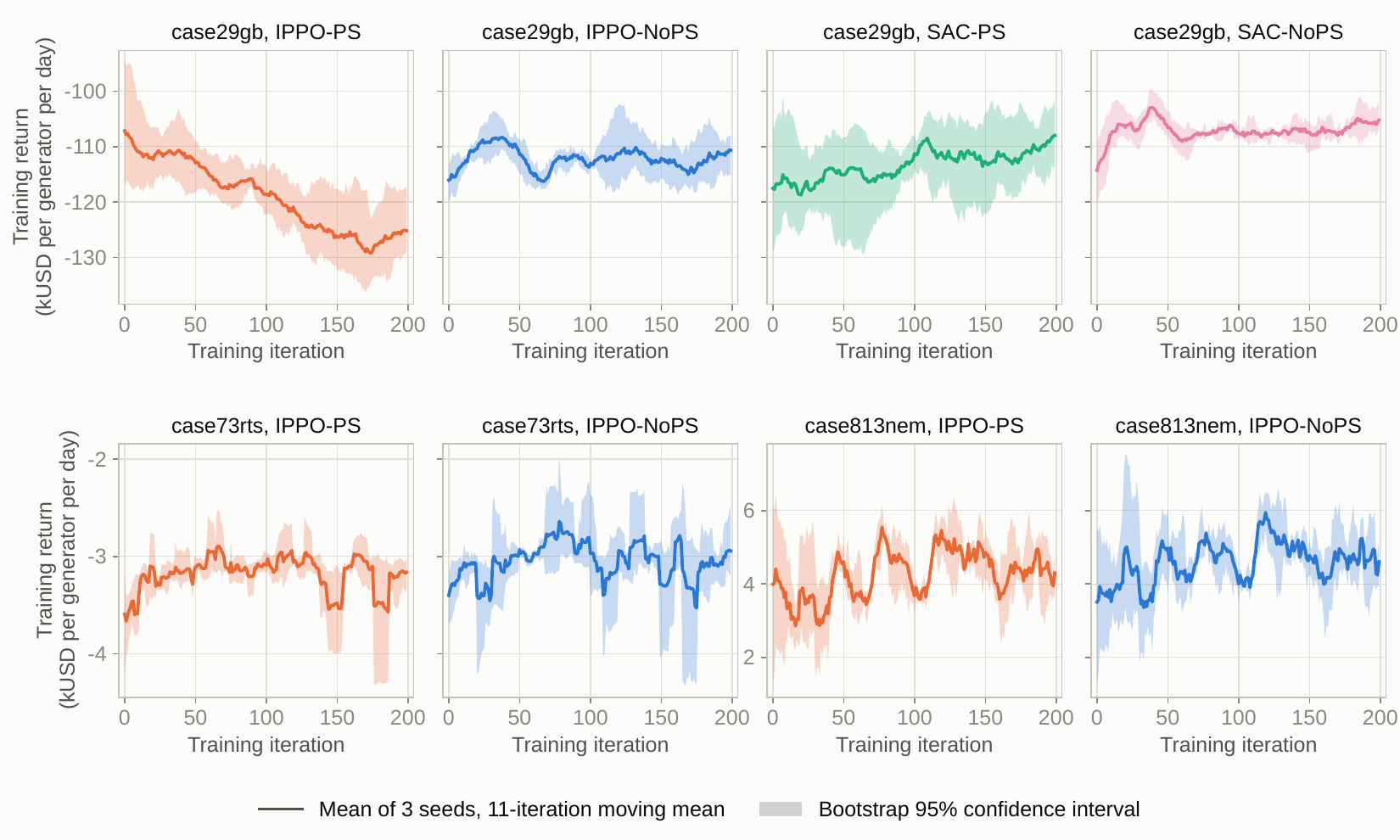}
\caption{
Training return on the three test systems. Top: the four learners on case29gb. Bottom: IPPO-PS and IPPO-NoPS on case73rts and case813nem. Return is the mean daily profit per generator in each iteration. Lines are the mean of three seeds, each smoothed over 11 iterations, and bands are bootstrap 95\% confidence intervals. On case29gb, only the return of IPPO-PS falls, and on the other two systems, both learners rise.
}
\label{fig.appendix.as.training}
\end{figure}

\subsubsection{Discussion of results}
\label{app:results:as:eval}

\begin{figure}[h!]
\centering
\includegraphics[width=\linewidth]{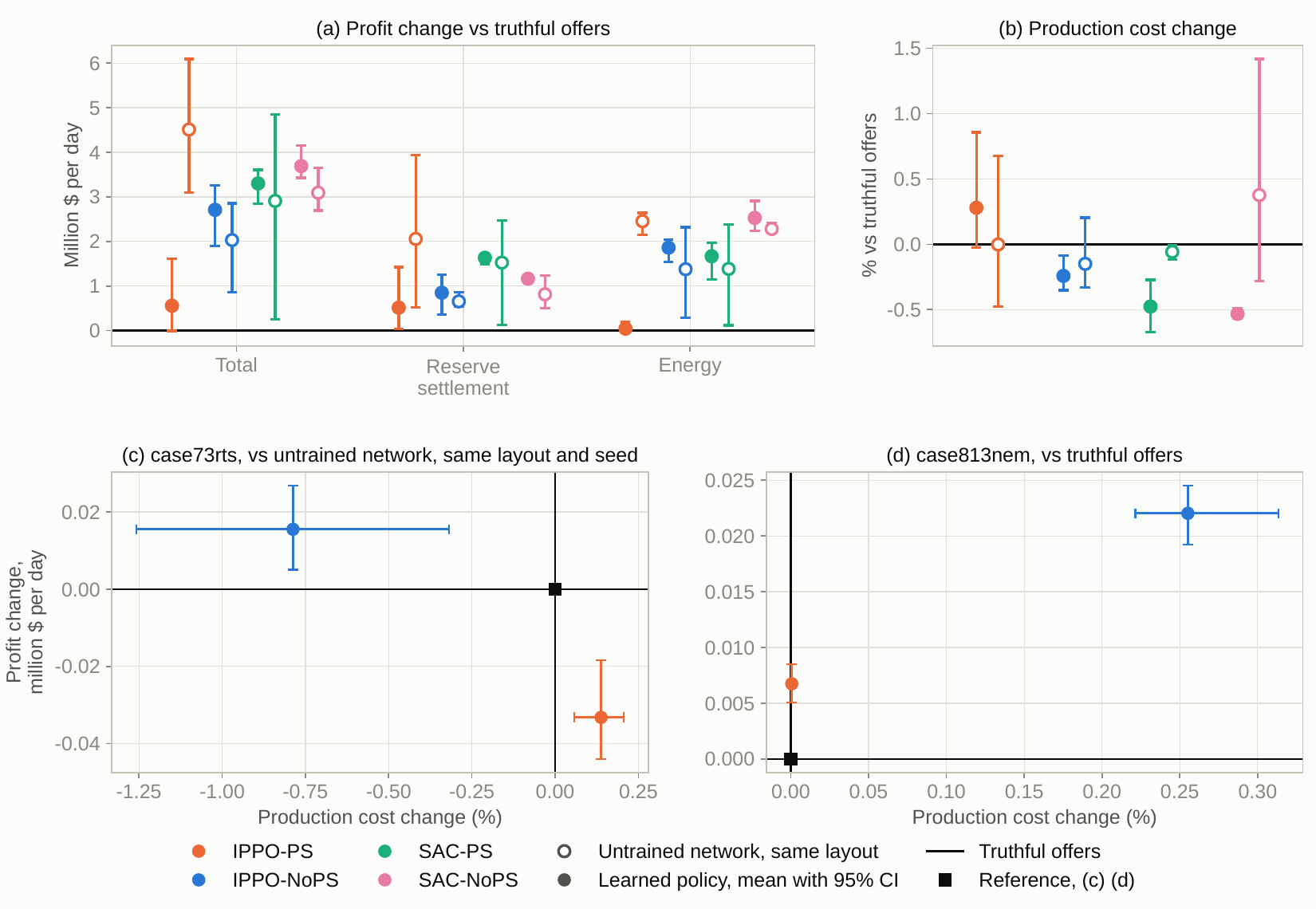}
\caption{
Evaluation on the 36 evaluation days. (a) case29gb: change in total profit of all 66 units relative to truthful offers (million dollars per day), split into reserve payments and energy. (b) case29gb: change in production cost. Filled markers are trained policies; hollow markers are untrained networks. (c) case73rts, relative to the untrained network (truthful offers not evaluated), and (d) case813nem, relative to truthful offers: change in total profit against change in production cost for IPPO-PS (orange) and IPPO-NoPS (blue). The black square marks the reference. Markers are seed means with bootstrap 95\% confidence intervals. On case29gb, all learners earn more than truthful offers at nearly unchanged cost, but only IPPO-PS earns less than its untrained network; on the other two systems, the differences are small.
}
\label{fig.appendix.as.eval}
\end{figure}

Training does not make the learners more profitable in this market. The untrained networks, with random reserve offers, already earn \pounds2 million to \pounds4.5 million per day more than truthful offers as shown in Figure~\ref{fig.appendix.as.eval}. IPPO-PS gives almost all of this back: after training, it earns only \pounds0.56 million per day more than truthful offers, \pounds3.95 million less than its untrained network. The other three learners stay near their starting point, at \pounds2.71 million to \pounds3.69 million per day above truthful offers. No learner receives more gains compared to its own untrained network.

A much larger gain is available, but no learner finds it. If all units raise their reserve offers together to \pounds150/MWh, total profit rises by \pounds9.05 million per day as shown in the subfigure c of Figure~\ref{fig.rt}. Only \pounds0.51 million to \pounds1.63 million per day of the learners' profit comes from reserve payments, and the rest comes from energy. Production cost stays within 0.6\% of its truthful level as shown in the subfigure b of Figure~\ref{fig.appendix.as.eval}.

The other two test systems show the same results, but the differences are small: all changes in profit are within a few tens of thousands of dollars per day as shown in subfigures c and d of Figure~\ref{fig.appendix.as.eval}. After training, IPPO-PS again earns the same as or less than before training, while IPPO-NoPS earns slightly more.

\subsubsection{Reserve offer analysis}
\label{app:results:as:markup}

\begin{figure}[h!]
\centering
\includegraphics[width=\linewidth]{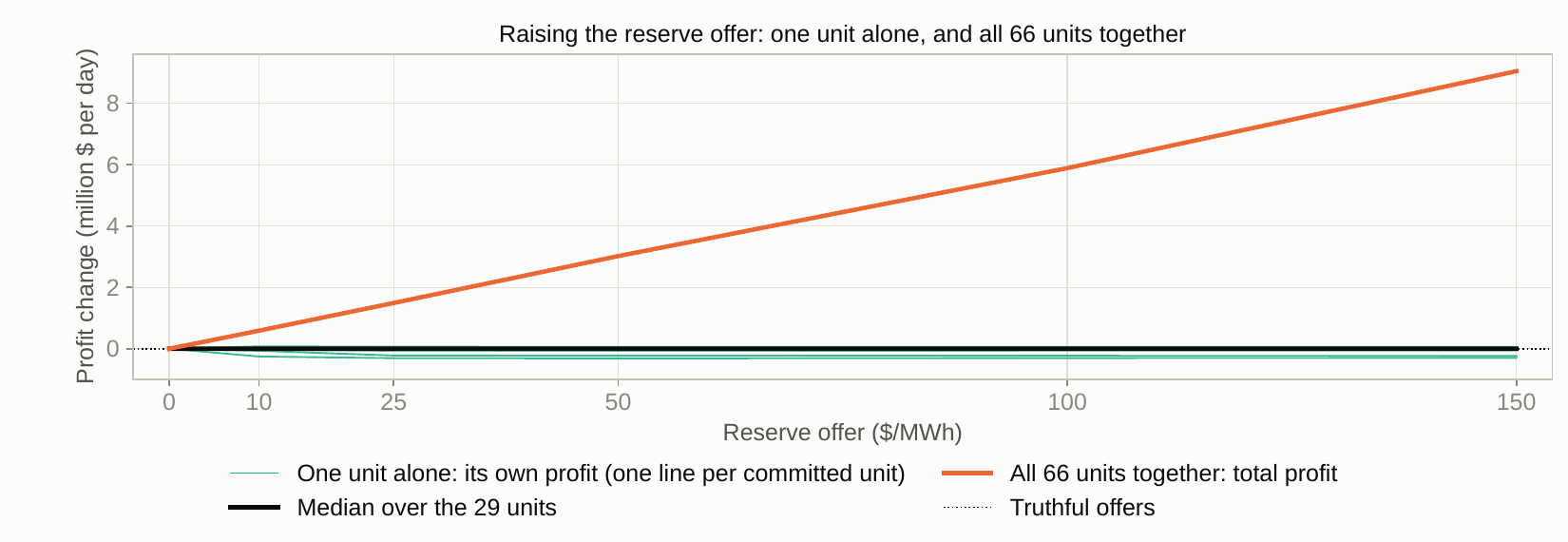}
\caption{
Raising the reserve offer on case29gb from 0 to \pounds150/MWh on both energy and reserve products. Teal: own profit change of each of the 29 committed units when it raises its offer alone, with the others truthful. Black: the median. Orange: total profit change of all 66 units when all raise their offers together. Values in million dollars per day. Raising alone gains almost nothing, while raising together gains \pounds9.05 million per day at \pounds150/MWh.
}
\label{fig.appendix.as.markup}
\end{figure}

In ancillary service markets, a unit gains almost nothing by raising its reserve offer alone, but all units gain a lot by raising it together as shown in Figure~\ref{fig.appendix.as.markup}. When one of the 29 committed units raises its reserve offer from 0 to \pounds150/MWh while the others offer truthfully, its own profit usually does not change. Across the 29 units, the median change is zero, the largest gain is \pounds0.08 million per day, and the largest loss is \pounds0.32 million per day. When all 66 units raise their reserve offers together, total profit rises linearly with the offer, by \pounds9.05 million per day at \pounds150/MWh, while production cost changes by less than 0.001\%.

\FloatBarrier

\subsection{M4: Peer-to-Peer Local Energy Market}
\label{app:results:p2p}

Section~\ref{sec.results.p2p} finds that an arbitrage that pays for a single household (charging at noon and discharging in the evening) cancels itself when many households adopt it, and that whether a learner finds it depends on the parameter layout. This section first describes the community and its metering data (Section~\ref{app:results:p2p:data}), then gives the training curves (Section~\ref{app:results:p2p:training}) and the evaluation results on held-out episodes (Section~\ref{app:results:p2p:eval}); the last section covers how the gain from arbitrage changes with the number of households arbitraging (Section~\ref{app:results:p2p:crowding}).

\subsubsection{Market setting}
\label{app:results:p2p:data}

\begin{figure}[h!]
\centering
\includegraphics[width=\linewidth]{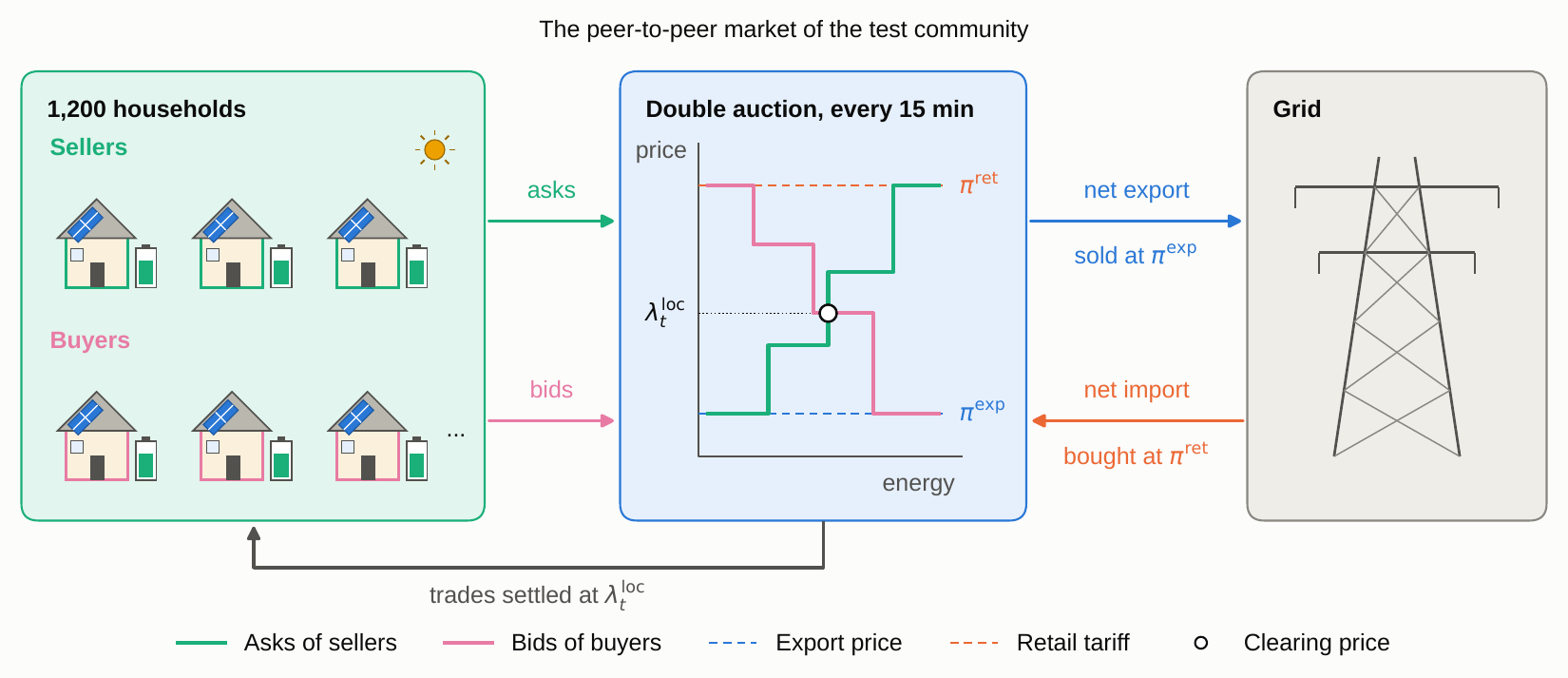}
\caption{The P2P market of the test community. Each of the 1\,200 households has PV and a battery. In every quarter-hour, sellers ask and buyers bid in a community double auction, which sets the clearing price $\lambda^{\mathrm{loc}}_t$ between the export price $\pi^{\mathrm{exp}}$ and the retail tariff $\pi^{\mathrm{ret}}$; trades in the community are settled at $\lambda^{\mathrm{loc}}_t$. Energy the auction leaves unmatched is settled with the grid.
}
\label{fig.appendix.p2p.scheme}
\end{figure}

The test system at community level has the 1\,200 households with PV in the Fluvius 2024 metering data, as shown in Figure~\ref{fig.appendix.p2p.scheme}. Each household is considered as one agent. Each period is a quarter-hour, and an episode is the 96 periods of one day. The smart meters record only the injection and the offtake of each household, and the market takes their difference as the net position of each household. The export price is $\pi^{\mathrm{exp}} = 73.0$~EUR/MWh and the retail tariff is $\pi^{\mathrm{ret}} = 333.4$~EUR/MWh. Each household has one battery with a capacity of 0.011~MWh, a round-trip efficiency of 0.85 and a battery degradation cost of 13.88~EUR per MWh of throughput; its state of charge (as a fraction of capacity) stays between 0.15 and 1.0 and starts at 0.5.

\begin{figure}[h!]
\centering
\includegraphics[width=\linewidth]{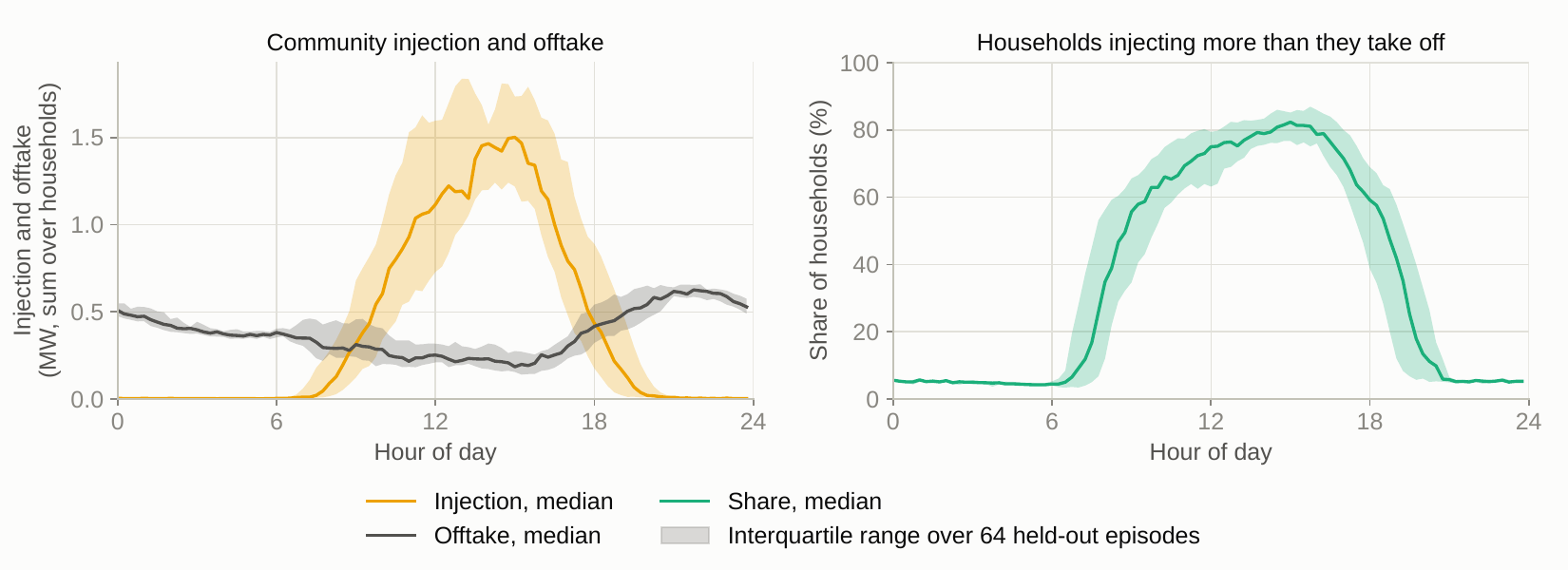}
\caption{The test community of 1\,200 households: community injection and offtake (left) and the share of households injecting more than they take off (right), by hour of day over the 64 held-out episodes.}
\label{fig.appendix.p2p.community}
\end{figure}

In every period, each household chooses a battery power (charge or discharge) and a price between the two grid prices. A household with a positive net position sells, and the price is its ask; one with a negative net position buys, and the price is its bid. As shown in Figure~\ref{fig.appendix.p2p.community}, at noon most households inject more than they take off and are sellers: around 15:00 about 80\% households are sellers, and at night injection is close to zero.

\begin{table}[h!]
  \caption{Data of the test community.}
  \label{tab:app-p2p-data}
  \centering
  \footnotesize
  \begin{tabularx}{\linewidth}{>{\raggedright\arraybackslash}p{0.19\linewidth}>{\raggedright\arraybackslash}p{0.12\linewidth}>{\raggedright\arraybackslash}p{0.15\linewidth}Y>{\raggedright\arraybackslash}p{0.12\linewidth}}
    \toprule
    Data source & Households & Period covered & Training and held-out & Licence \\
    \midrule
    Quarter-hour injection and offtake from Fluvius smart meters\textsuperscript{a} & 1\,200, all with PV & 2024-04-01 to 2024-10-26, 209 days & 3 consecutive days held out of every 15; 14\,702 training and 2\,702 held-out episode starts; evaluation on 1\,024 held-out episodes, of which the figures below use 64 & Fluvius open data licence \\
    \bottomrule
  \end{tabularx}
  \par\smallskip
  \parbox{\linewidth}{\scriptsize\raggedright \textsuperscript{a}\,\href{https://opendata.fluvius.be/explore/dataset/1_50-verbruiksprofielen-dm-elek-kwartierwaarden-voor-een-volledig-jaar/}{\texttt{opendata.fluvius.be/\allowbreak explore/\allowbreak dataset/\allowbreak 1\_50-\allowbreak verbruiksprofielen-\allowbreak dm-\allowbreak elek-\allowbreak kwartierwaarden-\allowbreak voor-\allowbreak een-\allowbreak volledig-\allowbreak jaar/}}.}
\end{table}

\subsubsection{Agent training}
\label{app:results:p2p:training}

\begin{figure}[h!]
\centering
\includegraphics[width=\linewidth]{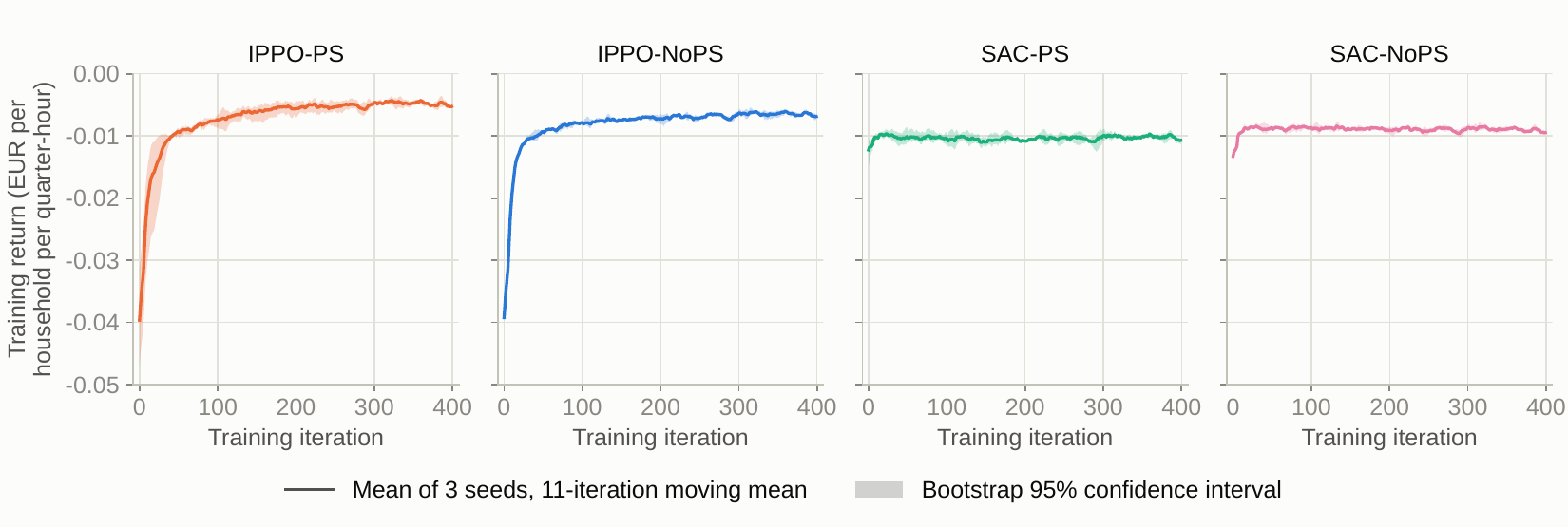}
\caption{Training return of the four learners up to iteration 400; the policy at iteration 400 is the one evaluated throughout this section. Shown is the mean over three seeds of the 11-iteration moving mean of the return on 64 sampled training episodes, with the shading the bootstrap 95\% confidence interval. Both IPPO learners rise fastest in the first 50 iterations and both SAC learners reach their level within a few dozen; after iteration 50 all four change little.}
\label{fig.appendix.p2p.training}
\end{figure}

The training setup of the four learners is that of Appendix~\ref{app:train}, and the two SAC learners differ from Table~\ref{tab:train-hp} only in using two hidden layers of 64 units, a replay buffer of 8\,192 environment steps, a batch size of 16 environment steps and one gradient step every 4 environment steps. The training return of both IPPO learners rises fastest in the first 50 iterations, and both SAC learners level off within the first few dozen iterations. After that the return changes little, as shown in Figure~\ref{fig.appendix.p2p.training}: from iterations 100–149 to iterations 351–400, the mean return of each IPPO learner rises by only 0.0009 to 0.0027 EUR per household per quarter-hour.

\subsubsection{Discussion of results}
\label{app:results:p2p:eval}

\begin{figure}[h!]
\centering
\includegraphics[width=\linewidth]{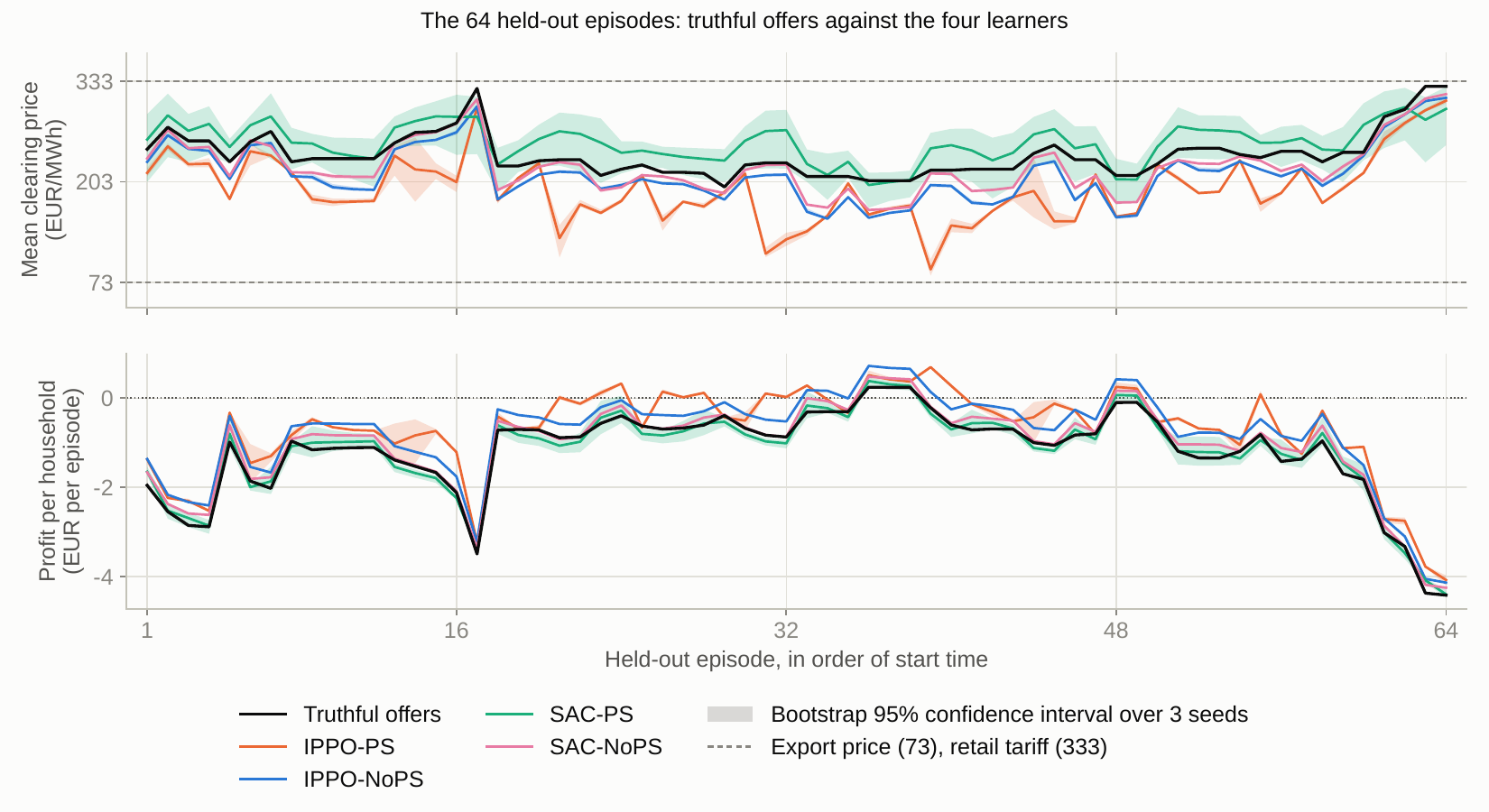}
\caption{Results on the 64 held-out episodes ordered by start time.
Top: mean clearing price per episode. Bottom: profit per household per episode. Black is truthful offers; each learner is the mean of three seeds with a bootstrap 95\% confidence interval. 
IPPO-NoPS lowers the price and raises the profit on every episode, while IPPO-PS and SAC-NoPS gain on 54 and 57 episodes. The seeds of SAC-PS disagree, and two of them raise the price.
}
\label{fig.appendix.p2p.heldout}
\end{figure}

We evaluate the results under truthful offers and under the considered learners. Truthful offers mean that households with batteries stay idle, and sellers ask the export price and buyers bid the retail tariff. As shown in Figure~\ref{fig.appendix.p2p.heldout}, IPPO-NoPS is the most consistent learner: all three of its seeds lower the clearing price and raise the profit per household on every one of the 64 episodes. IPPO-PS and SAC-NoPS are less consistent, with all three seeds earning more than under truthful offers on 54 and 57 episodes, respectively.
Averaged over the 64 episodes, the profit per household is $-1.201$~EUR per episode under truthful offers and $-0.769$ to $-0.685$~EUR over the three seeds of IPPO-PS, the highest of the four learners. The three seeds of SAC-PS disagree on the clearing price: the mean clearing price of two of them is 270.3 and 272.9, against 238.1~EUR/MWh under truthful offers.

\begin{figure}[h!]
\centering
\includegraphics[width=\linewidth]{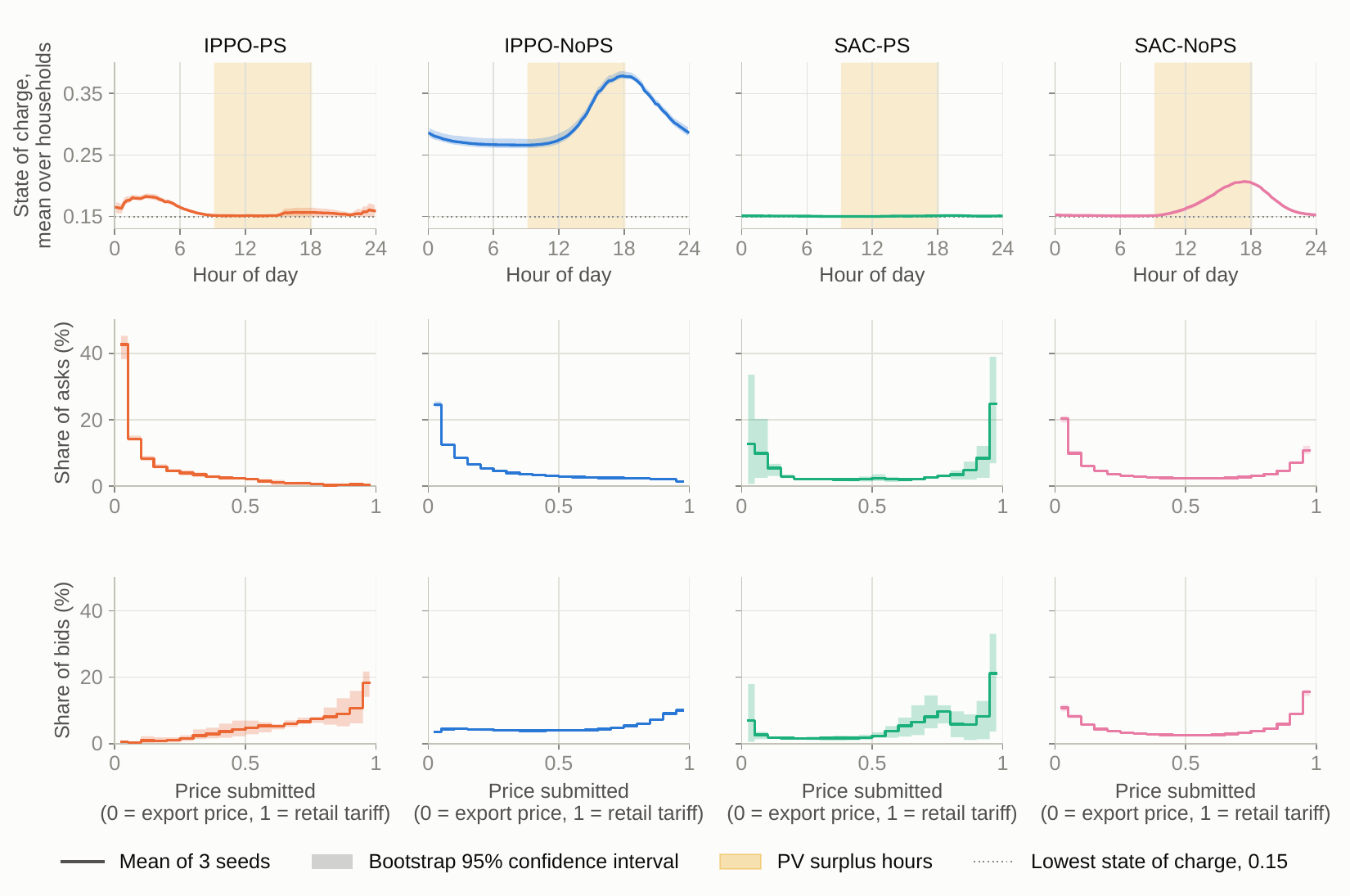}
\caption{Learned actions of the four learners on the 64 held-out episodes (mean of three seeds, bootstrap 95\% confidence interval). Top: state of charge averaged over households by hour of day (last 12 hours of each episode only). The band marks the PV surplus hours, those in which the community injects more than it draws on most days. Middle and bottom: asks and bids, scaled from the export price (0) to the retail tariff (1). 
}
\label{fig.appendix.p2p.actions}
\end{figure}

Figure~\ref{fig.appendix.p2p.actions} shows the actions each learner learns. The state of charge in each period of the day (the last 12 hours of each episode only: in every episode the learned policies first sell the energy stored at the initial state of charge of 0.5) splits the different learners into two behaviours according to whether they share parameters: IPPO-NoPS and SAC-NoPS charge in the hours of PV surplus and discharge in the evening, IPPO-NoPS by a large amount and SAC-NoPS by a small one; IPPO-PS and SAC-PS stay near the lowest state of charge. 
The learners also differ in their bidding prices. Under both IPPO learners, most sellers ask a price close to the export price. Buyers lean towards the retail tariff under IPPO-PS, but bid at all levels under IPPO-NoPS. Under SAC-NoPS, most asks and bids are close to one of the two grid prices, and few lie in between. The three seeds of SAC-PS learn very different prices, so SAC-PS has the widest confidence intervals in the figure.

\subsubsection{Arbitrage gain and participation}
\label{app:results:p2p:crowding}

\begin{figure}[h!]
\centering
\includegraphics[width=\linewidth]{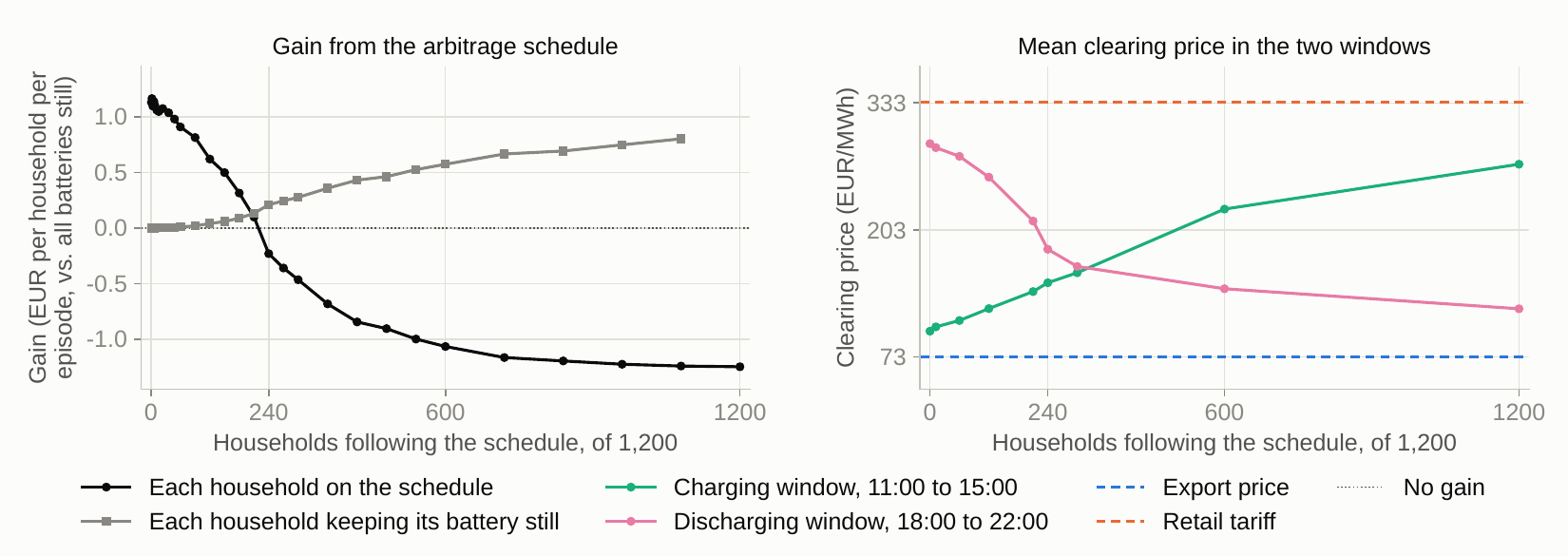}
\caption{
On the 64 held-out episodes, $k$ households follow a fixed arbitrage schedule (charge 11:00--15:00, discharge 18:00--22:00) and the rest keep their batteries idle. Left: gain per household, for those arbitraging and those idle, relative to all batteries idle. Right: mean clearing price in the charging and discharging windows. The horizontal axis of both panels is $k$. Arbitrage pays while few households take part and turns into a loss at 210 to 240 households, before the two window prices cross.
}
\label{fig.appendix.p2p.crowding}
\end{figure}

This section asks how the gain from arbitrage changes with the number of households arbitraging at the same time.
The fixed schedule of Section~\ref{sec.results.p2p} charges at 2.62~kW from 11:00 to 15:00 and discharges at 2.62~kW from 18:00 to 22:00, with all households offering truthfully; it is replayed on the 64 held-out episodes with $k$ households on the schedule and the rest idle (Figure~\ref{fig.appendix.p2p.crowding}).
The gain is the change in profit per household per episode (EUR) relative to all batteries idle.

Arbitrage pays only when few households take part, as shown in the left subfigure in Figure~\ref{fig.appendix.p2p.crowding}. A single arbitraging household gains 1.129~EUR per episode, but the gain falls as more households join, turns negative at 210 to 240 households, and reaches $-1.248$~EUR when all 1\,200 take part. The households that keep their batteries idle benefit instead: with 600 households arbitraging, each idle household gains 0.573~EUR. 

More specifically, arbitrage closes the price spread it relies on as shown in the right subfigure in Figure~\ref{fig.appendix.p2p.crowding}: charging at noon raises the noon price, and discharging in the evening lowers the evening price. Once about 240 households take part, the remaining spread no longer covers the round-trip losses and the degradation cost. This saturation comes from the market, not from the learner. For a single agent (household) the price is exogenous, but as more agents follow the same strategy, their actions change the price, which becomes endogenous. A profit measured for one agent, or against historical prices, therefore overstates what the strategy earns when many adopt it.

\FloatBarrier

\subsection{M5: Local Flexibility Market}
\label{app:results:flex}

In this appendix section, we provide detailed market simulation settings and the discussion of the performance of the four learners.

\subsubsection{Market setting}
\label{app:results:flex:data}

\begin{figure}[h!]
\centering
\includegraphics[width=\linewidth]{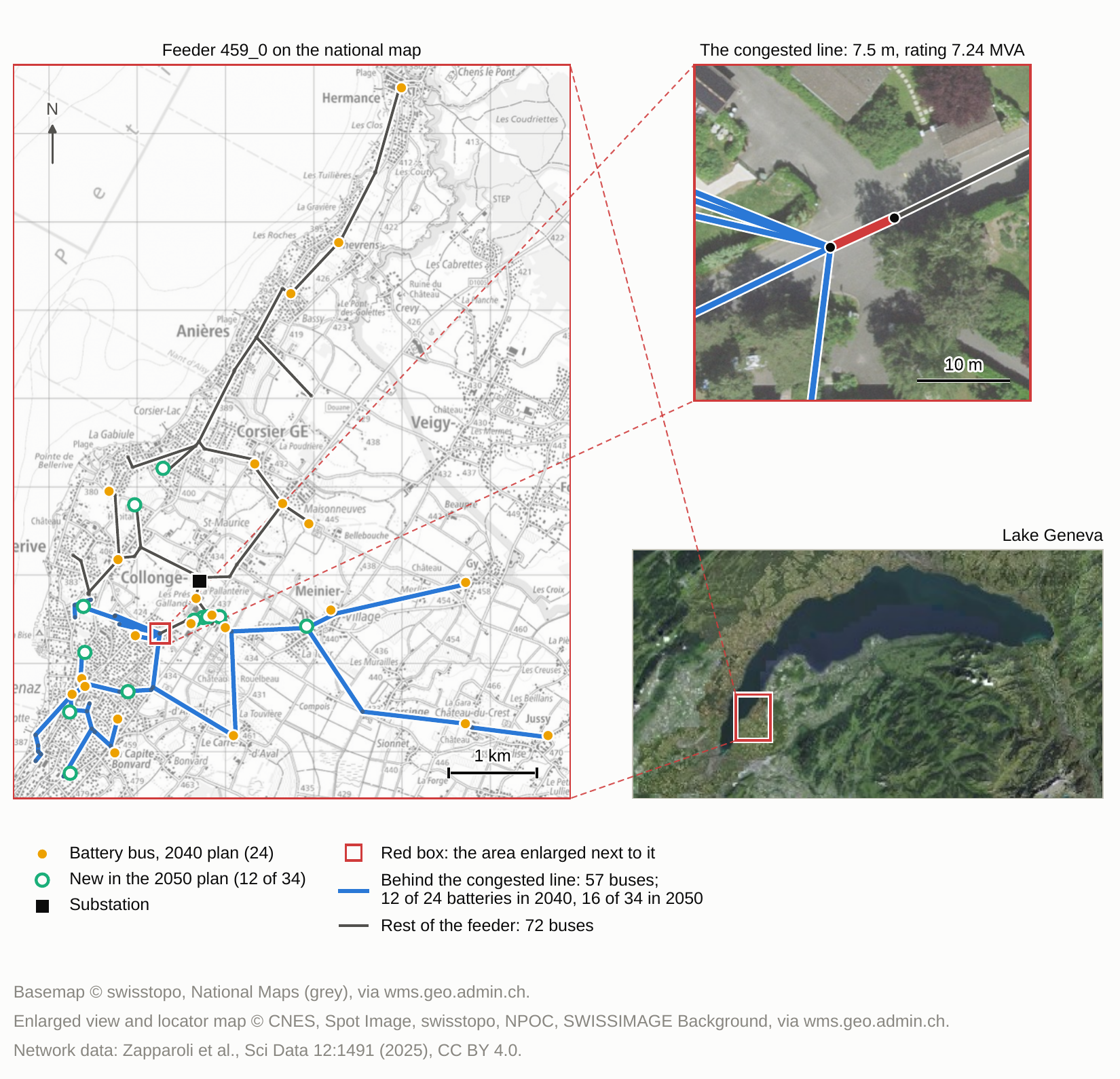}
\caption{
Feeder 459\_0 in Switzerland. Bottom right: its location in the Lake Geneva region. Left: the feeder, with a small red box marking the congested line. Top right: an aerial image of the congested line. Lines behind the congested line are blue, and only batteries on these lines can relieve it.
}
\label{fig.appendix.flex.map}
\end{figure}

The test system is the Swiss distribution feeder 459\_0, as shown in Figure~\ref{fig.appendix.flex.map}. The system has 129 buses and 128 lines, and we scale its load by 1.50. 
We consider an episode as one day, and the market is cleared hour by hour. Each aggregator operates one battery, and aggregator $i$ submits three quantity bids at period $t$, including a price $\pi_{i,t}$, an offered quantity, and planned charging. The distribution system operator (the operator) clears the market by minimising the cost of satisfying demand, as detailed in Appendix~\ref{app:flex}.
Each winning aggregator is paid its own offer. The price floor is set to the replacement cost $c^{\mathrm{rep}} = 149.88$~CHF/MWh.

The system requires that the bus voltage stays above 0.9 per unit and that the flow of every line keeps within its maximum value.
The operator buys flexibility to keep voltage within its range and the congested line within its rating. The requirement is how much the flow on the line would exceed 98\% of its rating if the operator bought nothing. This flow includes the load and the charging the aggregators plan, so the aggregators can raise the requirement themselves as detailed in Section~\ref{app:results:flex:requirement}. In the Swiss test system, voltage stays within its limits all year, so the line is the only reason to buy.

\begin{figure}[h!]
\centering
\includegraphics[width=\linewidth]{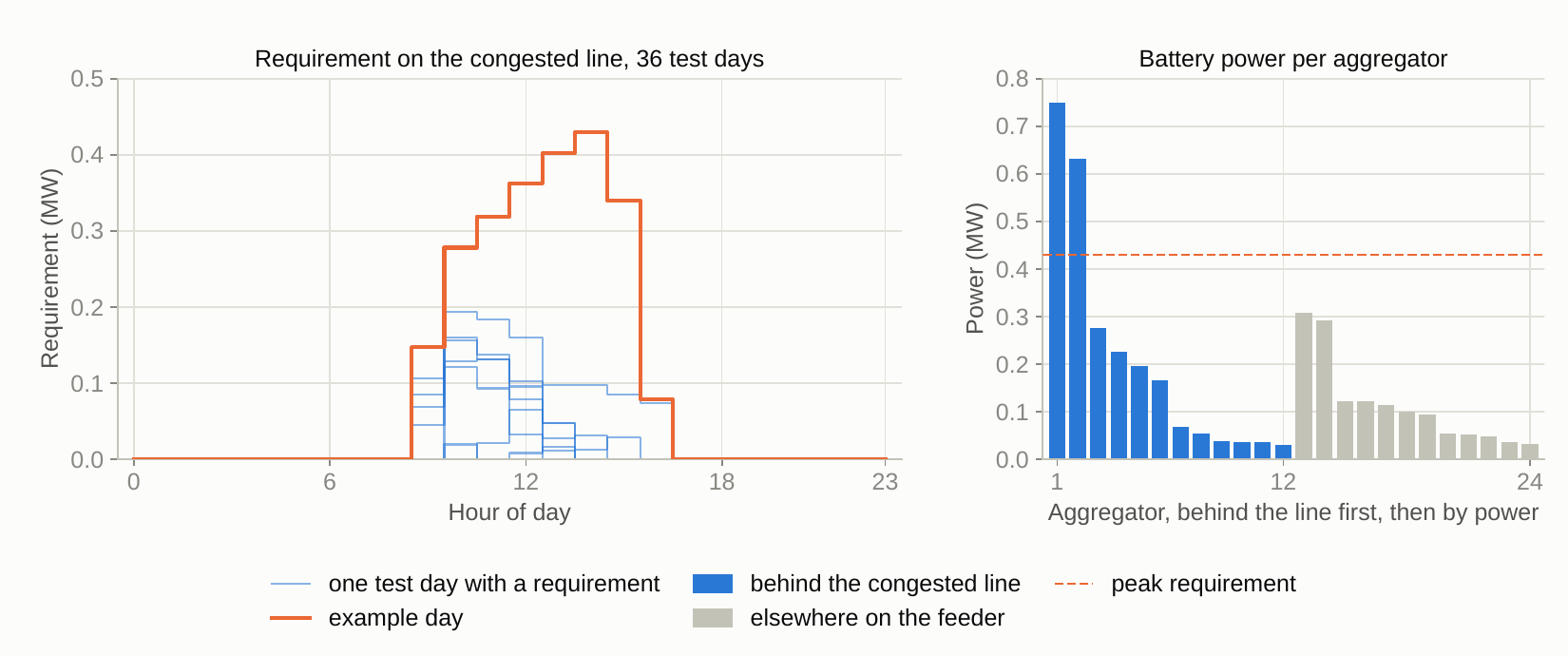}
\caption{
Demand and supply of flexibility. Left: the requirement in each hour of the 36 test days, with all batteries idle. Right: the power of each of the 24 batteries in the 2040 placement, with the 12 downstream of the congested line first. 
}
\label{fig.appendix.flex.requirement}
\end{figure}

The flexibility market trades much less than the P2P energy market, because it buys only what is needed to keep the network within its limits. With all batteries idle, it buys on only 9 of the 36 test days, and never more than 0.43~MW, as shown in the left subfigure of Figure~\ref{fig.appendix.flex.requirement}. Only batteries downstream of the line can relieve it: 12 of 24 in the 2040 placement and 16 of 34 in the 2050 placement. The 12 batteries alone can supply 2.51~MW, almost six times over the largest requirement as shown in the right subfigure of Figure~\ref{fig.appendix.flex.requirement}. Supply therefore far exceeds demand.

\begin{table}[h!]
  \caption{Overview of Swiss test system data.}
  \label{tab:app-flex-data}
  \centering
  \footnotesize
  \begin{tabularx}{\linewidth}{>{\raggedright\arraybackslash}p{0.12\linewidth}>{\raggedright\arraybackslash}p{0.14\linewidth}>{\raggedright\arraybackslash}p{0.15\linewidth}>{\raggedright\arraybackslash}p{0.08\linewidth}YY}
    \toprule
    Data & Source & Licence & Training days & Validation days & Test days \\
    \midrule
    Swiss distribution feeder 459\_0 & SwissDN\textsuperscript{a}, Zapparoli et al. (2025)\textsuperscript{b} & CC BY 4.0 & 293 days & 36 days (the 5th, 15th and 25th of each month) & 36 days (the 1st, 11th and 21st of each month) \\
    Energy price (category C2, 2026) & ElCom\textsuperscript{c} & opendata.swiss Open use & --- & --- & --- \\
    \bottomrule
  \end{tabularx}
  \par\smallskip
  \parbox{\linewidth}{\scriptsize\raggedright \textsuperscript{a}\,\url{https://doi.org/10.5281/zenodo.15056134}; \textsuperscript{b}\,\url{https://doi.org/10.1038/s41597-025-05830-y}; \textsuperscript{c}\,\url{https://energy.ld.admin.ch/elcom/electricityprice}.}
\end{table}

\subsubsection{Agent training}
\label{app:results:flex:training}
\label{app:results:flex:setting}

The training setup is that of Appendix~\ref{app:train}, and IPPO differs from Table~\ref{tab:train-hp} only in the following: each iteration takes 4 full-batch gradient steps, the learning rate of $3 \times 10^{-4}$ decays linearly to zero over 300 iterations, the log standard deviation of exploration is annealed linearly from $-0.5$ to $-3.0$ over the same span, and the initialisation scale of the actor output layer is 0.01 rather than 1.0. Training uses the training days of Table~\ref{tab:app-flex-data}; the stopping rule and the checkpoint evaluated on the test days are given in Table~\ref{tab:train-scale}. On the test days the selected checkpoint acts with sampled actions, IPPO at a log standard deviation of $-3.0$, the end point of its annealing, and SAC at the log standard deviation of its own actor.

\begin{figure}[h!]
\centering
\includegraphics[width=\linewidth]{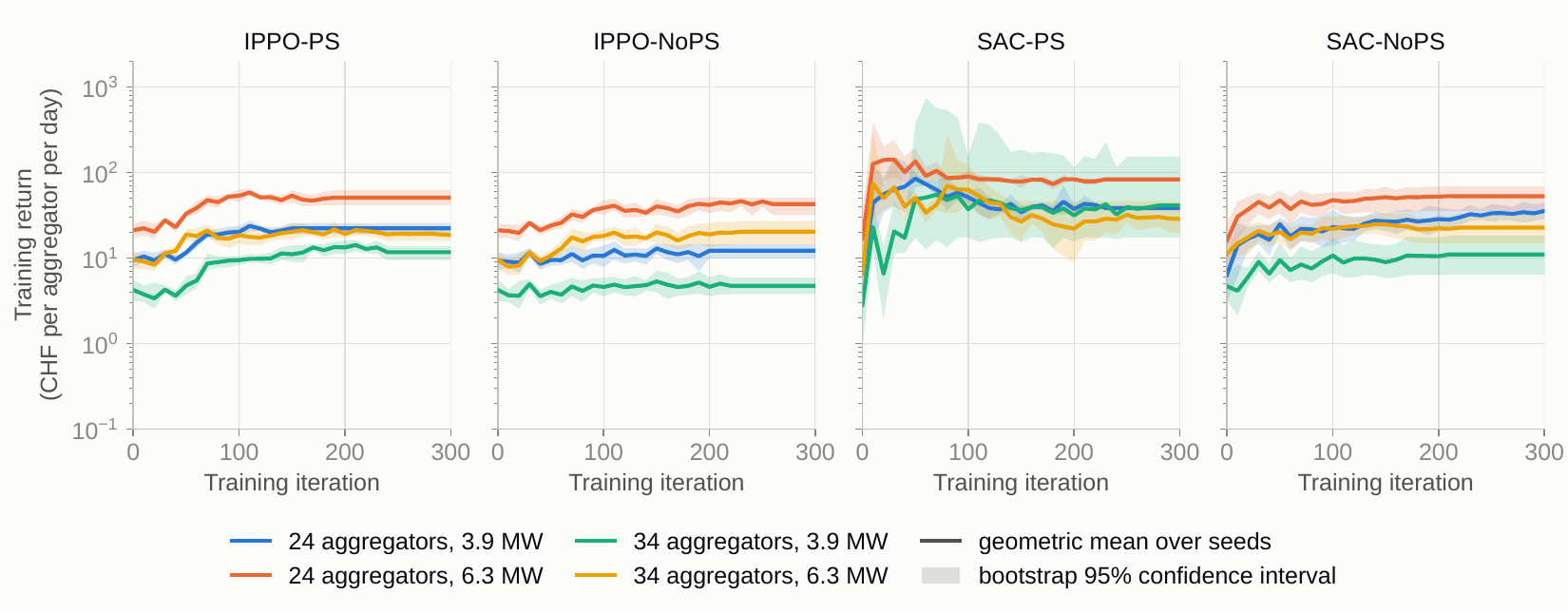}
\caption{Training return of the four learners against training iteration; lines are the geometric mean over the seeds of each configuration and shaded bands the bootstrap 95\% confidence interval, each seed held at its last value after it stops. All four learners reach a plateau within about 100 iterations.}
\label{fig.appendix.flex.training}
\end{figure}

\subsubsection{Analysis of the procurement requirement}
\label{app:results:flex:requirement}

\begin{figure}[h!]
\centering
\includegraphics[width=\linewidth]{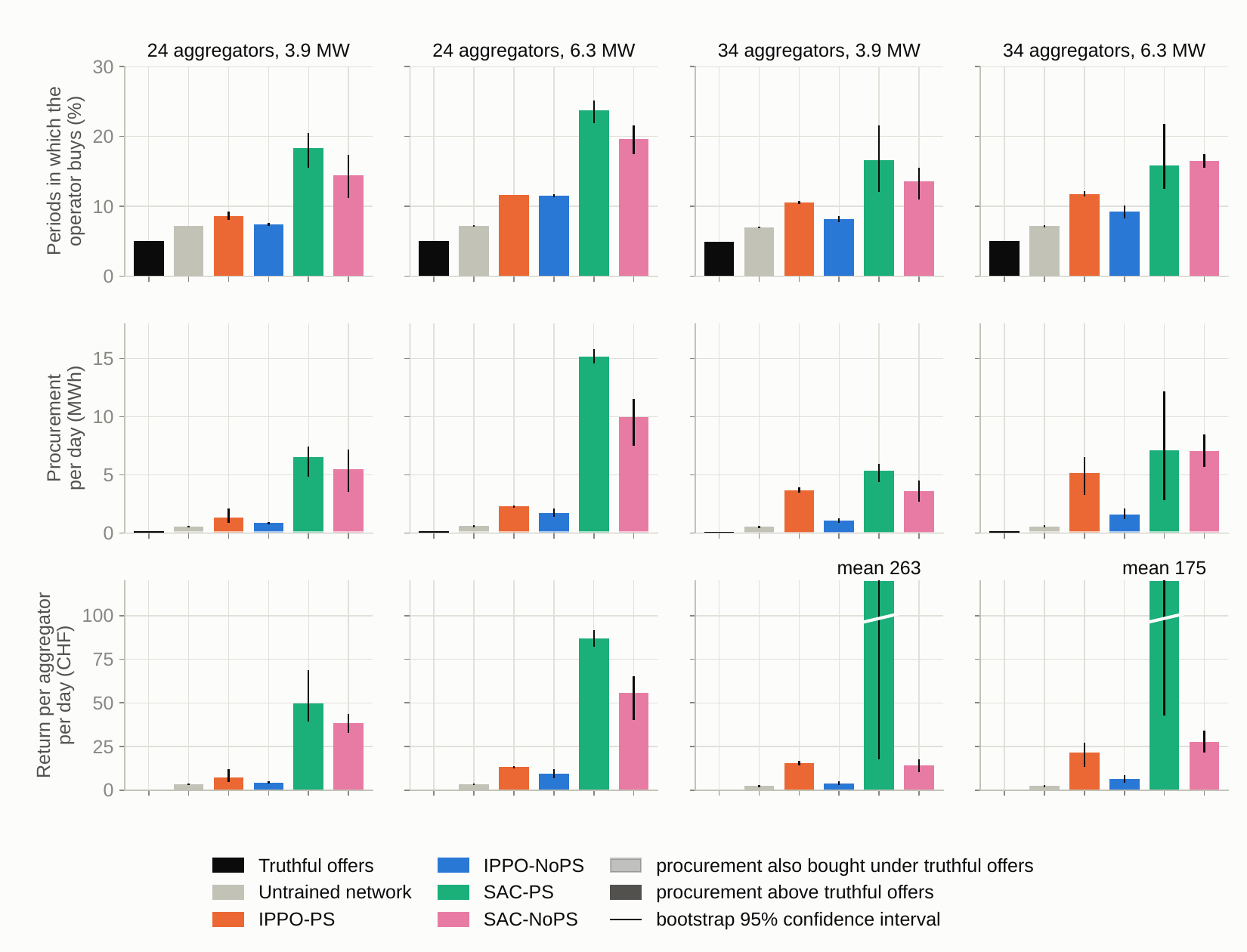}
\caption{
Source of the flexibility requirement over the 36 test days of the four configurations. Bars show truthful offers, the untrained shared and per-agent networks pooled in one bar, and the four learners at validation-selected checkpoints (seed mean and bootstrap 95\% confidence interval). Two SAC-PS return bars exceed the axis and are cut at its top, with their means shown above. Rows show the share of periods with procurement; daily procurement split into what truthful offers also buy and the excess; and daily return per aggregator. Learners raise procurement above the truthful-offer level, and their returns rise with it.
}
\label{fig.appendix.flex.source}
\end{figure}

Under truthful offers, the operator buys flexibility in about 5\% of periods: 43 of 864 hours across the 36 test days in the 2040 placement, as shown in Figure~\ref{fig.appendix.flex.requirement}. All four learners increase this share. More than 80\% of daily procurement lies above what truthful offers buy, and aggregator returns track procurement as shown in Figure~\ref{fig.appendix.flex.source}. The excess comes from planned charging, which raises baseline flow on the congested line above 98\% of its rating and leads the operator to buy flexibility from batteries behind it. Cleared prices remain near the floor: on the test days, all IPPO seeds and 75\% of SAC seeds clear within 1\% of it.

In the two 3.9~MW configurations, replacing truthful-offer aggregators with SAC-PS learners raises the share of periods with flexibility procurement from 5.0\% to 18.7\% with 24 aggregators, and from 4.9\% to 15.4\% with 34 aggregators (Figure~\ref{fig.flex}(c)). With the monitor on, all six paired runs in the 34-aggregator configuration show daily procurement falling from about 3.2 to 0.15~MWh and daily return per learner falling from about 118 to 1~CHF.

\subsubsection{Analysis of a unilateral markup}
\label{app:results:flex:unilateral}
\begin{figure}[h!]
\centering
\includegraphics[width=\linewidth]{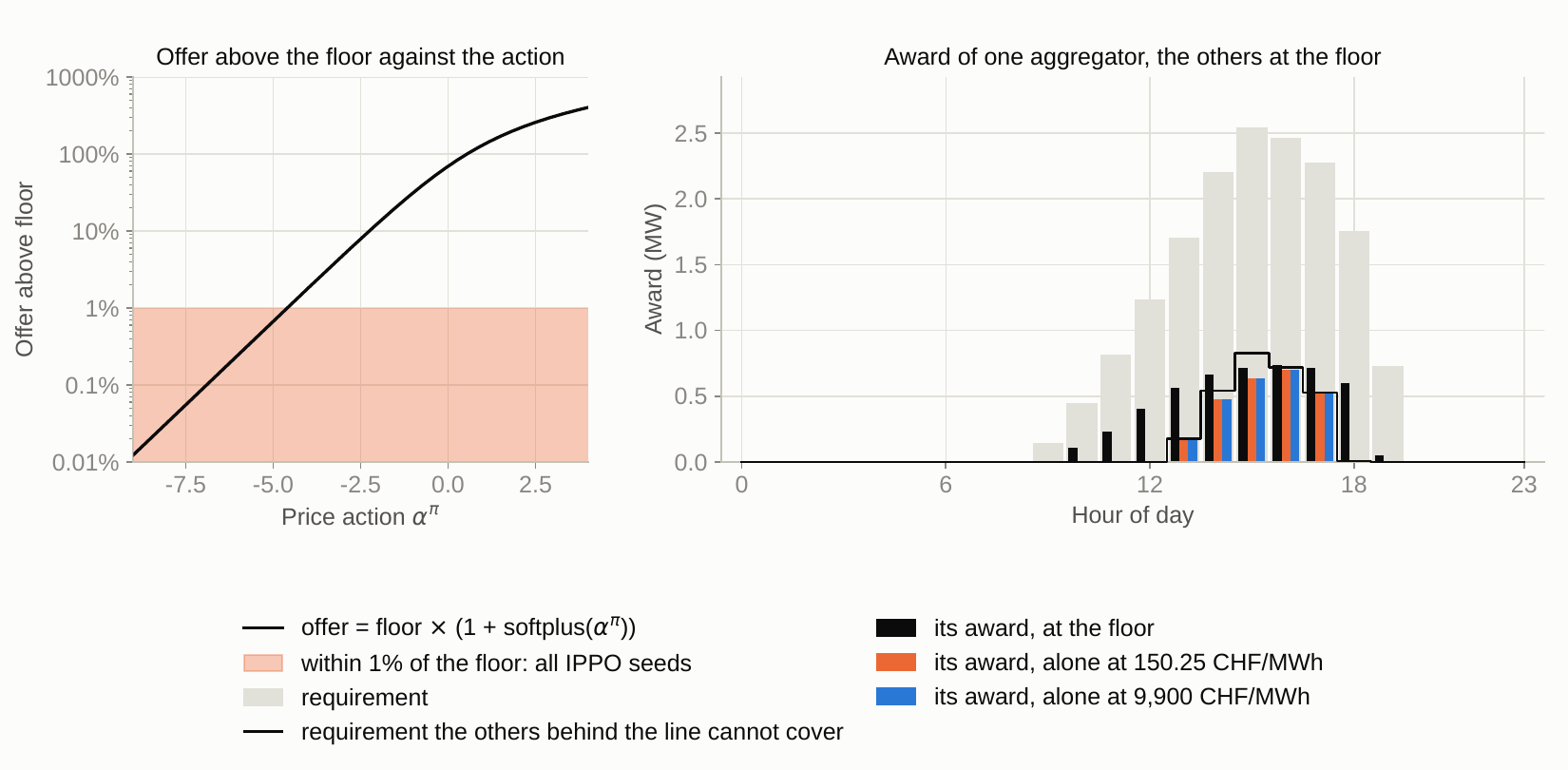}
\caption{
Unilateral offer increase. Left: offer above the floor against price action (log scale). Near the floor, the gap shrinks by about a factor of $e$ for each unit decrease in the action. Cleared prices for all IPPO seeds are within 1\% of the floor. Right: aggregator 7's hourly award on the example day at offers of the floor, 150.25 and 9\,900~CHF/MWh, with all other aggregators at the floor.
}
\label{fig.appendix.flex.cliff}
\end{figure}

In the 24-aggregator, 3.9~MW configuration, aggregator 7's return from a unilateral offer rises with the offer up to the highest offer swept, 9\,900~CHF/MWh, as shown in Figure~\ref{fig.flex}(b). With all others at the floor, raising its offer to 150.25~CHF/MWh (0.25\% above the floor) lets the cheaper aggregators behind the congested line clear first. On the example day, aggregator 7 is awarded only from 13:00 to 18:00, when their offers cannot meet the requirement  as shown in the right subfigure of Figure~\ref{fig.appendix.flex.cliff}. Its award over the 36 test days falls from 8.86 to 2.53~MWh and remains at 2.53~MWh even at 9\,900~CHF/MWh. Its return therefore drops sharply (just above the floor), then rises almost linearly with its offer. For the curves in Figure~\ref{fig.flex}(b), return exceeds its floor value again at about 481 and 489~CHF/MWh in the two 3.9~MW configurations, but only at about 4\,590~CHF/MWh in the 34-aggregator, 6.3~MW configuration.

Near the floor, the bidding offer becomes nearly flat: $\pi=c^{\mathrm{rep}}[1+\mathrm{softplus}(\alpha^\pi)]$ gives $d\pi/d\alpha^\pi=c^{\mathrm{rep}}\mathrm{sigmoid}(\alpha^\pi)$, which approaches zero as $\alpha^\pi$ decreases as shown in the left subfigure of Figure~\ref{fig.appendix.flex.cliff}. This weakens the return gradient reaching the price action and helps explain why learners stay near the floor.

\FloatBarrier

\end{document}